\documentclass[review]{elsarticle}
\usepackage{lineno,hyperref}
\usepackage{graphicx}
\usepackage{subfig}
\usepackage[linesnumbered,ruled,lined]{algorithm2e}
\usepackage{float}
\usepackage{tikz}
\usepackage{epstopdf}
\usepackage{amssymb}
\usepackage{amsmath}
\journal{Comput. Methods Appl. Mech. Engrg. }
\newcommand{\tr}{\mathop{\mathrm{tr}}}
\newcommand{\erfc}{\mathop{\mathrm{erfc}}}
\newcommand{\sign}{\mathop{\mathrm{sign}}}
\begin{document}
%
%
\begin{frontmatter}
		\title{An integrative smoothed particle hydrodynamics framework for modeling cardiac function}
		\author{Chi Zhang}
		\ead{c.zhang@tum.de}
		\author{Jianhang Wang}
		\ead{jianhang.wang@tum.de}
		\author{Massoud Rezavand}
		\ead{massoud.rezavand@tum.de}
		\author{Dong Wu}
		\ead{dong.wu@tum.de}
		\author{Xiangyu Hu \corref{mycorrespondingauthor}}
		\cortext[mycorrespondingauthor]{Corresponding author.}
		\ead{xiangyu.hu@tum.de}
		\address{Department of Mechanical Engineering,  Technical University of Munich, 85748 Garching, Germany}
\begin{abstract}
Mathematical modeling of cardiac function 
can provide augmented simulation-based diagnosis tool for complementing and extending human understanding of cardiac diseases 
which represent the most common cause of worldwide death. 
As the realistic starting-point for developing an unified meshless approach for total heart modeling, 
herein we propose an integrative smoothed particle hydrodynamics (SPH) framework for addressing  the simulation of 
the principle aspects of cardiac function, 
including cardiac electrophysiology, passive mechanical response and electromechanical coupling. 
To that end, several algorithms, 
e.g., 
splitting reaction-by-reaction method combined with quasi-steady-state (QSS) solver , 
anisotropic SPH-diffusion discretization  and total Lagrangian SPH formulation,
are introduced and exploited for dealing with the fundamental challenges of developing integrative SPH framework for simulating cardiac function, 
namely, (i) the correct capturing of the stiff dynamics of the transmembrane potential and the gating variables , 
(ii) the stable predicting of the large deformations and the strongly anisotropic behavior of the myocardium, 
and (iii)  the proper coupling of electrophysiology and tissue mechanics for electromechanical feedback. 
A set of numerical examples demonstrate the effectiveness and robustness of the present SPH framework, 
and render it a potential and powerful alternative that can augment current lines of total cardiac modeling and clinical applications. 
\end{abstract}
\end{frontmatter}
%
%
\section{Introduction}\label{sec:introduction}
The heart is one of our not only most vital, 
but also most complex organs.
The four chambers and four valves act precisely in concert to regulate the heart's filling, 
ejecting and overall pump function, 
by the interplay of electrical and mechanical (including solid and fluid) dynamics. 
Cardiac diseases which effect the heart through complex mechanisms represent one of the most important category of problems in public health, 
effecting millions of people each year according to the reports of World Health Organization (WHO) \cite{who}. 
Mathematical modeling of the heart and its function can complement and expand our understanding of cardiac diseases 
and the clinical practice of cardiology \cite{quarteroni2017cardiovascular}. 
In cardiac research, 
computational modeling and simulation have received tremendous efforts and are recognized as the community's \textit{next microscope, only better} \cite{trayanova2011whole}. 
However, the integrative model capable of simulating the fully coupled cardiac function is still in its infancy. 
While the state-of-art computational models are able to simulate the coupled electromechanics or the coupled fluid-solid dynamics, 
they face serious difficulties on integrating all the three dynamics due to the conflicts 
between their limited modeling flexibility with respect to the complex physical processes involved. 

To date, 
there are mainly two computational approaches \cite{hunter2003modeling} that 
developed fro integrative cardiac modeling, viz. 
the finite-element method (FEM) \cite{zienkiewicz1977finite} and the immersed-boundary method (IBM) \cite{peskin2002immersed}. 
However, 
the ability of FEM is hindered by the coupling between 
the solid and fluid mechanics. 
Typical difficulties include the treatment of the convective terms, 
the incompressibility constraint 
and the updating of the mesh; especially when the opening and closing of valves are taken into account. 
The IBM has been developed to compute FSI problems, 
which are the main difficulty for the FEM modeling. 
However, the fairly weak coupling formulation of distributing the forces computed on the deformable Lagrangian mesh 
to the Eulerian mesh using a kernel function in IBM can lead to 
the Lagrangian-Eulerian mismatches on the kinematics.
Such difficulty becomes very serious when the material properties and 
active stress are complex as the ones in cardiac myocardium.
An alternative approach, 
the meshless methods, 
e. g. smoothed particle hydrodynamics (SPH) \cite{lucy1977numerical, gingold1977smoothed, hu2006multi, zhang2017weakly}, 
has shown peculiar advantages in handling multi-physics and multi-scale problems \cite{zhang2017generalized, rezavand2020weakly, zhang2020dual, zhang2019multi, liu2019smoothed, bian2015multi}.
These advantages render the SPH method an interesting and potent alternative in the integrative simulation of cardiac function.

As a realistic starting-point for developing an integrative meshless approach for cardiac modeling, 
the main objective of this work is to present a SPH framework for simulating the fundamental and indispensable components,   
e. g. the cardiac electrophysiology, 
passive mechanical response and electromechanical coupling (active mechanical response), 
of cardiac function. 
The cardiac electrophysiology describes the myocardial electrical activation sequence, 
based on the ion currents and tissue conductivity \cite{quarteroni2017cardiovascular}. 
The cardiac fibers contract due to the propagation of electrical stimuli initiated in the sinoatrial node.
This electrical stimuli produces a sharp rise (depolarization) followed by a sudden fall (repolarization) of the transmembrane potential. 
This phenomenon can be mathematically modeled by means of a reaction-diffusion equation where the source term encapsulates the cellular ion exchange.
In this paper, 
we consider the simple monodomain approach derived with the assumption of equal anisotropic conductivities in the intra- and extra-cellular compartments. 
The monodomain approach has been widely used for three-dimensional simulations considering ionic models ranging from 
simple FitzHugh-Nagumo variants \cite{fitzhugh1961impulses, aliev1996simple} to the more complex Luo-Rudy model \cite{franzone2014mathematical}.
Regarding the cardiac passive mechanical response, 
finite elasticity models are needed to describe cardiac contraction and relaxation due to the fact that the cardiac cells change in length by up to $20-30\%$ during a physiological contraction. 
Furthermore, 
the suitable elasticity model should replicate the anisotropic passive behavior determined via a set of collagen fibers and sheets duo to the extremely complex and heterogeneous material property of cardiac myocardium. 
In this work,  
we consider the classic invariant-based presentation of the strain energy proposed by Holzapfel and Ogden \cite{holzapfel2009constitutive} for the characterization of the passive mechanical response of the cardiac myocardium. 
The Holzapfel-Ogden model in which the cardiac myocardium is treated as a non-homogeneous, thick-walled and nonlinearly elastic material is a structural based model that accounts for the muscle fiber direction and the myocyte sheet structure. 
The electromechanical coupling can be phenomenologically described by means of activation models. 
In general, 
two approaches, namely, active stress \cite{nash2004electromechanical} and active strain \cite{taber2000modeling}, 
can be followed for the definition of activation models. 
Here, we consider the active stress \cite{nash2004electromechanical} approach with an evolution equation for active cardiomyocite concentration stress \cite{wong2011computational}. 

As the first attempt towards a integrative meshless model for cardiac modeling, 
the proposed SPH framework should accurately characterize its interesting critical aspects, 
including electrophysiology, passive and active mechanical responses. 
At first, 
we adopt an operator splitting scheme for the reaction-diffusion equation to split reaction and diffusion to ensure numerical stability and accuracy. 
This consideration also leads to a much larger time step size compared to the simple forward Euler method \cite{quarteroni2017cardiovascular}.  
Then, 
we introduce a splitting reaction-by-reaction method \cite{wang2019split} 
combined with quasi-steady-state (QSS) solver 
to capture the stiff dynamics of the transmembrane potential and the gating variables 
of the ionic model governed by nonlinear ordinary differential equations (ODEs). 
Furthermore, 
an anisotropic SPH discretization for diffusion equation derived by Tran-Duc et al. \cite{tran2016simulation} is modified 
by introducing a linear operator to improve the computational efficiency and using a correction kernel matrix to improve the numerical accuracy.
The total Lagrangian SPH formulation is employed to predict the large deformations and the strongly anisotropic behavior of the myocardium. 
Ultimately, 
the proposed SPH discretization for monodomain equation is integrated to predict the active response of myocardium 
by implementing the active stress approach \cite{nash2004electromechanical}. 
A comprehensive set of numerical examples, 
viz. 
benchmarks on iso- and aniso-tropic diffusion, 
the transmembrane potential propagation in the manner of free-pulse and spiral wave, 
passive and active responses of myocardium, 
and electrophysiology and electromechanics in a generic biventricular heart, 
are computed to demonstrate the accuracy, robustness and feasibility of the proposed SPH framework. 
Base on the present developments and the previous achievements of the SPH method \cite{zhang2019multi, zhang2017weakly, rezavand2020weakly}, the proposed framework
will shed light on the multi-physics and multi-scale total cardiac modeling, in particular 
with regards to the fluid-electro-structure interactions. 
For a better comparison and future openings for in-depth studies, all the computational codes and data-sets accompanying this work are available online at \url{https://github.com/Xiangyu-Hu/SPHinXsys}. 

This manuscript is organized as follows. 
Section \ref{sec:kinematics} introduces the basic principles of the kinetics 
and the governing equations describing the evolution of the transmembrane potential and the mechanical response. 
Section \ref{sec:constitutiveequations} presents the constitutive laws with respect to the monodomain approach, 
the passive and the active electromechanical responses. 
In Section \ref{sec:method}, the proposed SPH framework is fully described. 
A comprehensive set of examples are included in Section \ref{sec:experiments} and 
the concluding remarks and a summary of the key contributions of this paper are given in Section \ref{sec:conclusion}. 

%
%
\section{Kinematics and governing equations}
\label{sec:kinematics}
To characterize the kinematics of the finite deformation, 
the deformation map $\varphi$ which maps a material point $\mathbf{\mathbf{r}^0}$ from the initial reference configuration $\Omega^0 \subset \mathbb{R}^d $ 
to the point $\mathbf{r} = \mathbf{\varphi}\left(\mathbf{r}^0, t\right)$ 
in the deformed configuration $\Omega = \mathbf{\varphi} \left(\Omega^0\right)$ is introduced. 
In this work, 
we will use superscript $\left( {\bullet} \right)^0$ to denote the quantities in the initial reference configuration. 
The deformation tensor $\mathbb{F}$ can then be defined by its derivative with respect to the initial reference configuration as 
\begin{equation} \label{eq:deformationtensor}
\mathbb{F} = \nabla^{0} {\varphi} =  \frac{\partial \varphi}{\partial \mathbf{r}^0}  = \frac{\partial \mathbf{r}}{\partial \mathbf{r}^0} .
\end{equation}
Note that the deformation tensor $\mathbb{F}$ can also be calculated from the displacement $\mathbf{u} = \mathbf{r} - \mathbf{r}^0$ through
\begin{equation} \label{eq:deformationtensor-displacement}
\mathbb{F} = \nabla^{0} {\mathbf{u}}  - \mathbb{I},
\end{equation}
where $\mathbb{I}$ represents the unit matrix. 
For an incompressible material, we have the constraint
\begin{equation} \label{eq:incompressible}
J = \text{det}\left(\mathbb{F}\right) \equiv 1. 
\end{equation}
Associated with $\mathbb{F}$ are the right and left Cauchy-Green deformation tensors defined by 
\begin{equation} \label{eq:cauchy-green}
\mathbb{C} = \mathbb{F}^{T} \cdot \mathbb{F}\quad \text{and}\quad \mathbb{B} = \mathbb{F} \cdot \mathbb{F}^{T}, 
\end{equation}
respectively. 
Then, four typical invariants of $\mathbb{C}$ (and also of $\mathbb{B}$) can be defined as
\begin{align}\label{eq:invariants}
\mathit{I}_{I} &= \tr \left( \mathbb{C} \right) , & \mathit{I}_{ff} &= \mathbf{f}^0 \cdot \left( \mathbb{C} \mathbf{f}^0 \right), \nonumber \\
\mathit{I}_{ss} &= \mathbf{s}^0 \cdot \left( \mathbb{C} \mathbf{s}^0 \right), & \mathit{I}_{fs} &= \mathbf{f}^0 \cdot \left( \mathbb{C} \mathbf{s}^0 \right) ,
\end{align}
where $\mathbf{f}^0$ and $\mathbf{s}^0$  are the undeformed myocardial fiber and sheet unit direction, respectively. 
Here, $\mathit{I}_{I} $ is the first principal invariant, 
structure-based invariants $\mathit{I}_{ff} $ and $\mathit{I}_{ss} $ are the isochoric fiber and sheet stretch squared as the squared lengths of the deformed fiber and sheet vectors, 
i. e. $\mathbf{f} = \mathbb{F} \mathbf{f}^0 $ and $\mathbf{s} = \mathbb{F} \mathbf{s}^0$, 
while $\mathit{I}_{fs}$ indicates the fiber-sheet shear \cite{holzapfel2009constitutive}.

We consider a coupled system of partial differential equations (PDEs) governing the motion of the material point $\mathbf{r}$ and the evolution of the transmembrane potential $V_m$.  
The time dependent evolution of the transmembrane potential in Lagrangian framework is characterized by the normalized monodomain equation \cite{quarteroni2017cardiovascular}
\begin{equation}\label{eq:monodomain}
C_m \frac{\text{d}V_m}{\text{d}t} = \nabla \cdot  \left(\mathbb{D} \nabla V_m\right)  + I_{ion}  \quad \text{in}  \quad \Omega^0 \times \left[0, T \right],
\end{equation}
where $C_m$ is the capacitance of the cell membrane and $I_{ion}$ the ionic current. 
Note that the conductivity tensor is defined by 
$\mathbb{D} = d^{iso} \mathbb{I} + d^{ani} \mathbf{f}_0 \otimes \mathbf{f}_0$ 
with $ d^{iso}$ denoting the isotropic contribution 
and $d^{ani}$ the anisotropic contribution to account for faster conductivity along fiber direction $\mathbf{f}^0$. 

In a total Lagrangian framework, 
the conservation of the mass and linear momentum corresponding to the cardiac mechanics can be expressed as
\begin{equation}\label{eq:mechanical-mom}
\begin{cases}
\rho =  {\rho_0} \frac{1}{J} \quad \\
\rho^0 \frac{\text{d} \mathbf{v}}{\text{d} t}  =  \nabla^{0} \cdot \mathbb{P}^T \quad  
\end{cases} \Omega^0 \times \left[0, T \right],
\end{equation}
where $\rho$ is the density and $\mathbb{P}$ the first Piola-Kirchhoff stress tensor 
and $\mathbb{P} =  \mathbb{F} \mathbb{S}$ with $\mathbb{S}$ denoting the second Piola-Kirchhoff stress tensor. 
Note that the body force is neglected in Eq. \eqref{eq:mechanical-mom}. 
%
%
\section{Constitutive equations}
\label{sec:constitutiveequations}
To close the systems of Eqs.\eqref{eq:monodomain} and \eqref{eq:mechanical-mom}, 
we specify herein the constitutive laws for the ionic current $I_{ion}$ and the first Piola-Kirchhoff stress $\mathbb{P}$. 
\subsection{Cardiac electrophysiolgoy: monodomain approach}
\label{subsec:monodomain}
To close the monodomain equation Eq. \eqref{eq:monodomain},  
a model for the ionic current is required. 
Following Refs \cite{fitzhugh1961impulses, franzone2014mathematical}, 
we consider the so-called reduced-ionic model in which $I_{ion}(V_m, w)$ 
is a function of the trasnmembrane potential $V_m$ and the gating variable $w$ 
which represents the percentage of the open channels per unit area of the membrane. 
The most widely used reduced-ionic model is the Fitzhugh-Nagumo model \cite{fitzhugh1961impulses} and the variant Aliev-Panfilow model \cite{aliev1996simple}
which only have two currents, viz.
inward and outward, one gating variable and no explicit ionic concentration variables. 

The Fitzhugh-Nagumo model reads \cite{fitzhugh1961impulses} 
\begin{equation}\label{eq:fhn}
\begin{cases}
I_{ion}(V_m, w) = -V_m(V_m - a)(V_m - 1) - w \\
\dot{w} = g(V_m, w) = \epsilon_0 ( \beta V_m - \gamma w - \sigma)
\end{cases},
\end{equation}
where $\epsilon_0$, $\beta$, $\gamma$ and $\sigma$ are suitable constant parameters will be given specifically. 

As a variant of  Fitzhugh-Nagumo model, 
the Aliev-Panfilow model \cite{aliev1996simple} has been successfully implemented in previous simulations of ventricular fibrillation in real geometries \cite{panfilov1999three}
and it is particularly suitable for applications where electrical activity of the heart is the main interest. 
The Aliev-Panfilow model reads
\begin{equation}\label{eq:a-p}
\begin{cases}
I_{ion}(V_m, w) = -k V_m(V_m - a)(V_m - 1) - w V_m \\
\dot{w} = g(V_m, w) = \epsilon(V_m, w)(-k V_m(V_m - b - 1) - w)
\end{cases},
\end{equation}
where $\epsilon(V_m, w) = \epsilon_0 + \mu_1 w / (\mu_2 + V_m)$ and $k$, $a $, $b$,  $\epsilon_0$, $\mu_1$ and $\mu_2$ are suitable constant parameters to be fixed later. 

Note that both Fitzhugh-Nagumo and Aliev-Panfilow models involve dimensionless variable $V_m$, $w$ and $t$. 
The actual transmembrane potential $E$ in dimension $mV$ and time $T$ in dimension $ms$ can be obtained through \cite{aliev1996simple}
\begin{equation}\label{eq:a-p-dim}
\begin{cases}
E = 100V_m - 80\\
T = 12.9t
\end{cases}.
\end{equation}
\subsection{Cardiac electromechanics: passive and active mechanical response}
Following the work of Nash and Panfilov \cite{nash2004electromechanical}, 
we couple the stress tensor with the transmembrane potential $V_m$ through the active stress approach 
which decomposes the first Piola-Kirchhoff stress $\mathbb{P}$ into passive and active parts 
\begin{equation}
\mathbb{P} = \mathbb{P}_{passive} + \mathbb{P}_{active}. 
\end{equation}
Here,  
the passive component $\mathbb{P}_{passive}$ describes the stress required to obtain a given deformation of the passive myocardium, 
and an active component  $\mathbb{P}_{active}$ denotes the tension generated by the depolarization 
of the propagating transmembrane potential. 

For the passive mechanical response,    
we consider the Holzapfel-Odgen model which proposed the following strain energy function, considering different contributions and taking the anisotropic nature of the myocardium into account.
To ensure that the stress vanishes in the reference configuration and encompasses the finite extensibility, 
we modify the strain-energy function as 
\begin{eqnarray}\label{eq:new-muscle-energy}
\mathbf{W}  & = &  \frac{a}{2b}\exp\left[b (I_1 - 3 )\right] - a \ln J  + \frac{\lambda}{2}(\ln J)^{2} \nonumber \\
& + & \sum_{i = f,s} \frac{a_i}{2b_i}\{\text{exp}\left[b_i\left(\mathit{I}_{ii}- 1 \right)^2\right] - 1\} \nonumber \\
& + & \frac{a_{fs}}{2b_{fs}}\{\text{exp}\left[b_{fs}\mathit{I}^2_{fs} \right] - 1\} ,
\end{eqnarray}
where $a$, $b$, $a_f$, $b_f$, $a_s$, $b_s$, $a_{fs}$ and $b_{fs}$ are eight positive material constants, 
with the $a$ parameters having dimension of stress and $b$ parameters being dimensionless. 
Here, the second Piola-Kirchhoff stress $\mathbb{S}$ being defined by
\begin{equation}\label{eq:second-PK}
\mathbb{S} = 2 \frac{\partial \mathbf{W}}{\partial \mathbb{C}} -p\mathbb{C}^{-1} = 2\sum_{j} \frac{\partial \mathbf{W}}{\partial \mathit{I}_j} \frac{\partial \mathit{I}_j}{\partial \mathbb{C}} -p\mathbb{C}^{-1} \quad j = I,ff,ss,fs,
\end{equation}
where 
\begin{align}\label{eq:second-PK-2}
\frac{\partial \mathit{I}_1}{\partial \mathbb{C}} &=  \mathbb{I},
& \frac{\partial \mathit{I}_{ff}}{\partial \mathbb{C}} &= \mathbf{f}_0 \otimes \mathbf{f}_0, 
& \frac{\partial \mathit{I}_{ss}}{\partial \mathbb{C}} &= \mathbf{f}_0 \otimes \mathbf{f}_0, 
& \frac{\partial \mathit{I}_{fs}}{\partial \mathbb{C}} &= \mathbf{f}_0 \otimes \mathbf{s}_0 + \mathbf{s}_0 \otimes \mathbf{f}_0 ,
\end{align}
and $p$ is the Lagrange multiplier arising from the imposition of incompressibility. 
Substituting Eqs. \eqref{eq:second-PK} and \eqref{eq:second-PK-2} into Eq.\eqref{eq:new-muscle-energy} the second Piola-Kirchhoff stress is given as
\begin{eqnarray}
\mathbb{S} & = & a \text{exp} \left[b\left({\mathit{I}}_{I} - 3 \right)\right] + \left\{ \lambda\ln J - a \right\}\mathbb{C}^{-1} \nonumber\\ 
& + & 2a_f \left({\mathit{I}}_{f}- 1 \right)  \text{exp}\left[b_f\left({\mathit{I}}_{f}- 1 \right)^2\right] \mathbf{f}_0 \otimes \mathbf{f}_0  \nonumber \\
& + & 2a_s \left({\mathit{I}}_{s}- 1 \right)  \text{exp}\left[b_s\left({\mathit{I}}_{s}- 1 \right)^2\right] \mathbf{s}_0 \otimes \mathbf{s}_0  \nonumber \\
& + & a_fs {\mathit{I}}_{fs} \text{exp}\left[b_fs\left({\mathit{I}}_{fs}\right)^2\right] \mathbf{fs}_0 \otimes \mathbf{fs}_0 .
\end{eqnarray}

The cardiac electrical activation stem from two processes: 
the generation of ionic currents which produces the transmembrane potential at the microscopic scales 
and the traveling of the transmembrane potential from cell to cell at the macroscopic scales. 
The propagation of the transmembrane potential can be described by means of PDEs, 
suitably coupled with ODEs governing the ionic currents. 
In particular, 
a monodomain equation can be defined with the continuum assumption of the coexistence of extra- and intra-cellular information 
at every point. 
Following the active stress approach proposed by Nash and Panfilov \cite{nash2004electromechanical}, 
the active component provides the internal active contraction stress by
\begin{equation}
\mathbb{P}_{active} = T_a \mathbb{F} \mathbf{f}_0 \otimes \mathbf{f}_0,
\end{equation}
where $T_a$ represents the active magnitude of the stress and its evolution is given by an ODE as 
\begin{equation}
\dot{T_a} = \epsilon\left(V_m\right)\left[k_a\left(V_m - {V}_r \right) - T_a\right], 
\end{equation}
where parameters $k_a$ and ${V}_r$ control the maximum active force, the resting transmembrane  potential 
and the activation function \cite{wong2011computational}
\begin{equation}
\epsilon\left(V_m \right) = \epsilon_0 + \left(\epsilon_{\infty} -\epsilon_{-\infty} \right) \text{exp}\{-\text{exp}\left[-\xi\left(V_m - \overline{V}_m\right)\right]\}.
\end{equation} 
Here, the limiting values $\epsilon_{-\infty}$ at $V_m \rightarrow -\infty$ and $\epsilon_{\infty}$ at $V_m \rightarrow \infty$, 
the phase shift $\overline{V}_m$ and the transition slope $\xi$ will ensure a smooth activation of the muscle traction. 
%
%
\section{SPH method for cardiac eletrophysiology and electromechanics}
\label{sec:method}
In this section, 
the proposed SPH method for cardiac eletrophysiology, passive mechanical response and the electromemchanical coupling is presented. 
\subsection{Fundamentals of SPH}
Before moving on to the SPH discretization, 
we first briefly summarize the theory and fundamentals of the SPH method. 
For more details the readers are referred to the comprehensive review in Ref. \cite{monaghan1992smoothed}.

By introducing a Dirac delta function $\delta( \mathbf{r} - \acute{\mathbf{r}})$ around $\mathbf{r}$ , 
a continuous function $f(\mathbf{r})$ and its approximation, 
i.e., the smoothing kernel function $W(\mathbf{r}-\acute{\mathbf{r}},h)$ with smoothing length $h$ defining the support domain, 
has the relation
\begin{equation}\label{integral}
f(\mathbf{r}) =\int_{\Omega} f(\acute{\mathbf{r}}) \delta(\mathbf{r}- \acute{\mathbf{r}})d\acute{\mathbf{r}} \approx \int_{\Omega} f(\acute{\mathbf{r}})W(\mathbf{r}-\acute{\mathbf{r}},h)d\acute{\mathbf{r}}, 
\end{equation}
where $\Omega$ denotes the volume of the integral domain.
Here, 
the introduction of smoothing kernel function \cite{hu2006multi, wendland1995piecewise} establishes a discrete model 
due to the finite size of the smoothed length $h$.
From Eq. \eqref{integral} the gradient of function $f$ can be approximated by 
\begin{equation}\label{grad-pe}
\nabla f(\mathbf{r}) \approx \int_{\Omega} \nabla f(\acute{\mathbf{r}})W(\mathbf{r}-\acute{\mathbf{r}},h) d V(\acute{\mathbf{r}}) .
\end{equation}
Integrating by parts of Eq. (\eqref{grad-pe}) and applying Gauss theorem yields
\begin{equation}\label{grad-pe-gaus}
\nabla f(\mathbf{r}) \approx \int_{\partial\Omega} f(\acute{\mathbf{r}}) W(\mathbf{r}
-\acute{\mathbf{r}},h) \mathbf{n} d S(\acute{\mathbf{r}})-
\int_{\Omega} f(\acute{\mathbf{r}})\nabla W(\mathbf{r}-\acute{\mathbf{r}},h) d V(\acute{\mathbf{r}}) .
\end{equation}
If the computational domain is discretized by a set of particles, 
the gradient of $f$ can be approximated as in SPH form as the first term vanishes due to compact support of the kernel function
in the right hand side of Eq. \eqref{grad-pe-gaus}
\begin{equation}\label{eq:grad-sph}
\nabla f(\mathbf{r})- \sum^{N}_{j=1} \frac{m_j}{\rho _j}f(\mathbf{r}_j) \nabla _i W({\mathbf{r}_i - \mathbf{r}_j}, h) .
\end{equation}
Note that ${m_i}/{\rho_i}$ is defined to express the differential volume element $dV_i$.
\subsection{SPH discretization of monodomain equation}
As mentioned in Section \ref{subsec:monodomain}, 
the monodomain equation consists of a coupled system of PDE and ODE. 
The former governs the diffusion of the transmembrane potential and the latter the reactive kinetics of the gating variable. 
In this paper, 
we employ the operator splitting method \cite{quarteroni2017cardiovascular} which results in a PDE of anisotropic diffusion
\begin{equation}\label{eq:diffusion}
C_m \frac{\text{d}V_m}{\text{d}t}  = \nabla \cdot (\mathbb{D} \nabla V_m), 
\end{equation}
and two ODEs 
\begin{equation}\label{eq:ode-system}
 \begin{cases}
 C_m \frac{\text{d}V_m}{\text{d}t}  = I_{ion}(V_m, w) \\
 \frac{\text{d}w}{\text{d}t} = g(V_m, w) 
 \end{cases},  
\end{equation}
where $I_{ion}(V_m, w)$ and $ g(V_m, w)$ are defined by FitzHugh-Nagumo Eq. \eqref{eq:fhn} or Aliev-Panfilow model Eq. \eqref{eq:a-p}. 
\subsubsection{Discretization of anisotropic diffusion equation}
Different from the previous strategies for the discretization of diffusion equation \cite{biriukov2018stable,rezavand2019fully}, 
we employ and modify the anisotropic SPH dicretization proposed by Tran-Duc et al. \cite{tran2016simulation}. 
Following Ref. \cite{tran2016simulation},  
the diffusion tensor $\mathbb{D}$ is considered to be a symmetric positive-definite matrix and can be decomposed by Cholesky decomposition as 
\begin{equation}\label{eq:chol}
\mathbb{D} = \mathbb{L} \mathbb{L}^T
\end{equation}
where $\mathbb{L}$ is a lower triangular matrix with real and positive diagonal entries and $\mathbb{L}^T$ denotes the transpose of $\mathbb{L}$. 
The diffusion operator in Eq. \eqref{eq:diffusion} can be rewritten to isotropic form by 
\begin{equation}\label{eq:diffusion-trans}
\nabla \cdot (\mathbb{D} \nabla) =  \nabla \cdot (\mathbb{\mathbb{L} \mathbb{L}^T} \nabla) 
=  (\mathbb{L}^T \nabla)^T \cdot (\mathbb{L}^T\nabla) = \widetilde{\nabla}^2, 
\end{equation}
where $ \widetilde{\nabla} = \mathbb{L}^T \nabla$.
Then, the new isotropic diffusion operator is approximated by the following kernel integral with neglecting the high-order term
\begin{equation}\label{eq:diffusion-int}
\widetilde{\nabla} \cdot (\widetilde{\nabla})  \Phi =   2 \int_{\Omega} \frac{\Phi (\widetilde{\mathbf{r}}) - \Phi (\widetilde{\mathbf{r}}^{'}) }{|\widetilde{\mathbf{r}} - \widetilde{\mathbf{r}}^{'}|} \frac{\partial W\left( \widetilde{\mathbf{r}} - \widetilde{\mathbf{r}}^{'}, \widetilde{h}\right) }{\partial | \widetilde{\mathbf{r}} - \widetilde{\mathbf{r}}^{'} | } d \widetilde{\mathbf{r}}, 
\end{equation}
where $\widetilde{\mathbf{r}} = \mathbb{L}^{-1} \mathbf{r}$ and $\widetilde{h} = \mathbb{L}^{-1} h$ . 
Upon the coordinate transformation, 
the kernel gradient can be rewritten as
\begin{equation}\label{eq:diffusion-kernel-trans}
 \frac{\partial W\left( \widetilde{\mathbf{r}} - \widetilde{\mathbf{r}}^{'}, \widetilde{h}\right) }{\partial \left( \widetilde{\mathbf{r}} - \widetilde{\mathbf{r}}^{'}\right) } =
 \frac{1}{|\mathbb{L}^{-1}||\mathbb{L}^{-1} \mathbf{e}_{\widetilde{\mathbf{r}\mathbf{r}}|}} \frac{\partial W \left( {\mathbf{r}} - {\mathbf{r}}^{'}\right)}{\partial |{\mathbf{r}} - {\mathbf{r}}^{'}|}
\end{equation}
with $\mathbf{e}_{{\mathbf{r}}^{'} \mathbf{r}} = \frac{{\mathbf{r}^{'}} - \mathbf{r}}{|{\mathbf{r}}^{'} - \mathbf{r}|}$. 
At this stage, 
Eq. \eqref{eq:diffusion-trans} can be discretized in SPH form as 
\begin{equation}\label{grad-laplace}
\begin{split}
\widetilde{\nabla}^2 \Phi  & \approx  2 \sum_{j}^{N} \frac{m_j}{\rho_j} \bigg(\Phi(\mathbf{r_i}) - \Phi(\mathbf{r_j}) \bigg) \frac{1}{(\overline{\mathbb{L}}_{ij}^{-1}\mathbf{e}_{ij})^2} \frac{1}{r_{ij}} \frac{\partial W_{ij}}{\partial r_{ij}}
\end{split},
\end{equation}
where $\mathbf{e}_{ij} = \frac{\mathbf{r}_{ij}}{{r}_{ij}}$, $\overline{\mathbb{D}}_{ij} = \overline{\mathbb{L}}_{ij} \overline{\mathbb{L}}_{ij}^T $ 
and $\overline{\mathbb{D}}_{ij} =\frac{\mathbb{D}_i \mathbb{D}_j}{\mathbb{D}_i + \mathbb{D}_j} $, 
which ensure the antisymmetric property of the physical diffusion phenomenon. 
Note that Eq. \eqref{grad-laplace} is excessive computational expensive due to the fact that one time of Cholesky decomposition and the corresponding matrix inverse is required for each pair of particle interaction. 
To optimize the computational efficiency, 
we modify Eq. \eqref{grad-laplace} by replacing the term $\mathbb{L}_{ij}^{-1}$  with its linear approximation 
$\widetilde{\mathbb{L}}_{ij}$ given by 
\begin{equation}
\widetilde{\mathbb{L}}_{ij} = \frac{\widetilde{\mathbb{L}}_{i}\widetilde{\mathbb{L}}_{j} }{\widetilde{\mathbb{L}}_{i} + \widetilde{\mathbb{L}}_{j} }
\end{equation}
where $\widetilde{\mathbb{L}}_{i}$ is defined as 
\begin{equation}
\widetilde{\mathbb{D}}_{i} = \left( \widetilde{\mathbb{L}}^{-1}_{i}\right)  \left( \widetilde{\mathbb{L}}^{-1}_{i}\right) ^T .
\end{equation}
In this case, 
the Cholesky decomposition and the corresponding matrix inverse are computed once for each particle before the simulation. 
Also note that Eq. \eqref{grad-laplace} can be further improved by introducing a kernel correction matrix to improve the numerical accuracy 
which will be detailed in the following section.  
\subsubsection{Reaction-by-reaction splitting}
The system of ODEs defined by Eq. \eqref{eq:ode-system} are generally stiff, 
therefore numerical instability occurs where the integration time step is not sufficiently small.  
In this work, 
we employ a reaction-by-reaction splitting method proposed by Wang et al. \cite{wang2019split}. 
The multi-reaction system can be decoupled, e.g.  second-order accurate Strange splitting, as
\begin{equation}\label{eq:ode-spliting-2rd}
R^{(\Delta t)} = R_V^{(\frac{\Delta t}{2})} \circ R_w^{(\frac{\Delta t}{2})} \circ R_w^{(\frac{\Delta t}{2})} \circ R_V^{(\frac{\Delta t}{2})},
\end{equation}
where the $ \circ $ symbol separates each reaction and 
indicates that the operator $R_V^{(\Delta t)}$ is applied after $R_w^{(\Delta t)}$. 
Note that the reaction-by-reaction splitting methodology can be extended to more complex ionic models, 
e.g. the Tusscher-Panfilov model \cite{ten2004model}. 
 
Following Ref. \cite{wang2019split}, we rewrite an ODE in Eq. \eqref{eq:ode-system} in the following form 
\begin{equation}\label{eq:ode-new-form}
  \frac{\text{d} y}{\text{d} t} = q(y,t) - p(y,t) y, 
\end{equation}
where $q(y,t)$ is the production rate and $ p(y,t) y$ is the loss rate \cite{wang2019split}.
The general form of Eq. \eqref{eq:ode-new-form}, where the analytical solution is not explicitly known or difficult to derive, 
can be solved by using the quasi-steady-state (QSS) method for an approximate solution as
\begin{equation}\label{eq:ode-qss}
 y^{n + 1} = y^n e^{-p(y^n, t) \Delta t } + \frac{q(y^n, t) }{p(y^n, t)} \left(1 - e^{-p(y^n, t) \Delta t} \right).
\end{equation}
Note that QSS-based method is unconditionally stable due to the analytic form, 
and thus a larger time step is allowed for the splitting method, 
leading to a higher computational efficiency. 
\subsection{Total Lagrangian formulation}
\label{sec:total-sph}
The elastic response of the soft myocardium is highly nonlinear and their deformation under working load are intrinsically large, 
therefore a robust numerical method is required. 
In this work, 
we adopt the total Lagrangian SPH formulation, 
i.e., the initial reference configuration is used for finding the neighboring particles and 
the set of neighboring particles is not altered, 
to ensure the first-order consistency and eliminates the tensile-instability, 

Following the work of Vignjevic et al. \cite{vignjevic2006sph}, 
a correction matrix $\mathbb{B}^0$ is first introduced as
\begin{equation} \label{eq:sph-correctmatrix}
 \mathbb{B}^0_i = \left( \sum_j V_j \left( \mathbf{r}^0_j - \mathbf{r}^0_i \right) \otimes \nabla^0_i W_{ij} \right) ^{-1} ,
\end{equation}
where 
\begin{equation}\label{strongkernel}
 \nabla^0_i W_{ij} = \frac{\partial W\left( |\mathbf{r}^0_{ij}|, h \right)}  {\partial |\mathbf{r}^0_{ij}|} \mathbf{e}^0_{ij}
\end{equation}
stands for the gradient of the kernel function evaluated at the initial reference configuration. 
Again, 
the correction matrix is computed in the initial configuration and therefore, 
it is calculated only once before the simulation. 
Using Eqs. \eqref{eq:grad-sph} and \eqref{eq:sph-correctmatrix}, 
the linear momentum conservation equation, Eq.\eqref {eq:mechanical-mom}, can be discretized in the following form 
\begin{equation}\label{eq:sph-mechanical-mom}
\ddot{\mathbf{r}}_i = \frac{2}{m_i} \sum_j V_i V_j \tilde{\mathbb{P}}_{ij} \nabla^0_i W_{ij},
\end{equation} 
where the inter-particle averaged first Piola-Kirchhoff stress $\tilde{\mathbb{P}}$ is given as
\begin{equation}
 \tilde{\mathbb{P}}_{ij} = \frac{1}{2} \left( \mathbb{P}_i \mathbb{B}^0_i + \mathbb{P}_j \mathbb{B}^0_j \right). 
\end{equation}
Here, 
the first Piola-Kirchhoff stress tensor is computed with the constitutive law where the deformation tensor $\mathbb{F}$ is computed by
\begin{equation}
\mathbb{F}_i = \left( \sum_j V_j \left( \mathbf{u}_j - \mathbf{u}_i \right) \otimes \nabla^0_i W_{ij}  \right) \mathbb{B}^0_i - \mathbb{I} .
\end{equation}

It worth noting that, we apply the renormalization kernel correction to improve the accuracy and consistency of  Eq. \ref{grad-laplace} as
\begin{equation}\label{grad-laplace-renormalize}
C_m \dot V_m  = 2 \sum_{j} \frac{m_j}{\rho_j} \bigg(\mathbb{B}^0_i V_m(\mathbf{r_i}) - \mathbb{B}^0_j V_m (\mathbf{r_j}) \bigg) \cdot \frac{\mathbf{e}_{ij} \cdot \mathbf{e}_{ij}}{(\widetilde{\mathbb{L}}_{ij} \mathbf{e}_{ij})^2} \frac{1}{r_{ij}} \frac{\partial W_{ij}}{\partial r_{ij}} .
 \end{equation}
\subsection{Implementation}
Here we describe the details of the implementation of the proposed SPH framework 
for integrating the monodomain equation with mechanical response of myocardium. 
To maintain the numerical stability, 
the time step size for solving the monodomain equation is restricted by the diffusion coefficient 
\begin{equation}\label{eq:dt_p}
\Delta t_{p} = 0.5\left( \frac{h^2}{d|\mathbb{D}|} \right), 
\end{equation}
where $d$ is the dimension number and $|\mathbb{D}|$ the trace of the diffusion tensor. 
For the passive elastic response, 
he Courant-Friedichs-Levy (CFL) condition is given as
\begin{equation}\label{eq:dt_m}
\Delta t_{m}   =  0.6 \min\left(\frac{h}{c + |\mathbf{v}|_{max}},
\sqrt{\frac{h}{|\frac{\text{d}\mathbf{v}}{\text{d}t}|_{max}}} \right) .
\end{equation}
The final time step size is chosen from
\begin{equation}\label{eq:dt}
\Delta t   = \text{min} \left(\Delta t_{p}, \Delta t_{m} \right). 
\end{equation}

We denote the values at the beginning of a time step by the superscript $n$, 
at the mid-point by $n + \frac{1}{2}$ and eventually at the end of the time-step by $n + 1$.
Following the splitting method,  
the transmembrane potential $V_m$ and the gate variable $w$ are first updated in sequence for a half time step as
\begin{equation}
\begin{cases}
V_m^{n + \frac{1}{2}} = V_m^{n} + \frac{1}{2} \Delta t_p \left( \frac{\text{d}V_m}{\text{d}t}\right)_{reaction} \\
w^{n + \frac{1}{2}} = w^{n} + \frac{1}{2} \Delta t_p \left( \frac{\text{d}w}{\text{d}t}\right)_{reaction} 
\end{cases}. 
\end{equation}
Here, 
the diffusive operator is applied and the transmembrane potential $V_m$ is updated for a time step 
\begin{equation}
V_m^{*} = V_m^{n+\frac{1}{2}} + \Delta t_p \left( \frac{\text{d}V_m}{\text{d}t}\right)_{diffusion}. 
\end{equation}
Then the transmembrane potential $V_m$ and the gate variable $w$ are updated in inverse sequence  for another half time step as 
\begin{equation}
\begin{cases}
w^{n + \frac{1}{2}} = w^{n} + \frac{1}{2} \Delta t_p \left( \frac{\text{d}w}{\text{d}t}\right)_{reaction}  \\
V_m^{n + \frac{1}{2}} = V_m^{*} + \frac{1}{2} \Delta t_p \left( \frac{\text{d}V_m}{\text{d}t}\right)_{reaction}
\end{cases}. 
\end{equation}
At this point, 
the active cardiomyocite contraction stress $T_a$ is updates for one time step if active response is taken into consideration. 
Following Ref. \cite{zhang2019multi}, 
a position-based Verlet scheme is applied for the time integration of the mechanical response. 
At first, the deformation tensor, density and particle position are updated to the midpoint as 
\begin{equation}\label{eq:verlet-first-half-solid}
\begin{cases}
\mathbb{F}^{n + \frac{1}{2}} = \mathbb{F}^{n} + \frac{1}{2} \Delta t \frac{\text{d} \mathbb{F}}{\text{d}t}\\
\rho^{n + \frac{1}{2}} = \rho^0 \frac{1}{J} \\
\mathbf{r}^{n + \frac{1}{2}} = \mathbf{r}^{n} + \frac{1}{2} \Delta t {\mathbf{v}}^n
\end{cases}. 
\end{equation}
Then the velocity is updated by
\begin{equation}\label{eq:verlet-first-mediate-solid}
\mathbf{v}^{n + 1} = \mathbf{v}^{n} +  \Delta t  \frac{d \mathbf{v}}{dt}. 
\end{equation}
Finally, the deformation tensor and position of solid particles are updated to the new time step of the solid structure with 
\begin{equation}\label{eq:verlet-first-final-solid}
\begin{cases}
\mathbb{F}^{n + 1} = \mathbb{F}^{n + \frac{1}{2}} + \frac{1}{2} \Delta t \frac{\text{d} \mathbb{F}}{\text{d}t}\\
\rho^{n + 1} = \rho^0 \frac{1}{J} \\
\mathbf{r}^{n + 1} = \mathbf{r}^{n + \frac{1}{2}} + \frac{1}{2} \Delta t {\mathbf{v}}^{n + 1}
\end{cases}. 
\end{equation}
An overview of the complete solution strategy is presented in Algorithm \ref{al:algorithm1}  in Appendix  A.
%
%
\section{Numerical experiments}\label{sec:experiments}
This section is devoted to present a comprehensive set of numerical examples for validating 
the integrative SPH framework for the simulation of cardiac function with respect to electrophysioldoy, 
passive mechanical response and the electromechanical coupling.  
We start with the benchmarks on both iso- and aniso-tropic diffusion process.
We then validate the present method for cardiac electrophysiology 
by solving the monodomain equation on regular and irregular computational domain with iso- and aniso-tropic diffusion coefficients. 
Then the accuracy, robustness and applicability of the total Lagrangian SPH method for modeling the passive and active mechanical responses of myocardium are validated.
Having the validation studies presented,
the excitation and excitation-contraction of three-dimensional generic biventricular heart are to show the potential of the proposed SPH framework. 
In all the following simulations,
the $5th$-order Wendland smoothing kernel function \cite{wendland1995piecewise} with a smoothing length of $h = 1.3~dp$ is employed, where $dp$ denotes the initial particle spacing.
\subsection{Isotropic diffusion}
\begin{figure}[htb!]
	\centering
	\includegraphics[trim = 1mm 50mm 1mm 50mm, clip, width=.85\textwidth]{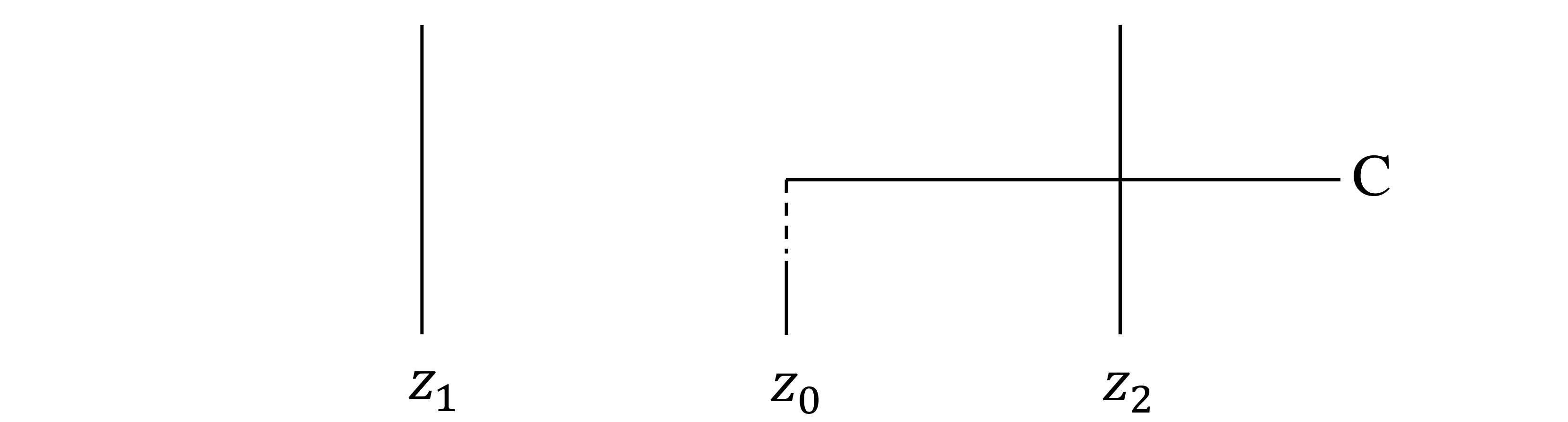}
	\caption{Schematic diagram for the one-dimensional diffusion problem with constant initial distribution of the pollutant concentration.}
	\label{figs:iso-diffusion-setup}
\end{figure}
\begin{figure}[htb!]
	\centering
	\subfloat[constant initial concentration]{\includegraphics[width=.49\textwidth]{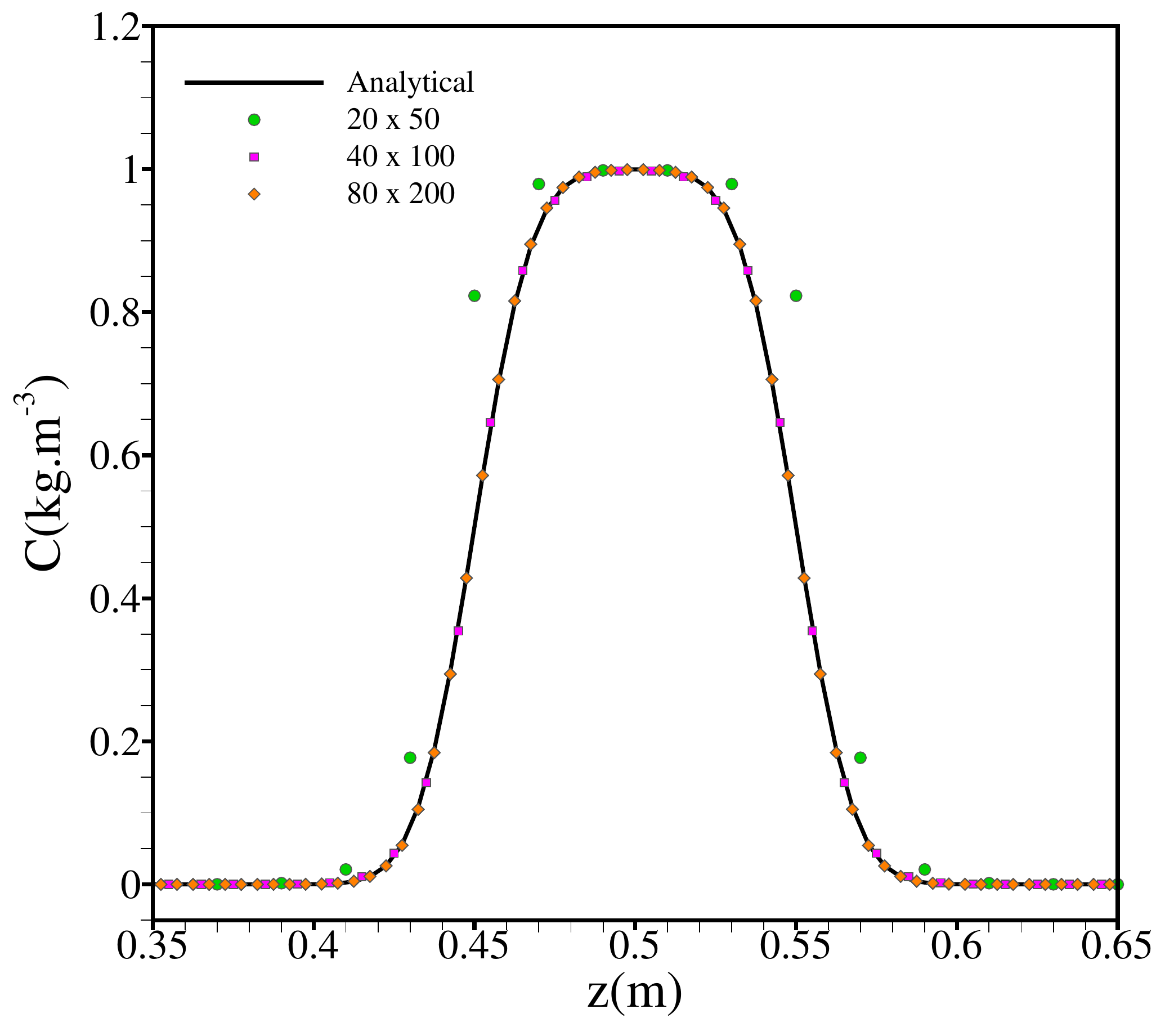} }
	\subfloat[exponential intial concentration]{ \includegraphics[width=.49\textwidth]{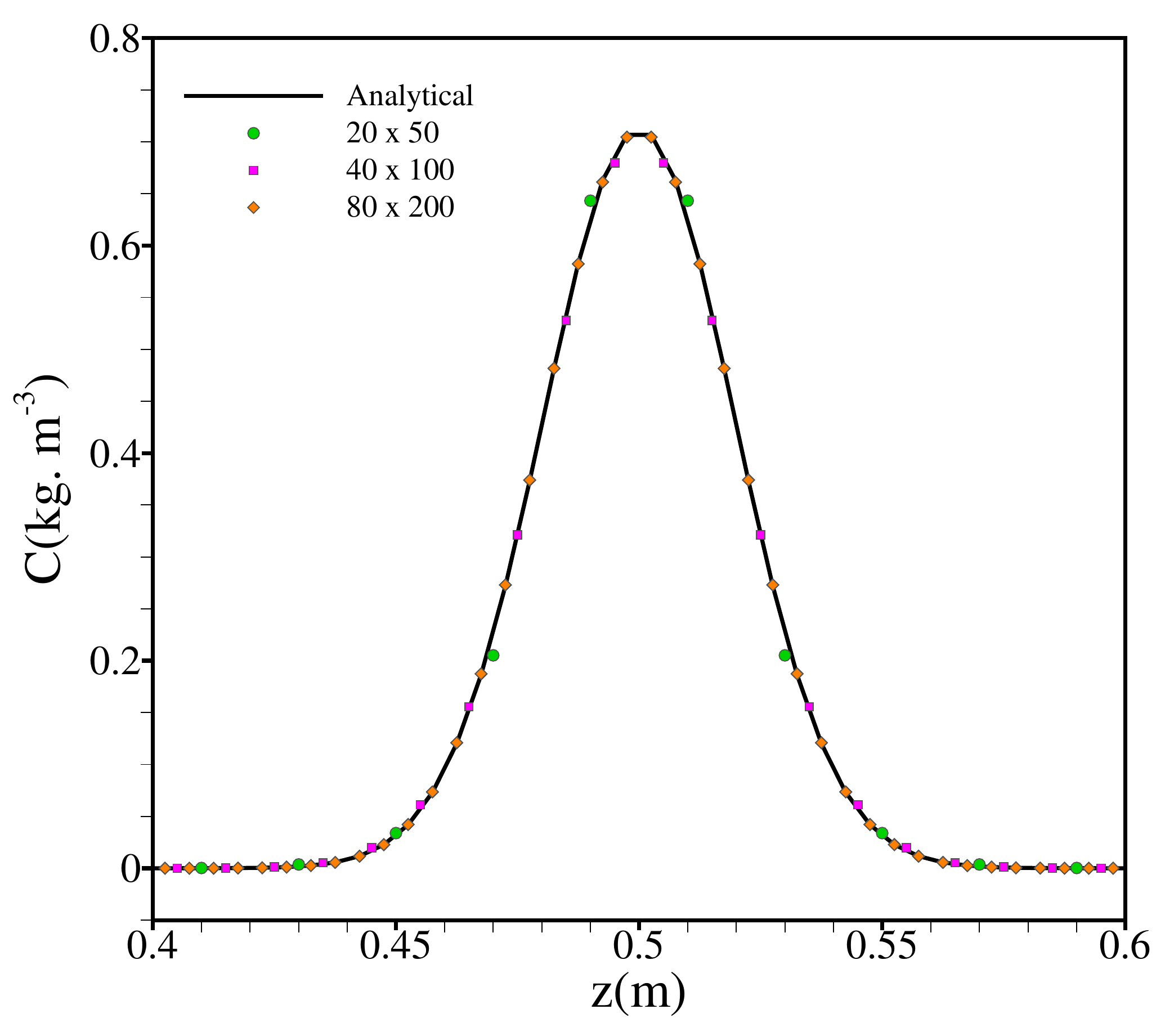}}
	\caption{Comparison of the concentration distribution between present results and the analytical solutions for isotropic diffusion processes with constant and exponential initial concentration distributions.}
	\label{figs:iso-diffusion-c}
\end{figure}

Following Refs \cite{zhu2001smoothed, rezavand2019fully}, 
the first problem studied herein of the one-dimensional isotropic diffusion is depicted in Figure \ref{figs:iso-diffusion-setup} 
where a $0.4 \times 1\text{m}$ rectangle is filled with water 
and a finite horizontal band of pollutant is located in the middle of the rectangle confined to the region of $z_1 \leq z \leq z_2$. 
The initial pollutant concentration is equal to $C_0 = 1 \text{kg}\cdot\text{m}^{-3}$ in the band and zero elsewhere, 
and the diffusion coefficient is set as $d^{iso} = 1.0 \times 10^{-4}$. 
According to Ref. \cite{crank1979mathematics}, 
the analytical solution is 
\begin{equation}\label{eq:diffusion-c-solution}
\begin{cases}
C(z,t) = \frac{C_0}{2} \erfc \left( \frac{z_1 - z}{\sqrt{4d^{iso}t}} \right)  \text{for} & z \leq z_0 \\
C(z,t) = \frac{C_0}{2} \erfc \left( \frac{z - z_2}{\sqrt{4d^{iso}t}} \right)  \text{for} & z > z_0
\end{cases}. 
\end{equation}
where $z_0 = 0.5 \text{m}$ $z_1 = 0.45 \text{m}$ and $z_2 = 0.55 \text{m}$. 
The numerical solution at $t = 1.0 s$ is shown in Figure \ref{figs:iso-diffusion-c} (a). 
The bell shaped distribution of the concentration is in agreement with the analytical solution. 
It can also be observed that the present results converges with increasing spatial resolution. 

This problem is further considered by setting an exponential initial pollutant concentration distribution as 
\begin{equation}\label{eq:diffusion-exp}
C(z,t = 0) = \exp \left( - \frac{(z - z_0)^2}{4d^{iso}t_0} \right), 
\end{equation}
where $t_0 = 1 \text{s}$ and $z_0 = 0.5 \text{m}$. 
Also, the analytical solution takes the following form \cite{crank1979mathematics}
\begin{equation}\label{eq:diffusion-exp-solution}
C(z,t = 0) = \frac{C_0}{\sqrt{t + t_0}} \exp \left( - \frac{(z - z_0)^2}{4d^{iso}(t + t_0)} \right). 
\end{equation}
Figure \ref{figs:iso-diffusion-c} (b) illustrates the comparison of the present predictions of the concentration distribution against the analytical solution at $t = 1.0 s$. 
Again, 
a good agreement is noted and the convergence of the concentration distributions with increasing resolution is observed.
%
\subsection{Anisotropic diffusion}
In this section, 
we consider the anisotropic diffusion process from a contaminant source in water. 
Following the work of Tran-Duc et al. \cite{tran2016simulation}, 
the contaminant source is located in a two dimensional $200 \text{m} \times 200 \text{m}$ square computation domain 
and the analytical solution of the contaminant distribution is
\begin{equation}\label{eq:anis-diffusion}
C(z,t) = \frac{1}{4\pi t \prod_{i = 1}^{2}\sqrt{\mathbb{D}_{ii}}} \prod_{i=1}^{2} \exp \left[-\frac{(x_i - x_{i,0})^2}{4t\mathbb{D}_{ii}} \right] .
\end{equation}
The initial condition for numerical solution is set at time $t = 120 \text{s}$. 

\begin{figure}[htb!]
	\centering
	\includegraphics[trim = 10mm 30mm 5mm 10mm, clip, width=.95\textwidth]{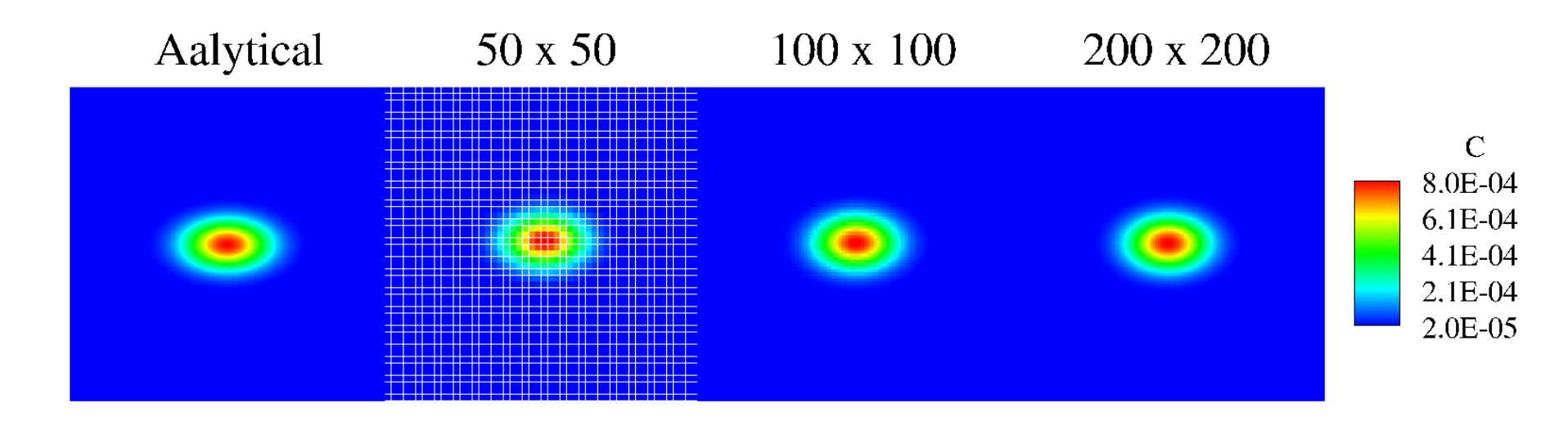}
	\caption{Concentration distribution at $1920s$ diffusion with the anisotropic diffusion tensor $\mathbb{D}_1$.}
	\label{figs:anis-diffusion-particle-1}
\end{figure}

\begin{figure}[htb!]
	\centering
	\subfloat[at $x = 100m$ \label{figs:anis-diffusion-1-s1}]{\includegraphics[width=.485\textwidth]{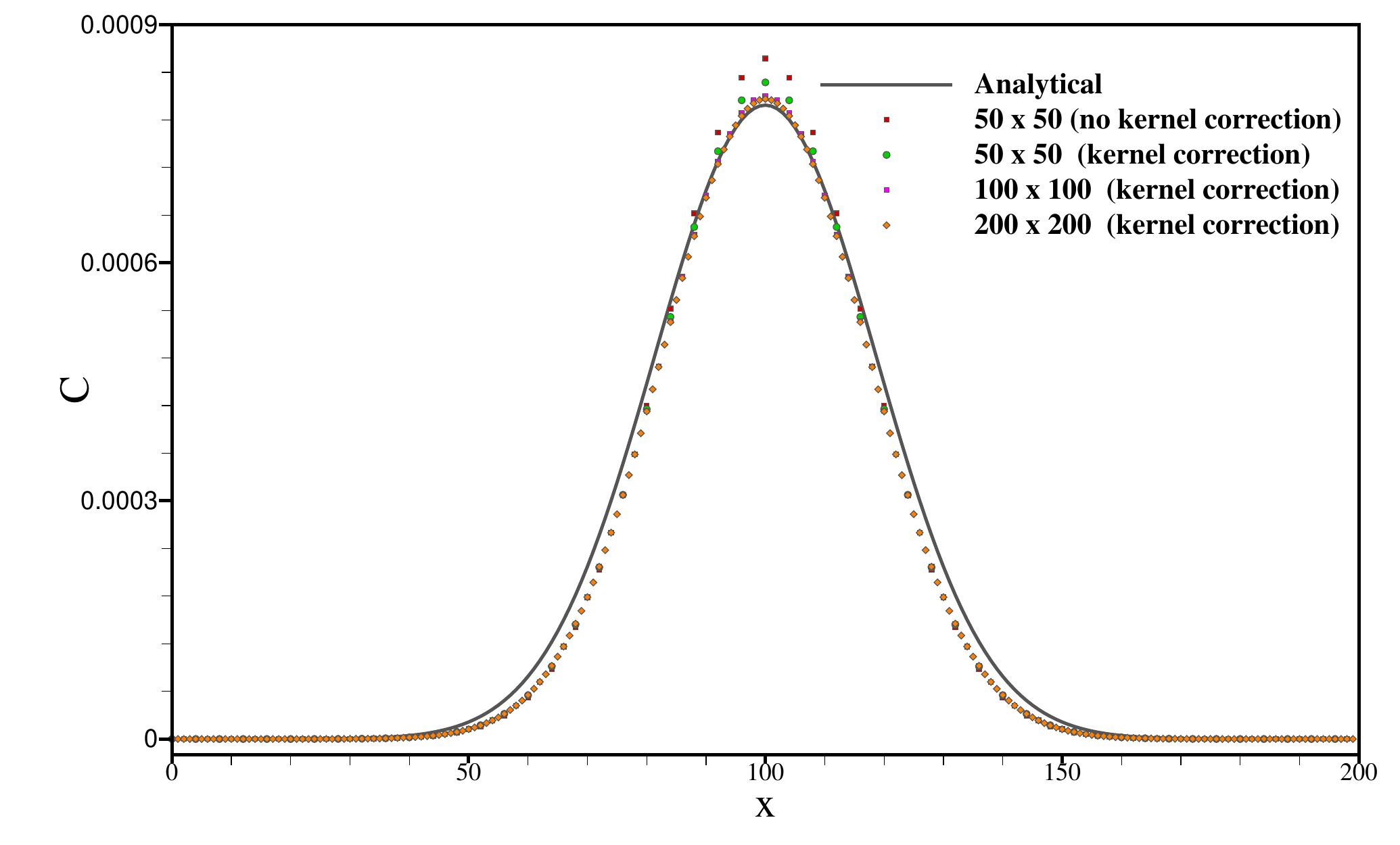} }
	\subfloat[at $y = 100m$ \label{figs:anis-diffusion-1-s2}]{ \includegraphics[width=.485\textwidth]{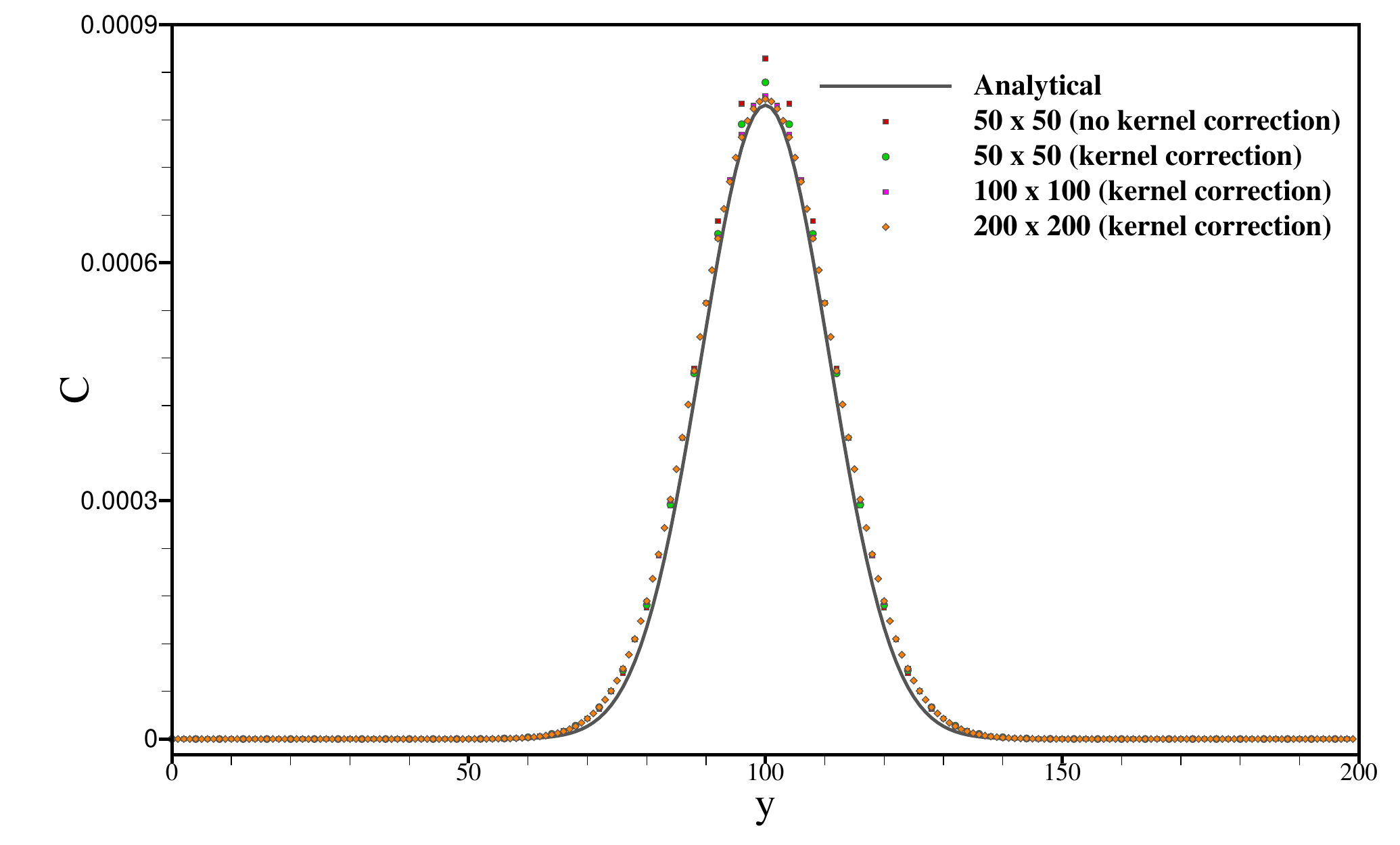}}
	\caption{Concentration distribution after $1800s$ diffusion with the anisotropic diffusion tensor $\mathbb{D}_1$: comparison with analytical solution by using three different spatial resolutions.}
	\label{figs:anis-diffusion-1}
\end{figure}

In the first case, 
the anisotropic diffusion tensor is 
\begin{equation}
\mathbb{D}_1 = \left[ \begin{array}{cc} 0.09 & 0 \\ 0 & 0.03 \end{array} \right]\left(\text{m}^2 \cdot \text{s}^{-1}\right) . 
\end{equation}
Figure \ref{figs:anis-diffusion-particle-1} shows the numerical and analytical distributions at time $t = 1920 s$. 
It can be observed that introducing the renormalized kernel correction can improve the computational accuracy. 
In general, 
the concentration distributions are like ellipses with major axis in $x$-direction and minor axis in $y$-direction 
due to the fact that the diffusion rate in $x$-direction is larger than that in $y$-direction. 
Also, the numerical solution is in agreement with the analytical one in both shape and value profiles. 
Figure \ref{figs:anis-diffusion-1} gives the present numerical concentration distributions 
at horizontal cross section at $x = 100 \text{m}$ (Figure \ref{figs:anis-diffusion-1-s1}) and 
vertical cross section at $y = 100 \text{m}$ (Figure \ref{figs:anis-diffusion-1-s2}) and the corresponding comparison with analytical solutions. 
The present SPH approximated concentration profiles are in good agreement with the analytical solution. 
Also, the present SPH results converges to the analytical solution as the spatial resolution increases. 
Compared with the results obtained by Tran-Duc et al. \cite{tran2016simulation} with resolution $400 \times 400$ (see Figure 3 in their work), 
present results shows similar accuracy even as lower resolution $200 \times 200$ is used with the introduction of the renormalized kernel correction. 
Similar to Ref. \cite{tran2016simulation}, 
the present results reduce the anisotropy level of diffusion process and show a bit less anisotropic compared with the analytical one. 
This discrepancy induced by the isotropic property of kernel function in SPH which averages and smoothes the concentration function independent of direction.

\begin{figure}[htb!]
	\centering
	\includegraphics[trim = 10mm 30mm 5mm 10mm, clip, width=.95\textwidth]{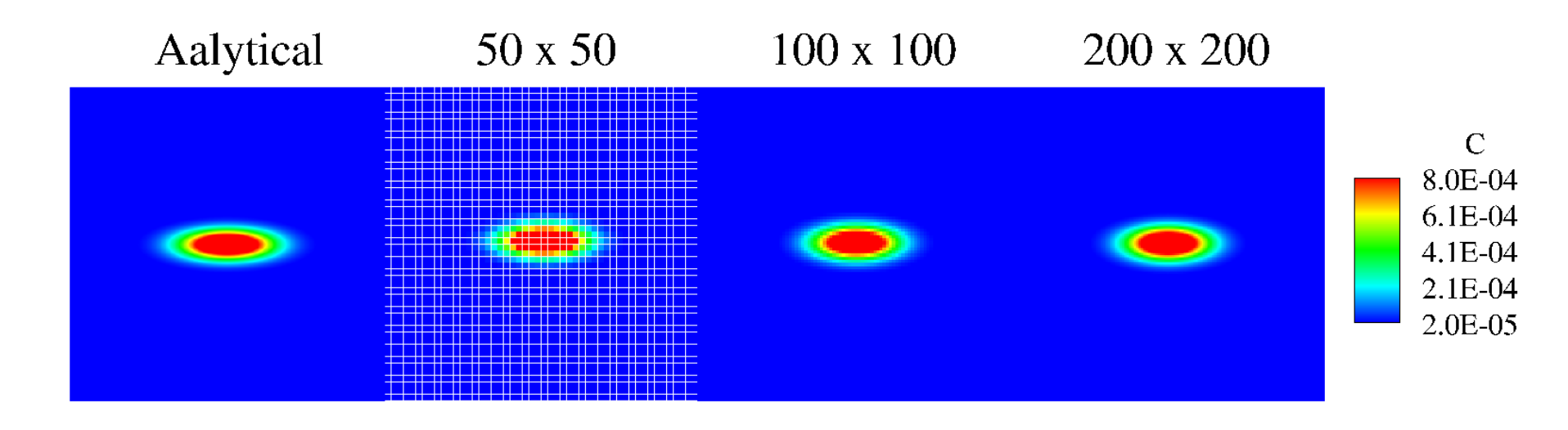}
	\caption{Concentration distribution at $1920s$ diffusion with the anisotropic diffusion tensor $\mathbb{D}_2$.}
	\label{figs:anis-diffusion-particle-2}
\end{figure}

\begin{figure}[htb!]
	\centering
	\subfloat[at $x = 100m$ \label{figs:anis-diffusion-2-s1}]{\includegraphics[width=.495\textwidth]{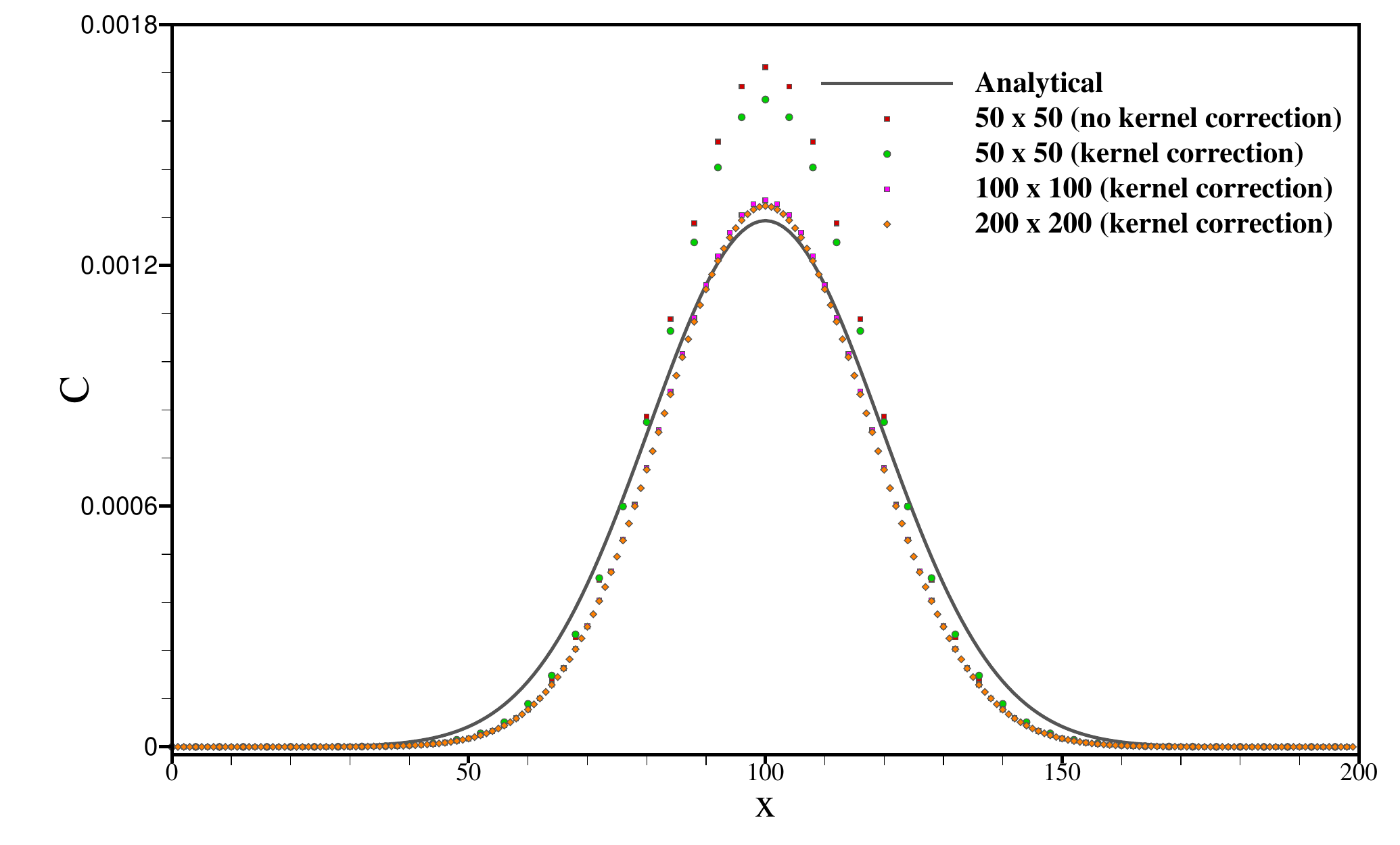} }
	\subfloat[at $y = 100m$ \label{figs:anis-diffusion-2-s2}]{ \includegraphics[width=.495\textwidth]{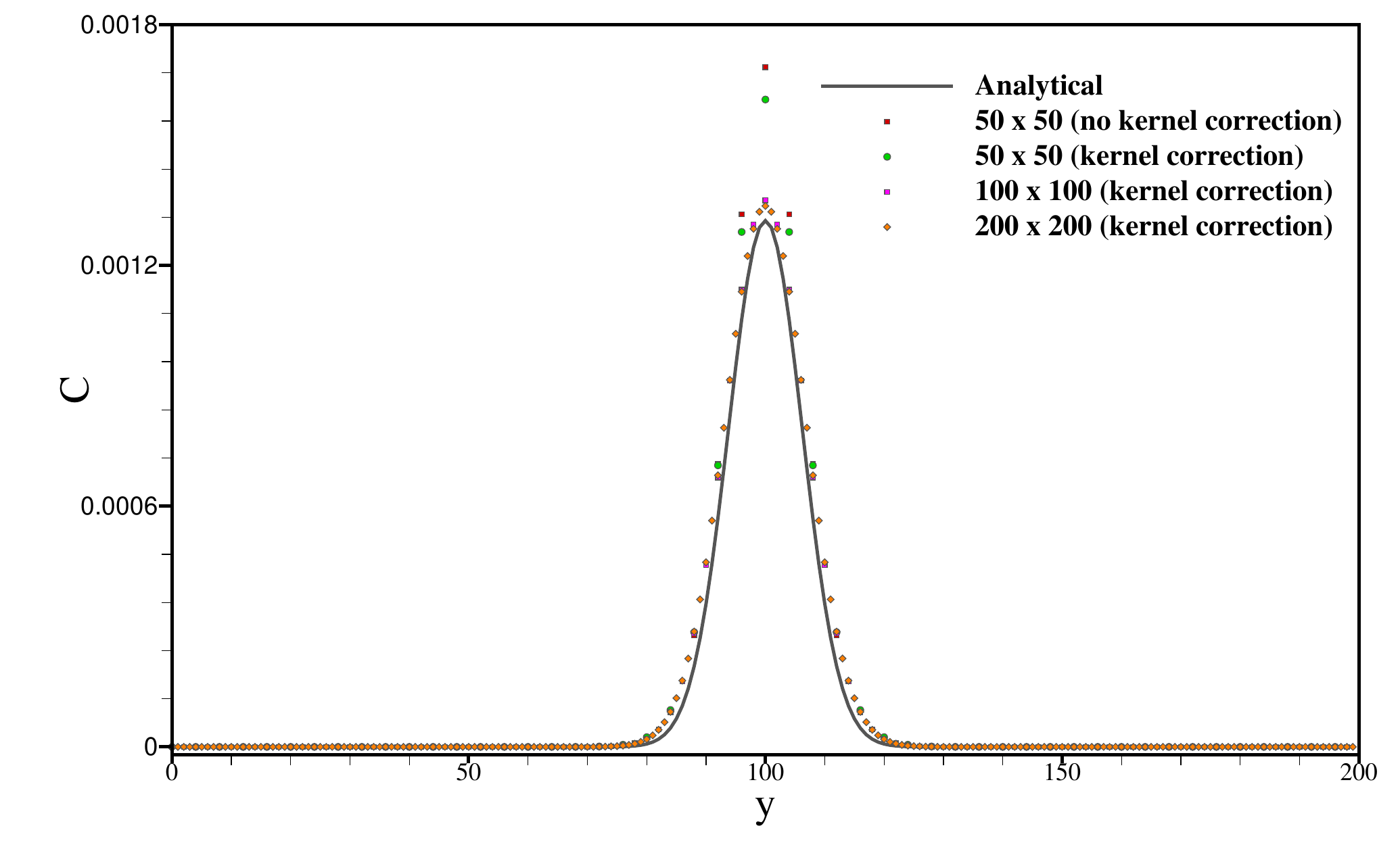}}
	\caption{Concentration distribution after $1800s$ diffusion with the anisotropic diffusion tensor $\mathbb{D}_2$: comparison with analytical solution by using three different spatial resolutions.}
	\label{figs:anis-diffusion-2}
\end{figure}

In the second test, 
a higher anisotropic ratio is considered by setting the diffusion tensor as 
\begin{equation}
\mathbb{D}_2 = \left[ \begin{array}{cc} 0.1 & 0 \\ 0 & 0.01 \end{array} \right]\left(\text{m}^2 \cdot \text{s}^{-1}\right). 
\end{equation}
Figure \ref{figs:anis-diffusion-particle-2} shows the comparison between the simulated and analytical concentration distributions at time $t = 1920 \text{s}$. 
Again, the renormalized kernel correction shows improved computational accuracy. 
As expected, the concentration distributions are also ellipses but with a higher ratio of major axis to minor compared with the results depicted in Figure \ref{figs:anis-diffusion-particle-1}. 
Again, the simulated distributions are in consistent with the analytical one in both shape and value profiles. 
The present numerical concentration distributions 
at horizontal cross section at $x = 100 \text{m}$ and vertical cross section at $y = 100 \text{m}$ are given in Figs. \ref{figs:anis-diffusion-2-s1}  and \ref{figs:anis-diffusion-2-s2}, respectively. 
%
\subsection{Propagation of transmembrane potential}
\label{sec:actionpotential-simple}
Following the work of Ratti and Verani \cite{ratti2019}, 
we consider a problem on the propagation of transmembrane potential. 
It is assumed that the transmembrane potential propagates in a two dimensional isotropic tissue 
in a square domain of $(0, 1)^2$ and the nondimensional time interval is set as $(0, 16)$. 
The transmembrane potential and gate variable are initialized by 
 \begin{equation}
 \begin{cases}
V_m  = exp\left[ -\frac{(x - 1.0)^2 + y^2}{0.25} \right] \\
w = 0.0
 \end{cases}, 
 \end{equation}
 and the nondimensional diffusion coefficients are $d^{iso} = 1.0$ and $d^{ani} = 0.0$. 
Here, we consider the Aliev-Panfilow model  with the constant parameters given in Table \ref{tab:ap-1}. 
\begin{table}[htb!]
	\centering
	\caption{Parameters for the Aliev-Panfilow model. }
	\begin{tabular}{cccccc}
		\hline
		k	& a & b  & $\epsilon_0$ & $\mu_1$ & $\mu_2$  \\
		\hline
		8.0	& 0.15   & 0.15      & 0.002 & 0.2 & 0.3  \\
		\hline	
	\end{tabular}
	\label{tab:ap-1}
\end{table}

\begin{figure}[htb!]
 	\centering
 	\includegraphics[trim = 2mm 2mm 2mm 2mm, clip, width=.45\textwidth]{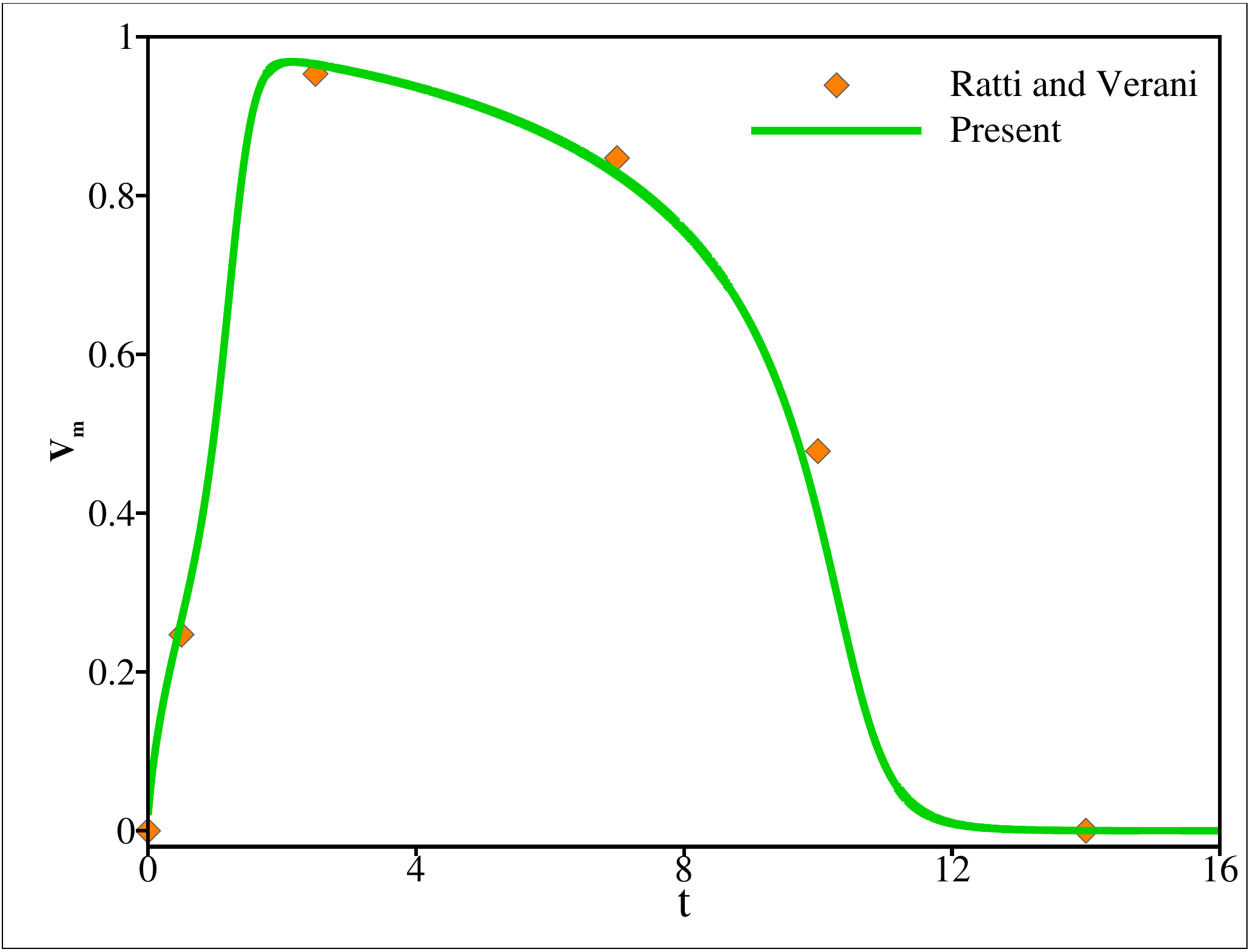}
 	\caption{Time evolution of the transmembrane potential at point $(0.3, 0.7)$ in comparison with the results reported by Ratti and Verani \cite{ratti2019}.}
 	\label{figs:depolarization}
 \end{figure}

Figure \ref{figs:depolarization} reports the predicted evolution profile of the transmembrane potential at point $(0.3, 0.7)$, 
and the comparison with those from Ratti and Verani \cite{ratti2019}. 
In general, a good agreement is noted. 
It is observed that in accordance with the previous numerical estimation \cite{ratti2019} and experimental observation \cite{franzone2014mathematical}, 
the quick propagation of the stimulus in the tissue and the slow decrease in the transmembrane potential after a plateau phase are well predicted by the present method. 
 %
 \subsection{Two dimensional spiral wave}
 \label{sec:2dspiralwave}
 We now validate the SPH method in reproducing the spiral waves by solving the monodomain equations with the FitzHugh-Nagumo model. 
 The spiral waves, 
which consists of complicated patterns of the transmembrane potential along with simple unidirectionally propagating pulses, 
 are suitable choices for validating the numerical solution of solving the reaction-diffusion equation. 
 In this work, 
 we consider both isotropic and anisotropic tissues in two dimensional rectangular and circular geometries with the given parameters of the FitzHugh-Nagumo model in Table \ref{tab:fn-model}
 \begin{table}[htb!]
 	\centering
 	\caption{Parameters for the FitzHugh-Nagumo model.}
 	\begin{tabular}{ccccc}
 		\hline
 		a	& $\epsilon_0$ & $\beta$ & $\gamma$ & $\sigma$  \\
 		\hline
 		0.1	& 0.01  & 0.5 & 1.0 & 0.0 \\
 		\hline	
 	\end{tabular}
 	\label{tab:fn-model}
 \end{table}
 %
 \subsubsection{Spiral waves in rectangular geometry}
Following Wang et al. \cite{wang2019simulation} and Liu et al. \cite{liu2012numerical}, 
the rectangular computational region is set as $\left[0, 2.5\right] \times \left[0, 2.5\right]$ and the transmembrane potential and gate variable are initialized by 
\begin{equation}
V_m  = 
\begin{cases} 1.0, & 0< x \leq 1.25; 0 < y < 1.25 \\
0.0, & elsewhere
\end{cases}, 
\end{equation}
and
\begin{equation}
w  = 
\begin{cases} 
0.1, & 0< x \leq 1.25; 1.25 \leq y < 2.5 \\
0.1, & 1.25 \leq x < 2.5; 0 < y < 2.5 \\
0, & elsewhere
\end{cases}. 
\end{equation}

\begin{figure}[htb!]
 	\centering
 	\includegraphics[trim = 5mm 25mm 5mm 15mm, clip, width=\textwidth]{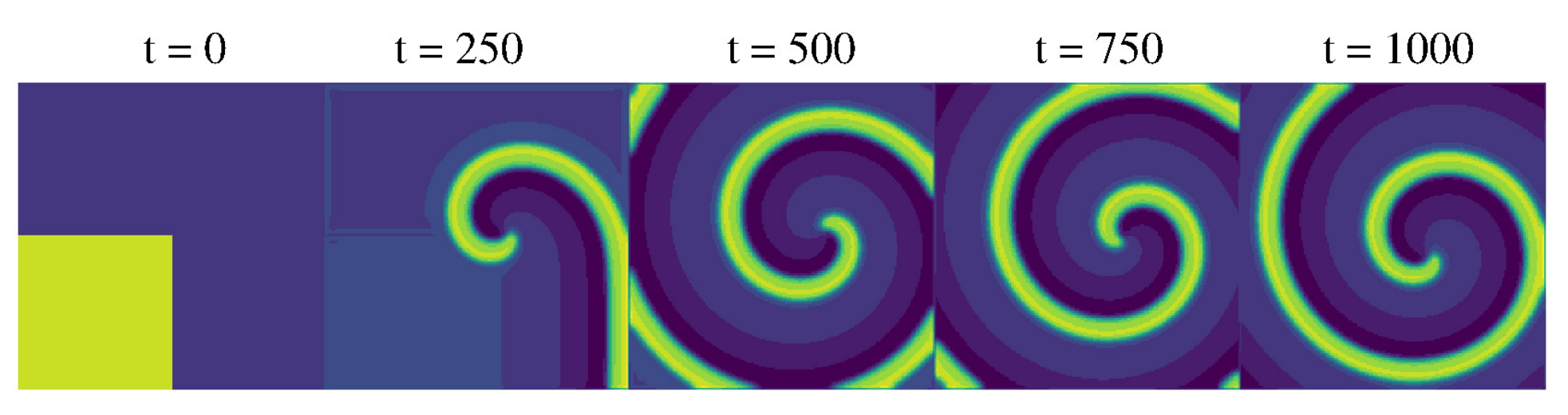}\\
 	\includegraphics[trim = 5mm 25mm 5mm 25mm, clip, width=\textwidth]{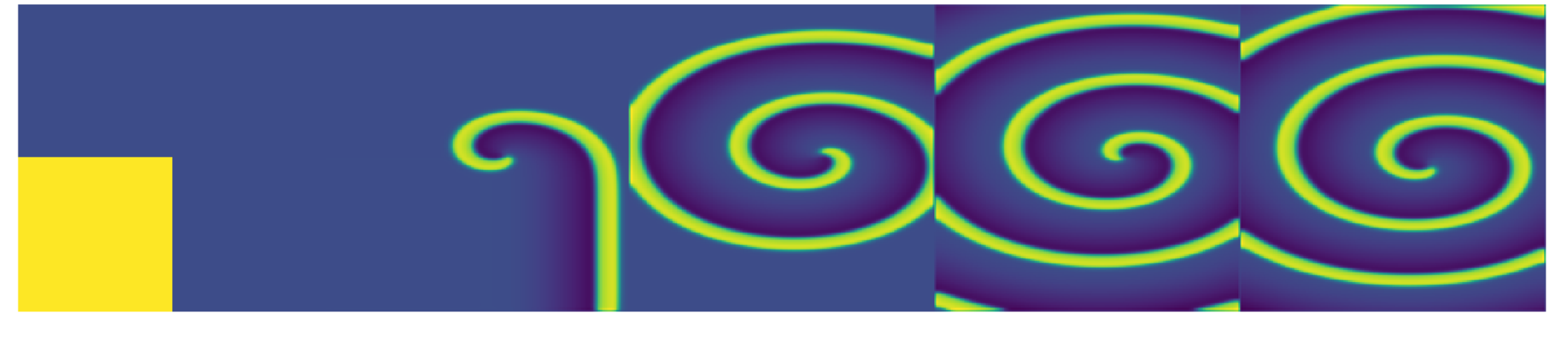}\\
 	\includegraphics[trim = 5mm 25mm 5mm 25mm, clip, width=\textwidth]{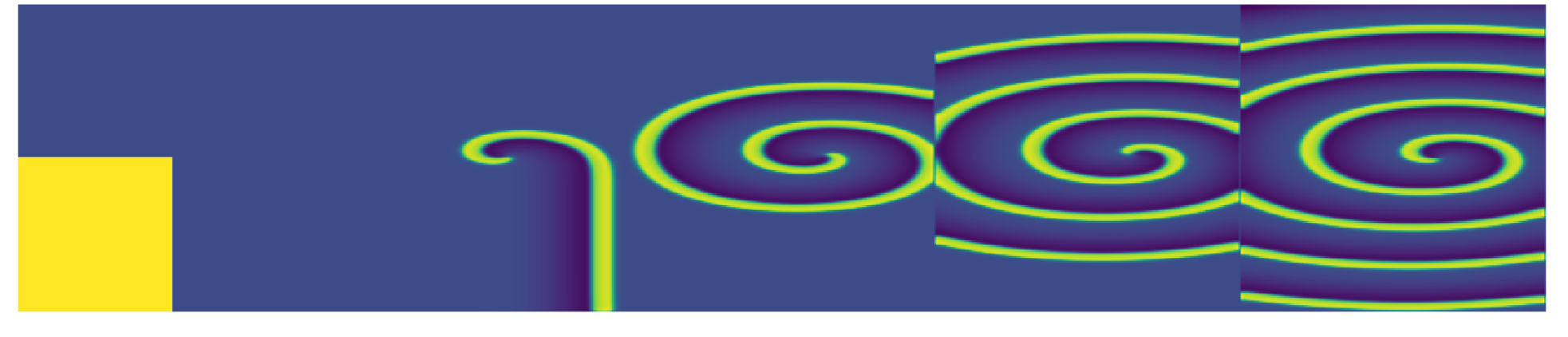}
 	\caption{Spiral waves of the FitzHugh-Nagumo model in rectangular computational domain with isotropic diffusion tensor $\mathbb{D}_1$(upper panel), 
 			$\mathbb{D}_2$ (middle panel) and $\mathbb{D}_3$ (lower panel) at five time instants.}
 	\label{figs:spiral-wave-square}
\end{figure}

In the first test, we consider an isotropic tissue with the nondimensional diffusion $\mathbb{D}_1$ where $d^{ios} = 1.0 \times 10^{-4}$  and $d^{ani} = 0.0$.
Figure \ref{figs:spiral-wave-square} (upper panel) shows a spiral wave of the stable rotation solution at five different time instants. 
As observed, 
the spiral wave generates a clockwise rotation curve in the rectangular region. 
Note that the spiral wave profiles obtained by the present SPH method is consistent with those obtained by Wang et al. \cite{wang2019simulation} and Liu et al. \cite{liu2012numerical} 
(see Figure 1 (a) and (b) in Ref. \cite{wang2019simulation}).
 
Figure \ref{figs:spiral-wave-square} (middle panel) shows the numerical results with an anisotropic diffusion tensor 
\begin{equation}\label{eq:spiral-wave-d2}
\mathbb{D}_2 = \left[ \begin{array}{cc} 1.0 \times 10^{-4} & 0 \\ 0 & 2.5 \times 10^{-5} \end{array} \right].
\end{equation}
It can be observed that the spiral wave now propagates with elliptical patterns as also seen in Refs. \cite{wang2019simulation, liu2012numerical}. 
Again, 
a good agreement with those given in Refs. \cite{wang2019simulation, liu2012numerical} (see Figure 2 (a) and (b) in Ref. \cite{wang2019simulation}) is noted. 

Another test with a larger diffusion ratio, 
given by
\begin{equation}\label{eq:spiral-wave-d3}
\mathbb{D}_3 = \left[ \begin{array}{cc} 1.0 \times 10^{-4} & 0 \\ 0 & 1.0 \times 10^{-5} \end{array} \right] ,
\end{equation}
is reported in Figure \ref{figs:spiral-wave-square} (lower panel). 
It can be found that the spiral wave now has a slightly smaller width compared with the one shown in Figure \ref{figs:spiral-wave-square} (middle panel). 
Also, 
the elliptical propagation shape of the spiral wave has a higher ratio between the major and minor axes. 
%
\subsubsection{Spiral waves in circular geometry}
In this part, 
the computational domain is changed to a nonuniform geometry, i.e. a circle of radius $R = 1.25$ and centered at $\mathbf{r}_0 = (R, R)$. 
The transmembrane potential and gate variable are initialized by \cite{wang2019simulation, liu2012numerical}
 \begin{equation}
 V_m  = 
 \begin{cases} 
 1.0, & \begin{array}{c} 
 				R - \sqrt{R^2 - \left(R - y\right)^2} < x \leq R \\
 				R - \sqrt{R^2 - \left(R - x\right)^2} < y \leq R 
 			\end{array} \\
 0.0, & elsewhere
 \end{cases}, 
 \end{equation}
 and
 \begin{equation}
 w  =  
\begin{cases} 
0.1 , & \begin{array}{c} 
				R - \sqrt{R^2 - \left(R - y\right)^2} < x < R + \sqrt{R^2 - \left(R - y\right)^2}  \\
				R \leq y < R +  \sqrt{R^2 - \left(R - x\right)^2}
			\end{array} \\
0.0, & elsewhere
\end{cases}. 
 \end{equation}

\begin{figure}[htb!]
 	\centering
 	\includegraphics[trim = 5mm 25mm 5mm 15mm, clip, width=\textwidth]{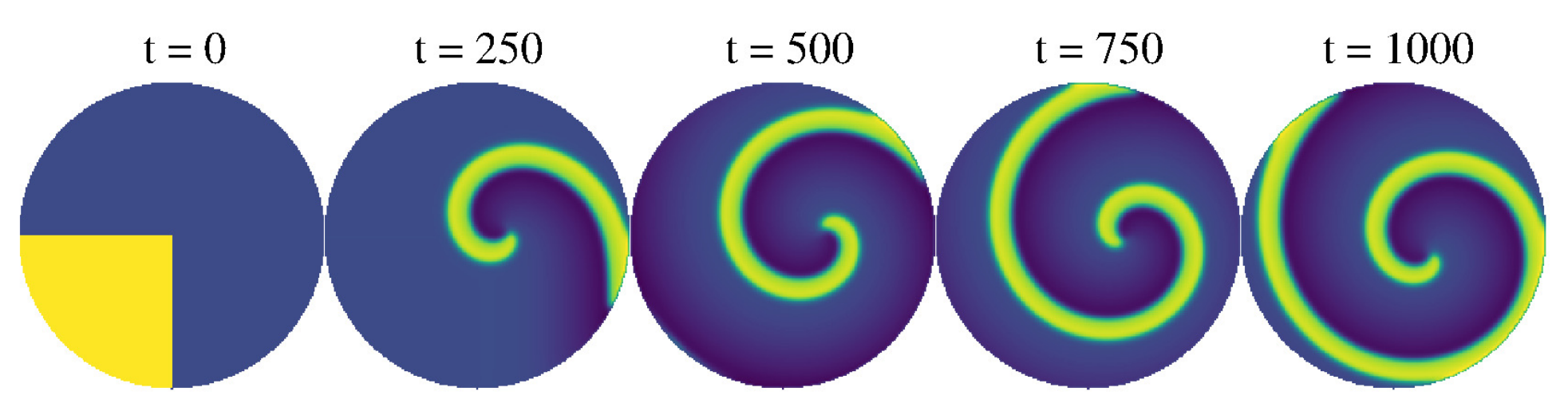}\\
 	\includegraphics[trim = 5mm 25mm 5mm 25mm, clip, width=\textwidth]{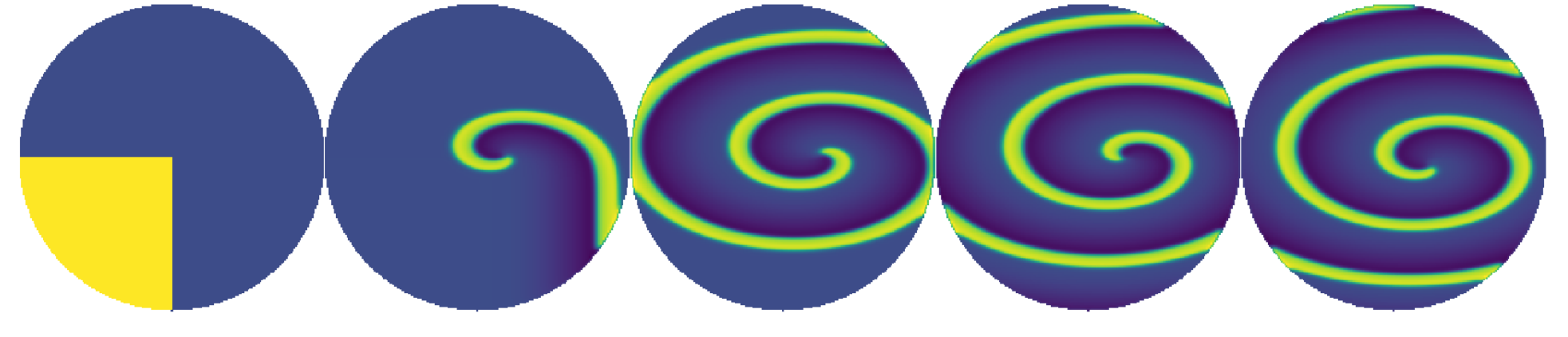}\\
 	\includegraphics[trim = 5mm 25mm 5mm 25mm, clip, width=\textwidth]{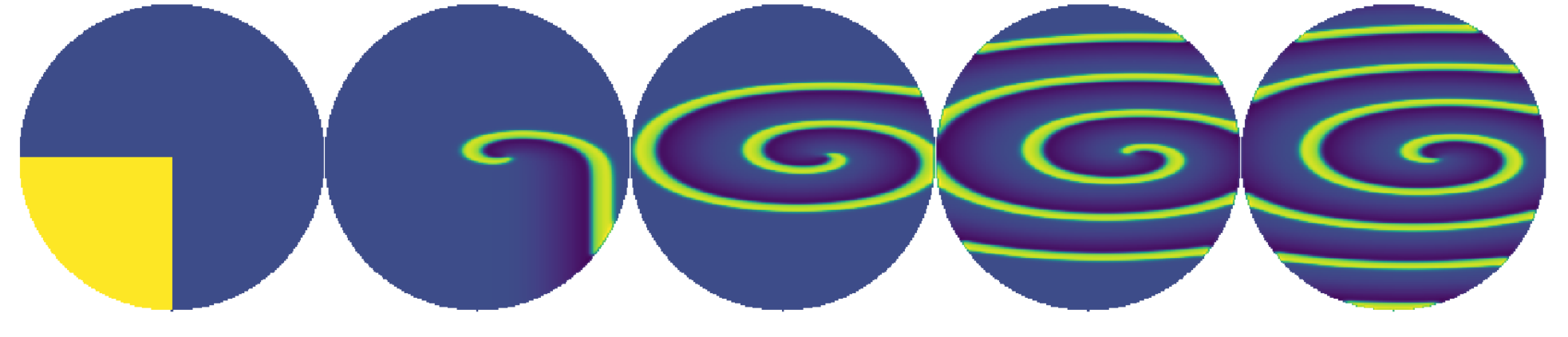}
 	\caption{Spiral waves of the FitzHugh-Nagumo model in circular computational domain with the isotropic diffusion tensor $\mathbb{D}_1$ (upper panel),
 		$\mathbb{D}_2$ (middle panel)	and $\mathbb{D}_3$ (lower panel) at five time instants.}
 	\label{figs:spiral-wave-circle}
\end{figure}

Figure \ref{figs:spiral-wave-circle} (upper panel) shows 
the contours of the stable rotating spiral wave with isotropic diffusion tensor $\mathbb{D}_1$ given by previous section. 
As expected, 
the spiral wave generates a curve and rotates clockwise as reported in \cite{wang2019simulation, liu2012numerical}. 
Again, 
a good agreement with those of Refs. \cite{wang2019simulation, liu2012numerical} is noted (see Figure 3 (a) and (b) in Ref. \cite{wang2019simulation}).

For the anisotropic diffusion tensor $\mathbb{D}_2$ given in Eq. \eqref{eq:spiral-wave-d2}, 
the transmemberane potential propagation at four different time instants are shown in Figure \ref{figs:spiral-wave-circle} (middle panel). 
Now, 
the spiral wave follows an elliptical pattern in the circular region.  
Figure \ref{figs:spiral-wave-circle} shows the propagation pattern of the spiral wave 
with anisotropic diffusion tensor $\mathbb{D}_3$ given in Eq. \eqref{eq:spiral-wave-d3}. 
With larger diffusion ratio, 
the propagation pattern of the spiral waves is effected and the width of the spiral wave is also slightly smaller than the one reported in Figure \ref{figs:spiral-wave-circle} (lower panel). 
%
\subsection{Mechanical response of myocardium}
\label{sec:response}
In this section, 
two benchmarks are investigated to validate the accuracy, robustness and applicability of 
the present SPH framework for modeling the passive and active mechanical responses of the cardiac myocardium.
\subsection{Passive mechanical response}
\label{sec:passive}
\begin{figure}[htb!]
	\centering
	\includegraphics[trim = 6cm 8cm 6cm 0.5cm, clip, width=.3\textwidth]{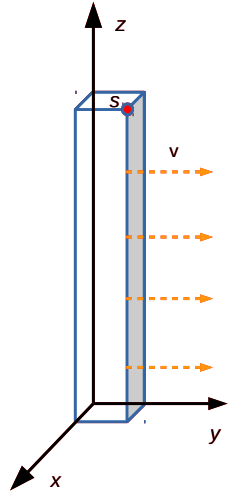}
	\caption{Passive response of a three-dimensional bending cantilever: initial configuration.}
	\label{figs:passive-setup}
\end{figure}

In this part, 
we consider the passive mechanical response of the myocardium in the form of bending cantilever. 
Following Aguirre et al.\cite{aguirre2014vertex}, 
a three-dimensional rubber-like cantilever whose 
bottom face is clamped to the ground and its body is allowed to bend freely by imposing an initial uniform velocity $\mathbf{v} = (5\sqrt{3}, 5, 0)^T \text{m} \cdot \text{s}^{-1}$ is considered (see Figure \ref{figs:passive-setup}). 
For in-depth comparisons, 
both neo-Hookean and Holzapfel-Odgen material models are applied.
For the neo-Hookean model, 
the strain-energy density function \cite{ogden1997non} is defined as
\begin{eqnarray}\label{Neo-Hookean-energy}
W  =  \mu \tr \left(\mathbb{E}\right) - \mu \ln J + \frac{\lambda}{2}(\ln J)^{2} ,
\end{eqnarray}
where $\lambda$ and $\mu$ are Lam$\acute{e}$ parameters, 
$K = \lambda + (2\mu/3)$ is the bulk modulus and $G = \mu$ is the shear modulus. 
The relation between the two modulus is given by
\begin{equation}\label{relation-modulus}
E = 2G \left(1+2\nu\right) = 3K\left(1 - 2\nu\right) ,
\end{equation}
with $E$ denoting the Young's modulus and $\nu$ the Poisson ratio.
Here, the Youngs' modulus is $E=1.7\times10^7 ~\text{Pa}$, Poisson ratio $\nu = 0.45$ and density $\rho = 1.1 \times 10^3 ~\text{kg} \cdot \text{m}^{-3}$. 
For the Holzapfel-Odgen model, the material parameters are given in Table \ref{tab:ho-simple} and the anisotropic terms are varying  accordingly. 
\begin{table}[htb!]
	\centering
	\caption{Parameters for the Holzapfel-Ogden constitution model (For the isotropic material, the anisotropic terms are set to zero).}
	\begin{tabular}{cccc}
		\hline
		$a = 5.860$ MPa	& $a_f = k a$ & $a_s = 0$  & $a_{fs} = 0.0$  \\
		\hline
		$b = 1.0$ 	& $b_f = 0$    & $b_s = 0$      & $b_{fs} = 0$   \\
		\hline	
	\end{tabular}
	\label{tab:ho-simple}
\end{table}

\begin{figure}[htb!]
	\centering
	\includegraphics[trim = 1.75cm  1mm 12cm  1mm, clip, height=0.32\textwidth]{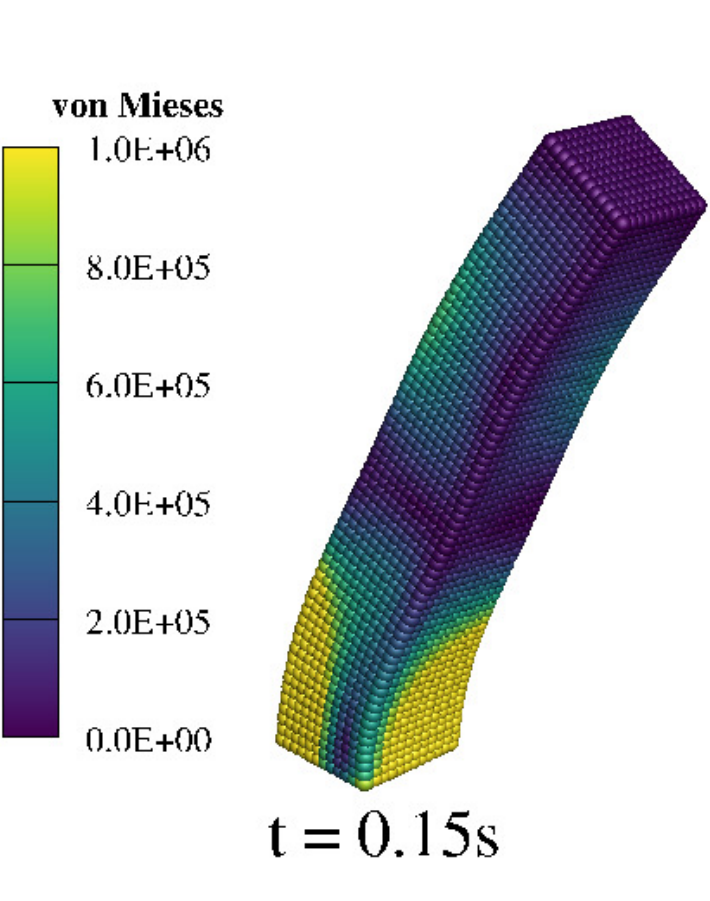}
	\includegraphics[trim = 4.5cm  1mm 10.5cm 1mm, clip, height=0.32\textwidth]{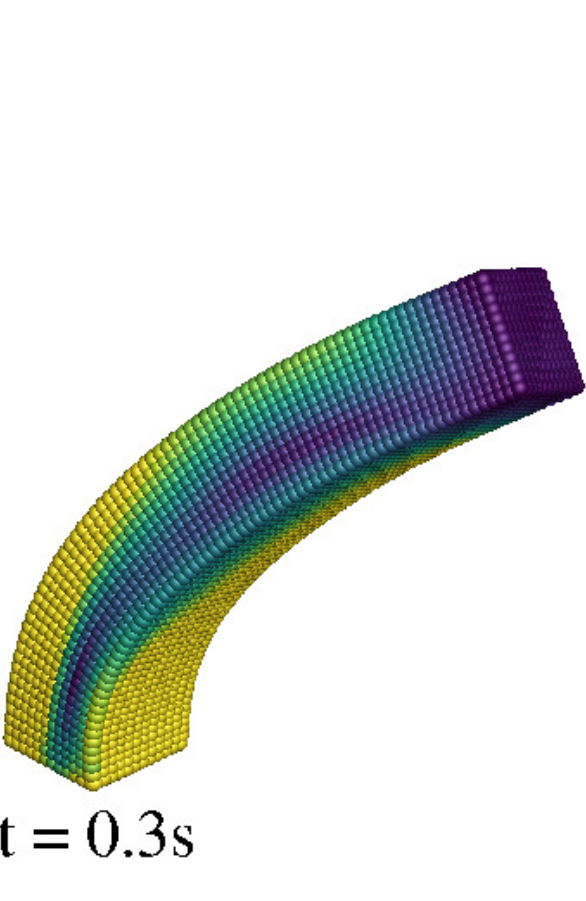}
	\includegraphics[trim = 4.5cm  1mm 9.5cm  1mm, clip, height=0.32\textwidth]{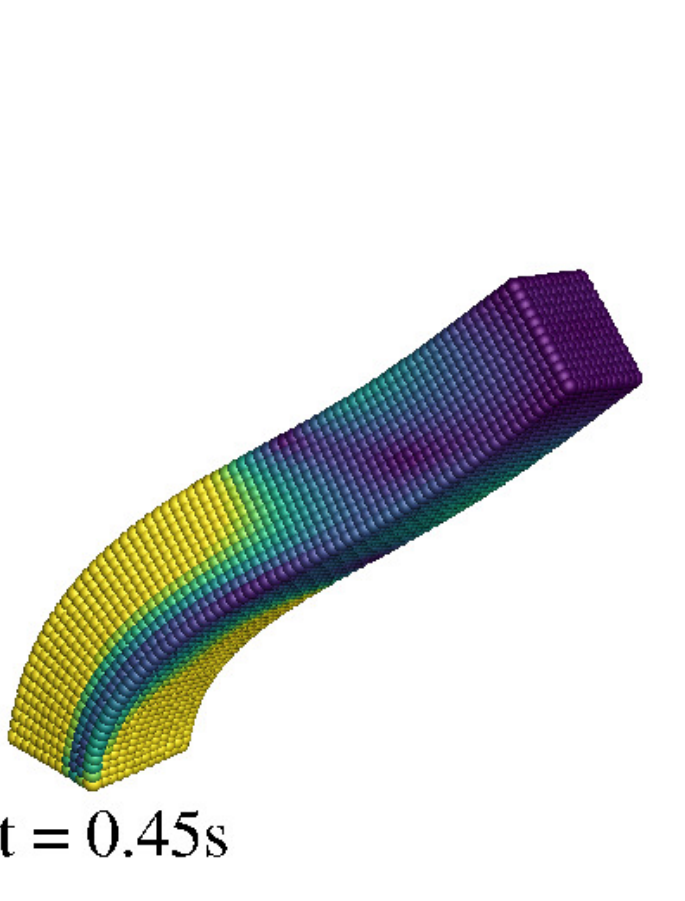}
	\includegraphics[trim = 4.5cm  1mm 9.5cm  1mm, clip, height=0.32\textwidth]{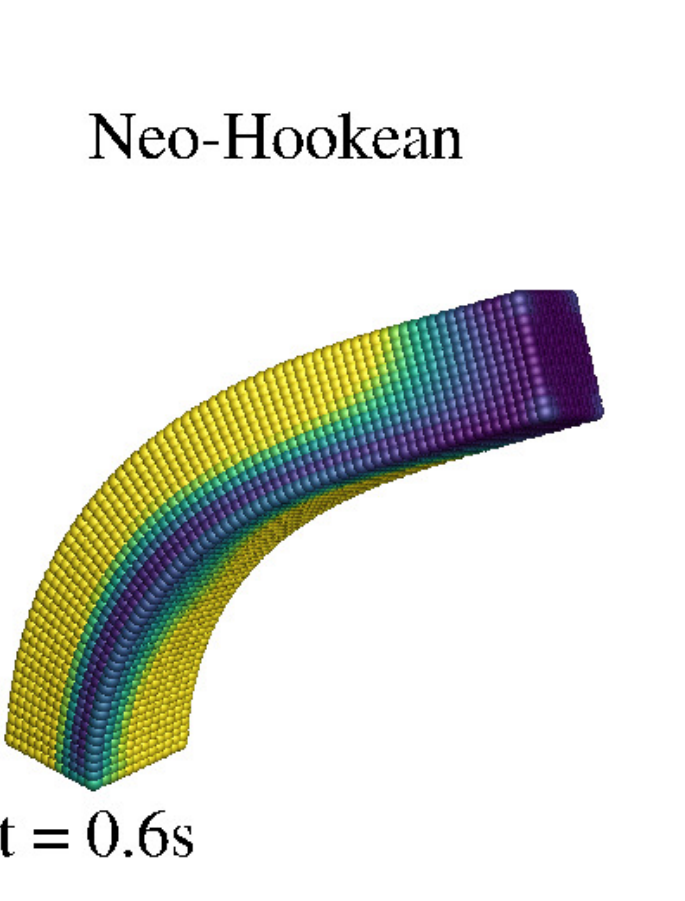} \\
	\includegraphics[trim = 1.75cm  1mm 12cm  1mm, clip, height=0.32\textwidth]{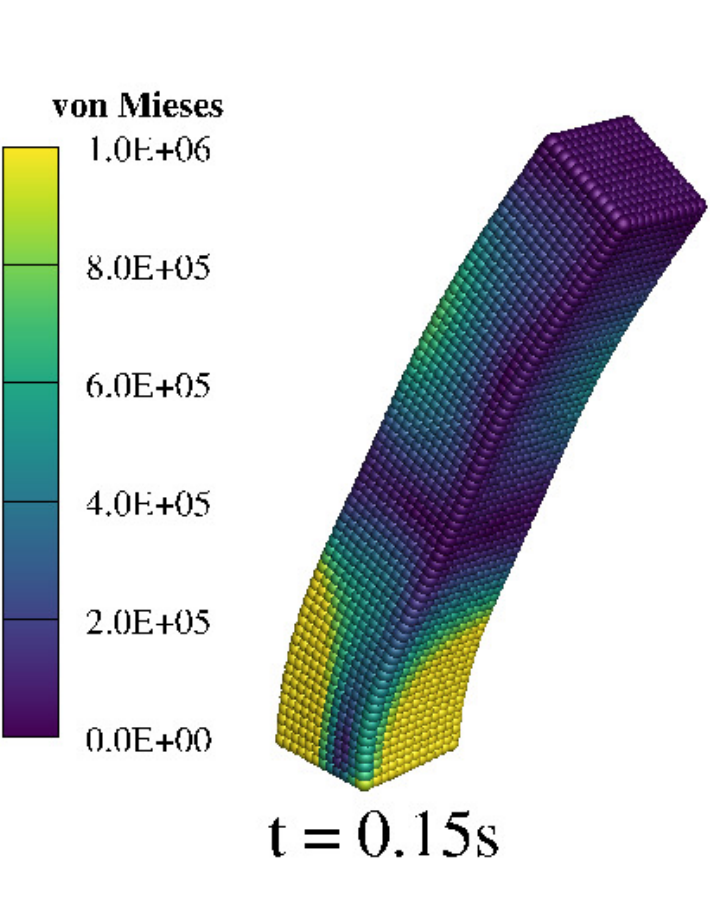}
	\includegraphics[trim = 4.5cm  1mm 10.5cm 1mm, clip, height=0.32\textwidth]{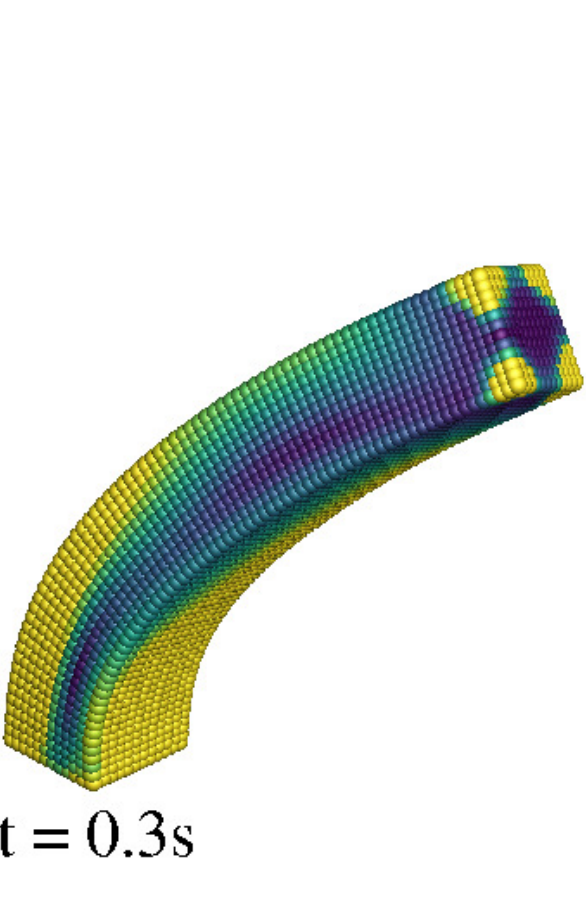}
	\includegraphics[trim = 4.5cm  1mm 9.5cm  1mm, clip, height=0.32\textwidth]{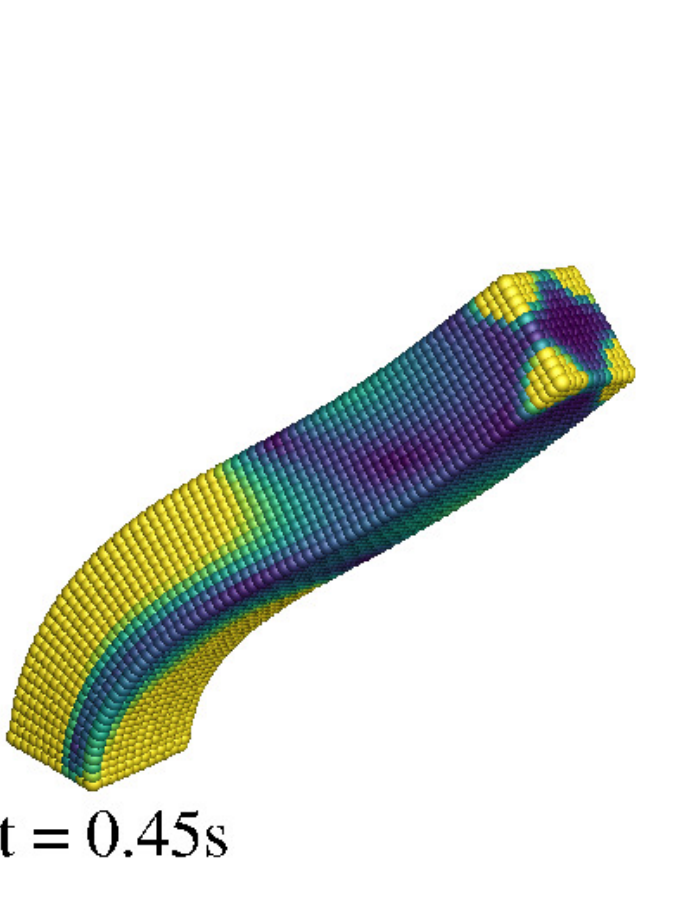}
	\includegraphics[trim = 4.5cm  1mm 9.5cm  1mm, clip, height=0.32\textwidth]{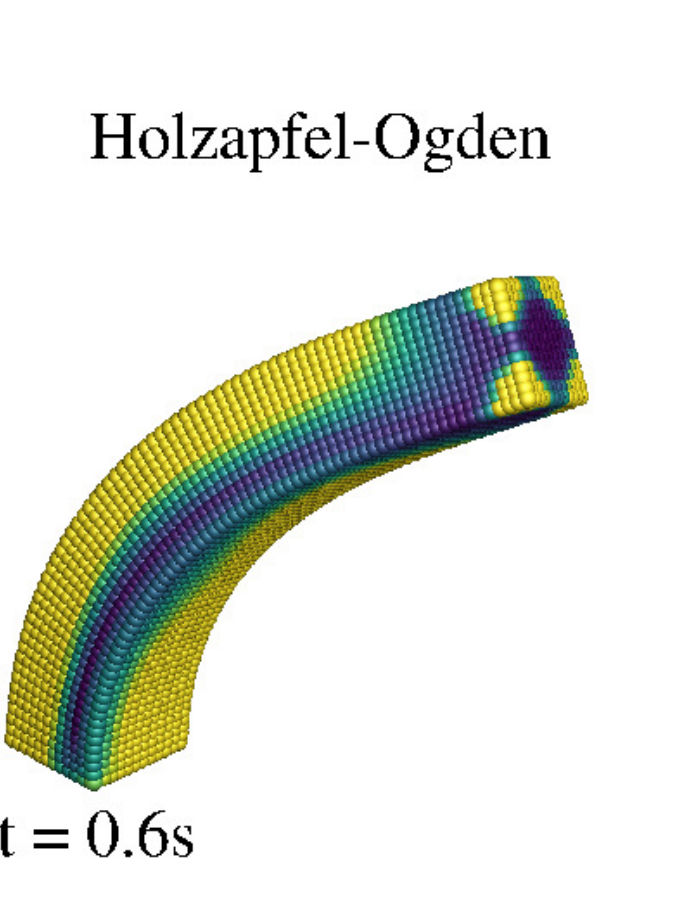}	
	\caption{Passive response of a three-dimensional bending cantilever:  
		time evolution of the von Mieses stress distribution in the deformed configuration for Neo-Hookean and Holzapfel-Ogden materials. The spatial particle discretization is $h/dp = 12$.}
	\label{figs:passive-iso-particle}
\end{figure}

\begin{figure}[htb!]
	\centering
	\includegraphics[trim = 2mm 2mm 2mm 2mm, clip, width=.85\textwidth]{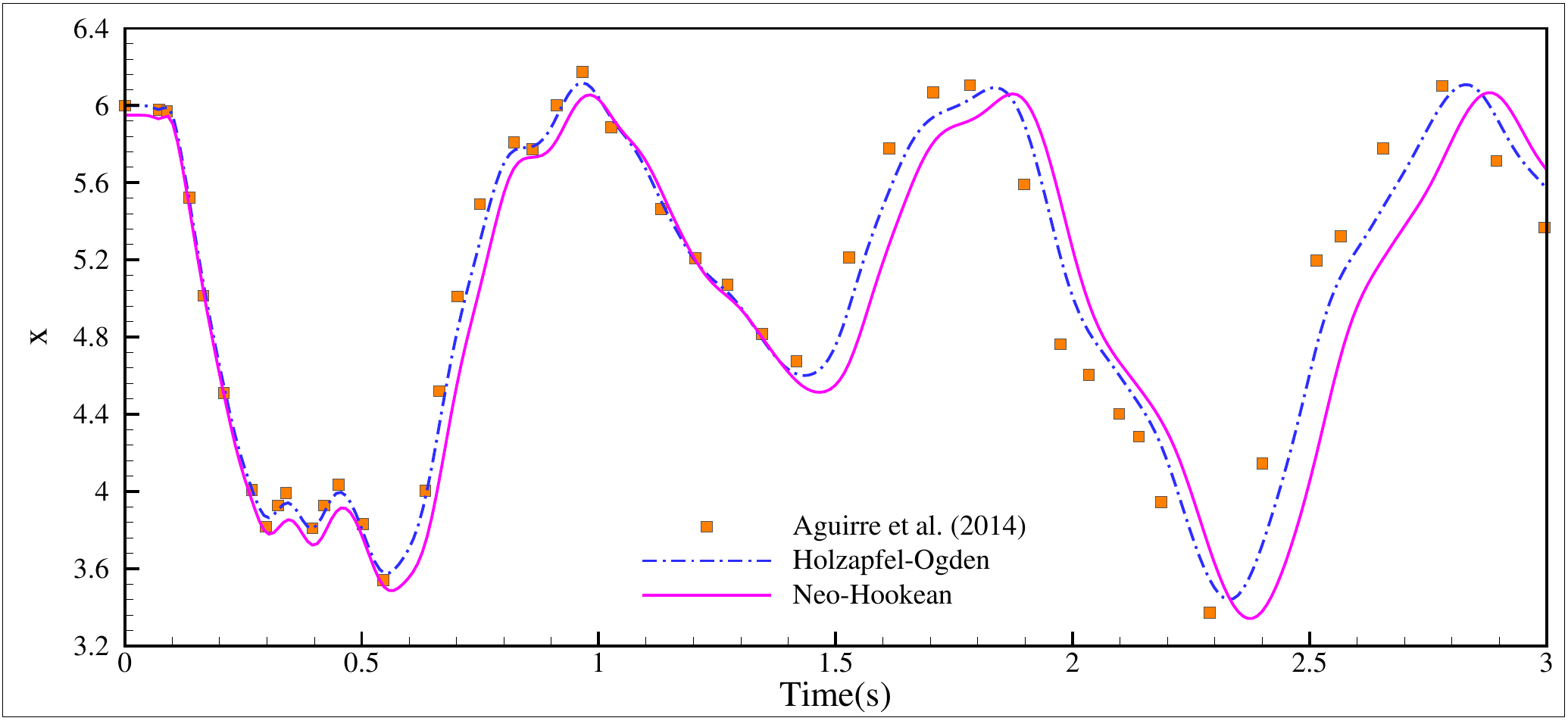}
	\caption{Passive response of a three-dimensional bending cantilever:  
		time history of the vertical position at node $S$. Results of isotropic Neo-Hookean and Holzapfel-Ogden materials are compared with that of Aguirre et al. \cite{aguirre2014vertex}. 
		The spatial particle discretization is $h / dp = 12$.}
	\label{figs:passive-iso-comparison}
\end{figure}

\begin{figure}[htb!]
	\centering
	\includegraphics[trim = 2mm 2mm 2mm 2mm, clip, width=.85\textwidth]{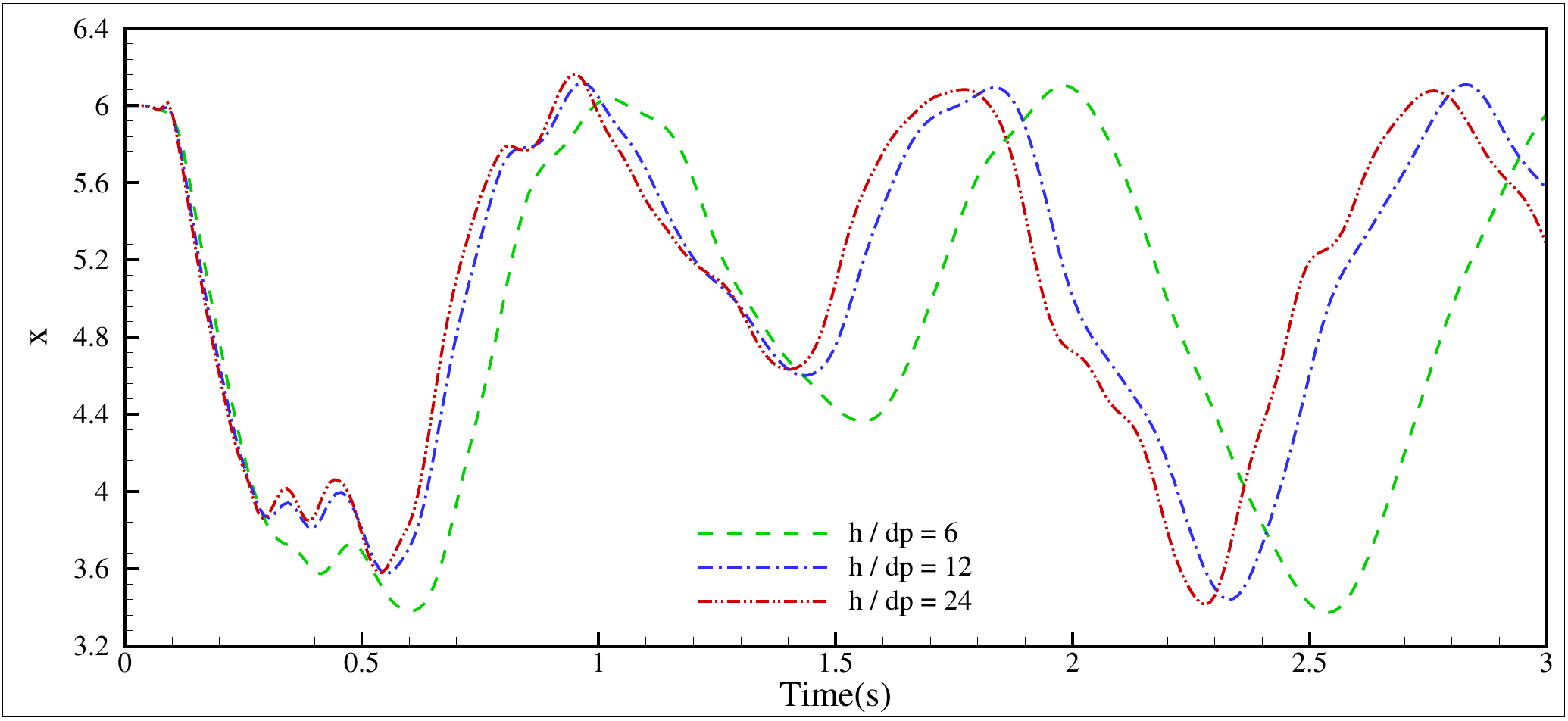}
	\caption{Passive response of a three-dimensional bending cantilever: time history of the vertical position at node S and the convergence study for the isotropic Holzapfel-Ogden material.}
	\label{figs:passive-iso-convergence}
\end{figure}

Figure \ref{figs:passive-iso-particle} shows the deformed configuration colored with von Mieses stress contours. 
Compared with the results reported in Ref. \cite{aguirre2014vertex} (see Figure 24 in their work), 
good agreement in the deformation is observed. 
Also note that both material models predict almost the same deformed configurations. 
Quantitative comparisons are given in Figure \ref{figs:passive-iso-comparison} which plots the time history of the vertical displacement of point $S$ and a good agreement is noted.
Figure \ref{figs:passive-iso-convergence} shows the convergence study of this example with isotropic Holzapfel-Odgen material model.
The convergence of the solution is illustrated with increased spatial resolution. 

\begin{figure}[htb!]
	\centering
	\includegraphics[trim = 1.75cm 1mm 9.5cm 1mm, clip, height=0.35\textwidth]{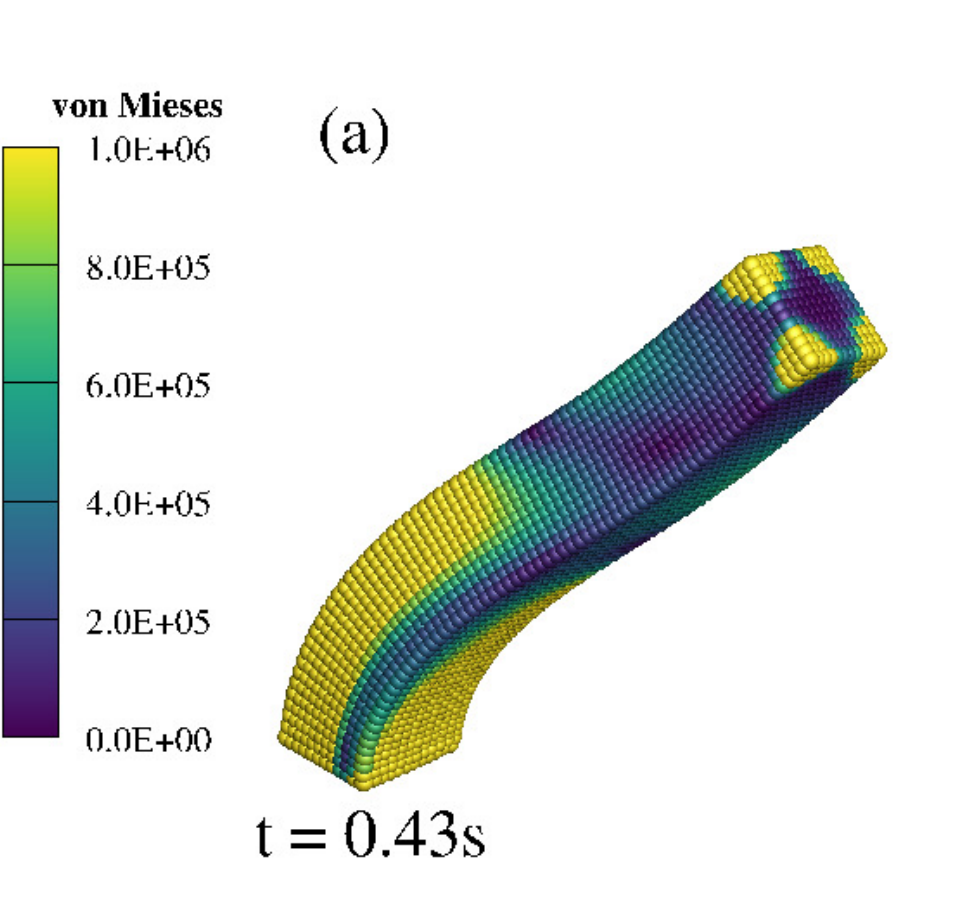}
	\includegraphics[trim = 4cm  1mm 9.5cm 1mm, clip, height=0.35\textwidth]{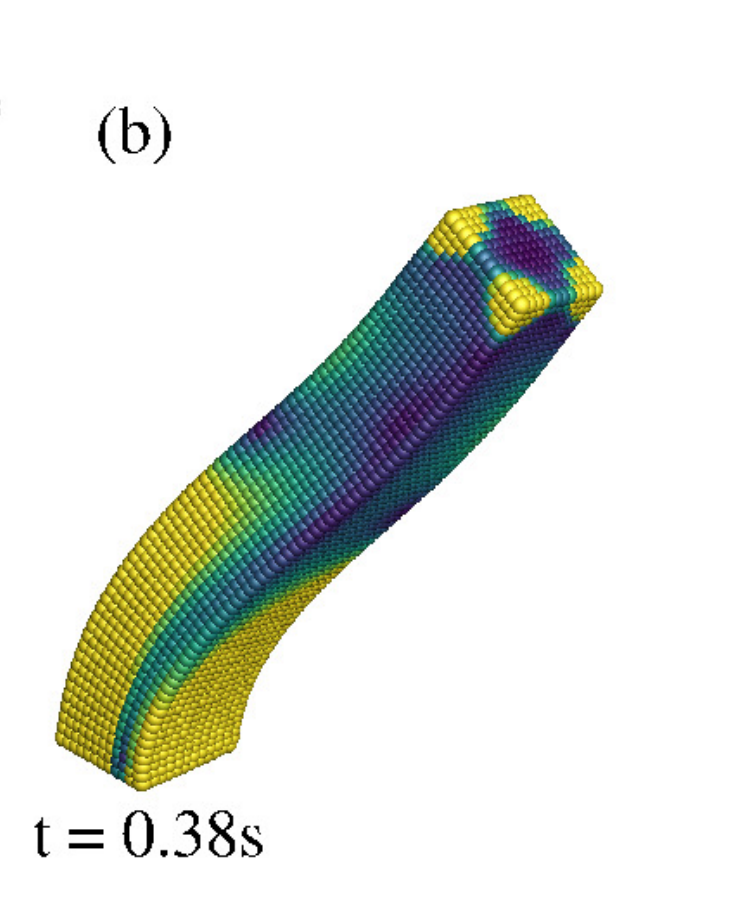}
	\includegraphics[trim = 4cm  1mm 9.5cm 1mm, clip, height=0.35\textwidth]{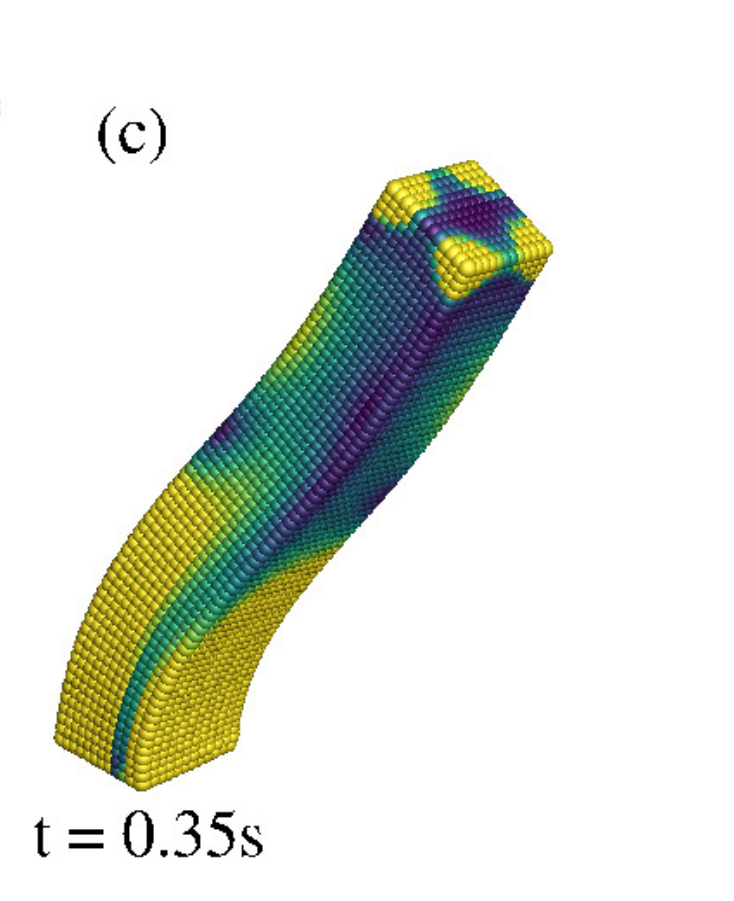}\\
	\includegraphics[trim = 1.75cm 1mm 9.5cm 1mm, clip, height=0.35\textwidth]{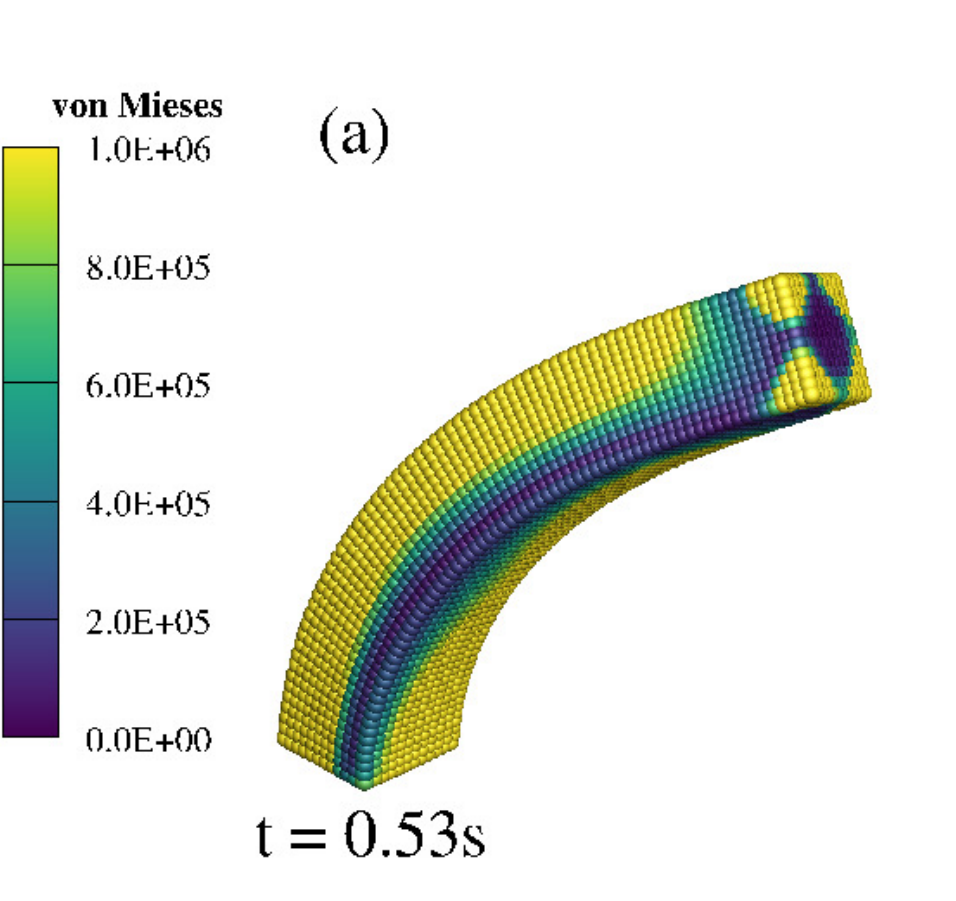}
	\includegraphics[trim = 4cm  1mm 9.5cm 1mm, clip, height=0.35\textwidth]{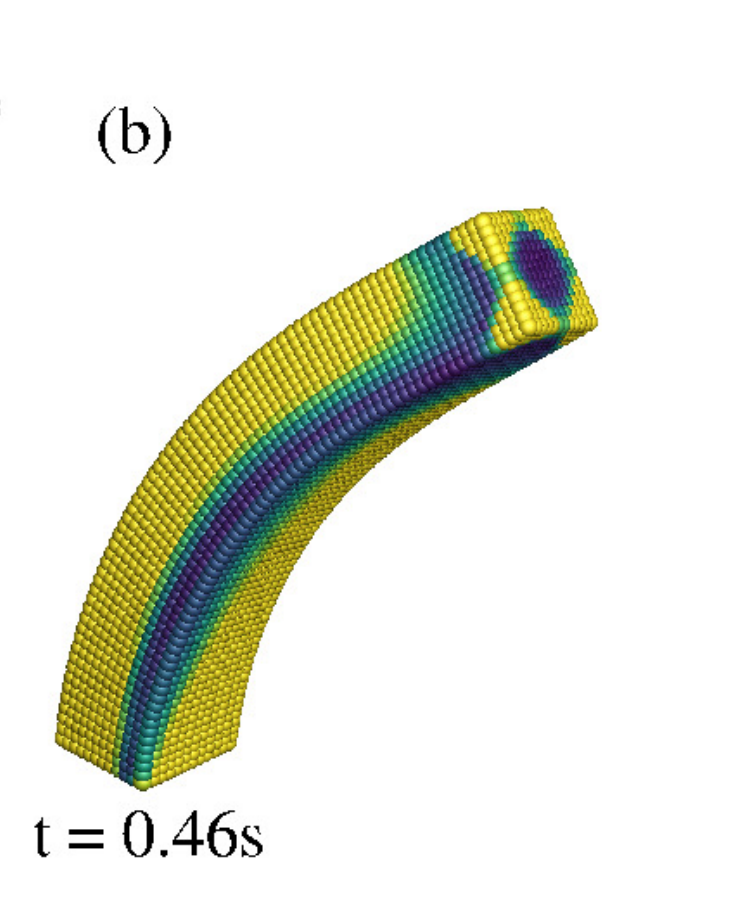}
	\includegraphics[trim = 4cm  1mm 9.5cm 1mm, clip, height=0.35\textwidth]{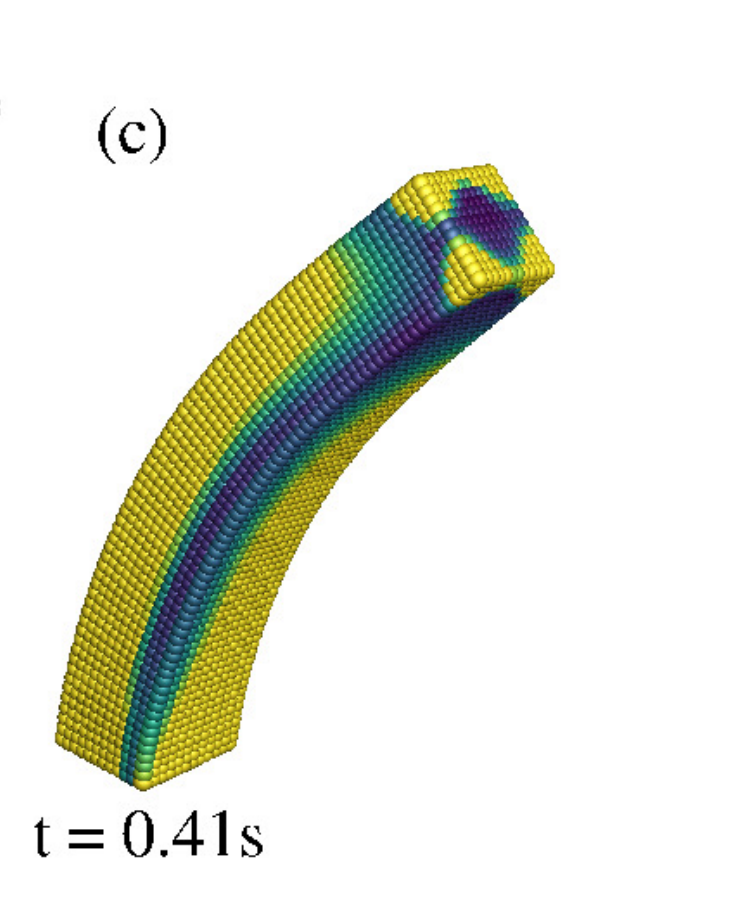}\\	
	\caption{Passive response of a three-dimensional bending cantilever: 
		time evolution of the von Mieses stress distribution in the deformed configuration for anisotropic Holzapfel-Ogden material with an spatial particle discretization of $h/dp = 12$. 
	   	The aniostropy ratio is set to (a) $a_f / a = 0.1$ , (b) $a_f / a = 0.5$ and (c) $a_f / a = 1.0$. 
   		Note that the corresponding vertical displacement of point $S$ is ploted in Figure \ref{figs:passive-ani}. }
	\label{figs:passive-ais-particle}
\end{figure}

\begin{figure}[htb!]
	\centering
	\includegraphics[trim = 2mm 2mm 2mm 2mm, clip, width=.85\textwidth]{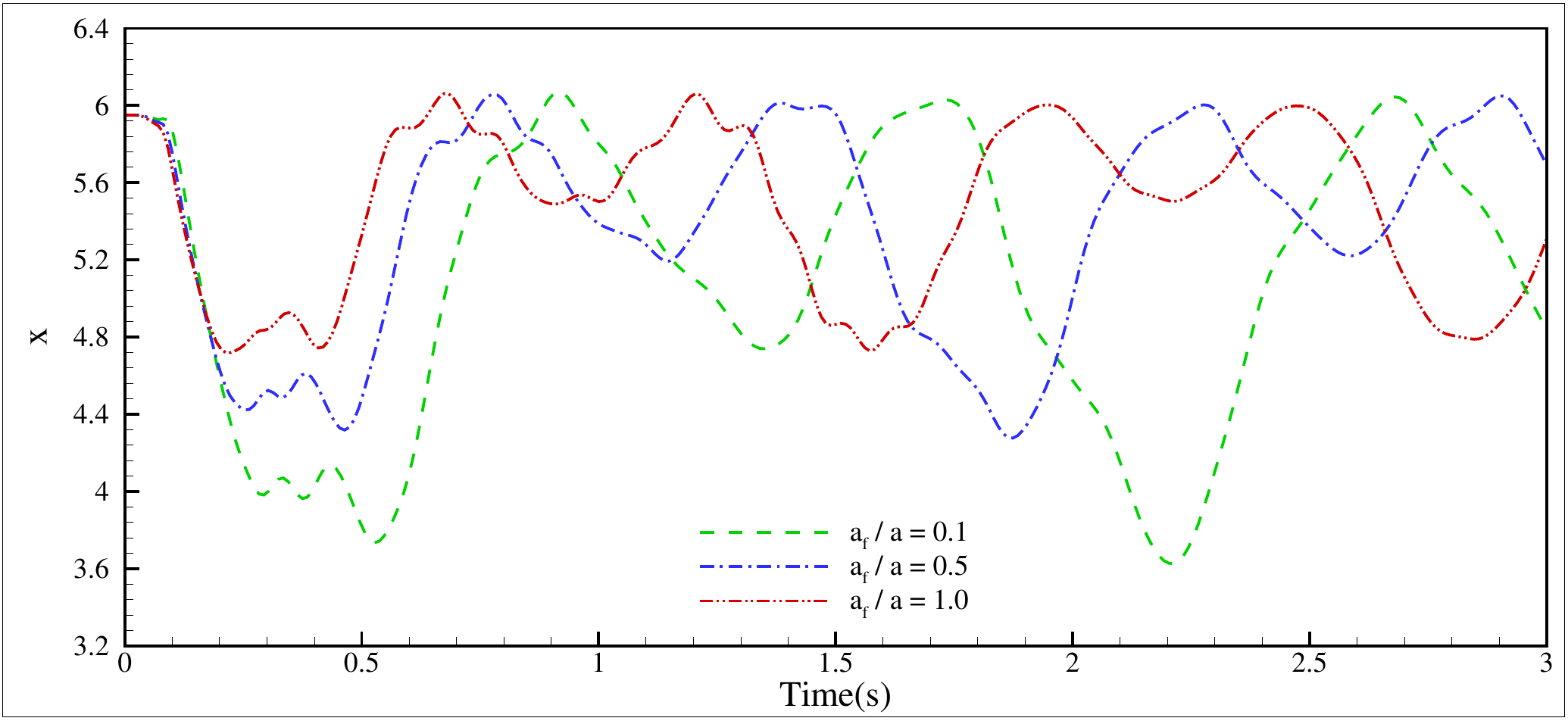}
	\caption{Passive response of a three-dimensional bending cantilever: time history of the vertical position at node $S$. 
		Study for anisotropic properties of the Holzapfel-Ogden material model with three aniostropic ratios : $a_f / a = 0.1$, $a_f / a = 0.5$ and $a_f / a = 1.0$. 
		The spatial particle discretization is $h/dp = 12$.}
	\label{figs:passive-ani}
\end{figure}

We further demonstrate the applicability of the present method by studying this example considering the anisotropic Holzapfel-Odgen material model.
For the anisotropic material, we set he fibre and sheet directions $\mathbf{f}_0$ aligned with $x$ and $y$ directions, respectively. 
Three tests with different aniostropic ratios, viz.  $a_f / a = 0.1$, $a_f / a = 0.5$ and $a_f / a = 1.0$, are studied. 
Figure \ref{figs:passive-ais-particle} shows the deformed configuration while 
Figure \ref{figs:passive-ani} plots the time history of the vertical displacement of point $S$. 
It can be observed that the deformation is reduced as the aniostropic ratio increases. 
%
\subsection{Active mechanical response}
\label{sec:active}
Following Garcia-Blanco et al. \cite{garcia2019new}, 
we consider a unit cube of myocardium with an orthogonal material direction.
The myocardium has the fiber and sheet directions parallel to the global coordinates 
and the constitutive law describing the passive response is the Holzapfel-Ogden model with the material parameters given in Table \ref{tab:ho-model1}. 
To initiate the excitation-induced response \cite{garcia2019new}, 
the transmembrane potential is linearly distributed along the vertical direction with $V_m = 0$ and $V_m = 30$ at bottom and top faces, respectively. 
For simplicity, 
the time variation of transmembrane potential is neglected and an ad-hoc activation law of active stress is given by
\begin{equation}
T_a = -0.5 V_m.
\end{equation}

\begin{figure}[htb!]
	\centering
	\includegraphics[trim = 2mm 2mm 2mm 2mm, clip, width=.95\textwidth]{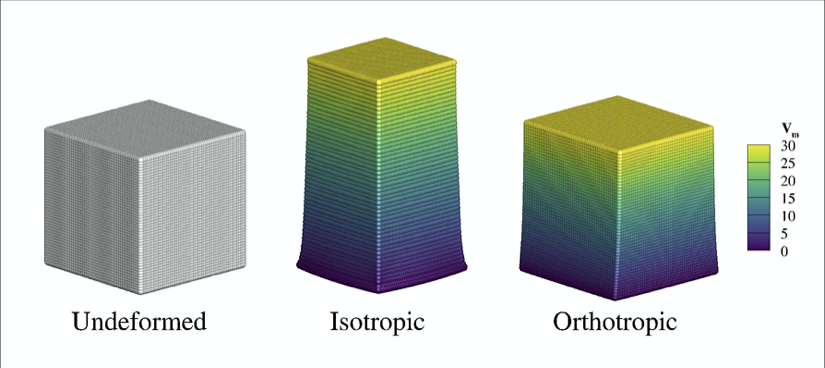}
	\caption{Active response of the unit cubic myocardium with both isotropic and anisotropic material properties: contour plot of the transmembrane potential.}
	\label{figs:active-response}
\end{figure}

Two different tests with iso- and aniso-tropic models are considered herein. 
Figure \ref{figs:active-response} shows the deformed configuration of the cubic myocardium.
Compared with the results reported in Ref. \cite{garcia2019new} (see Figure 7 in their work), 
a qualitative good agreement is noted for the isotropic test. 
Furthermore, 
the present simulation shows that the displacement of the top face is $0.53$, which is in good agreement with that of $0.535$ given in Ref. \cite{garcia2019new}. 
For the anisotropic test, 
the deformation is reduced due to the existence of the fiber and the sheet. 

\begin{table}[htb!]
	\centering
	\caption{Parameters for the Holzapfel-Ogden constitution model (For the isotropic material, the anisotropic terms are set to zero).}
	\begin{tabular}{cccc}
		\hline
		$a = 0.059$ kPa	& $a_f = 18.472$ kPa & $a_s = 2.841$ kPa  & $a_{fs} = 0.216$ kPa  \\
		\hline
		$b = 8.023$ 	& $b_f = 16.026$    & $b_s = 11.12$      & $b_{fs} = 11.436$   \\
		\hline	
	\end{tabular}
	\label{tab:ho-model1}
\end{table}
%
\subsection{Generic biventricular heart}
\label{subsec:biventricular-electromechanics}
To demonstrate the abilities of the present SPH framework in total cardiac simulation, 
we consider the transmembrane potential propagation as free pulses together with scroll waves 
and the corresponding excitation-contraction in three-dimensional generic biventricular heart. 

Following the work of \cite{sermesant2005simulation}, 
the inner surface of the left and right ventricles of the generic biventricular heart are described by two ellipsoids
\begin{equation}
\frac{x^2}{a_{lv}} + \frac{y^2}{b_{lv}} + \frac{z^2}{c_{lv}} = 1, \frac{x^2}{a_{rv}} + \frac{y^2}{b_{rv}} + \frac{z^2}{c_{rv}} = 1
\end{equation}
where $a_{lv} = 45 mm$, $b_{lv} = 54 mm$, $c_{lv} = 24 mm$ and $a_{rv} = 18 mm$, $b_{rv} = 58 mm$, $c_{rv} = 18 mm$. 
The ellipsoids are truncated from apex-to-centroid as shown in Figure \ref{figs:biventricular} (a). 
We impose a wall thickness of $6$ and $12$ on the left and right ventricle, respectively. 
To discretize the generic biventricular, particles are initialized through a relaxation-based algorithm, 
and the fiber and sheet directions are computed approximately by a coupling level-set and rule-based algorithm.
\subsubsection{Particle initialization}
\label{subsubsec:biventricular-solidmodel}
Before moving onto the simulation of biventricular heart, 
we introduce a relaxation-based technique to generate isotropic initial particle distribution. 
A coupled level-set and rule-based algorithm is also introduced for fiber and sheet reconstruction.

\begin{figure}[htb!]
	\centering
	\includegraphics[trim = 1mm 5mm 1mm 1mm, clip, width=.95\textwidth]{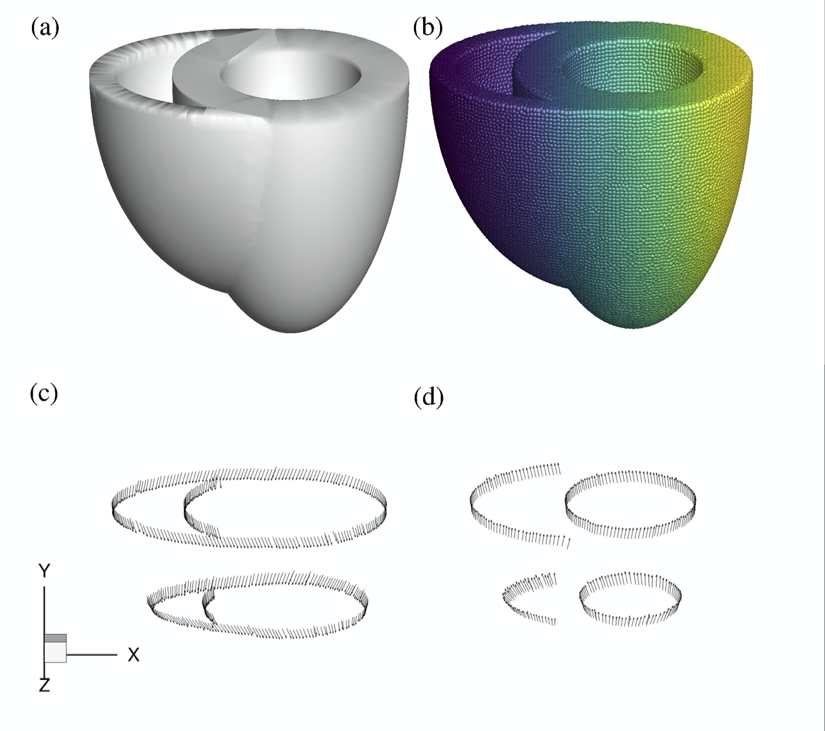}
	\caption{Generic biventricluar heart: (a) visualization of an STL file, (b) particle distribution, (c) fibre orientations of epicardium and (d) endocardium at cross sections located at $y = -10$ and $y = -30$.}
	\label{figs:biventricular}
\end{figure}

For solid dynamics, 
two approaches, 
viz, direct particle generation based on a lattice structure \cite{lacome2004smoothed} 
and particle generation based on a volume element mesh \cite{johnson2003modelling}, 
are commonly used in the SPH community. 
In the former approach, 
particles are positioned directly on a cubic lattice and equispaced particle distribution is obtained. 
Accurate surface description, in particular complex geometries, requires a fine resolution in this approach thereby rendering it limited to rather simple geometries \cite{hedayati2013new}.
The second approach convertes each volume element of a tetra or hexahedral mesh into a particle.
This approach shows advantages in describing complex geometries, 
however, 
yields significantly non-uniform particle distributions regarding the particles spacing and size. 
In this work, 
we introduce an approach initialized from standard triangle language (STL) input files, which uses the relaxation-based algorithm, proposed by Fu et al. \cite{fu2019isotropic} 
for mesh generation, to generate the initial particle distribution of the biventricular heart. 
Following Ref. \cite{fu2019isotropic}, 
a level-set field on a Cartesian background mesh is required for particle relaxation. 
In the present work, 
the geometry is described in the STL format as shown in Figure \ref{figs:biventricular} 
and a passer is used for reading data from the STL files.
Then the geometry surface is represented by the zero level-set of a signed-distance function, 
\begin{equation}
\Gamma = \{\left(x, y, z \right) | \phi\left(x, y, z\right) = 0 \}. 
\end{equation}
Here,  
the distance from a mesh point to the geometry surface is determined by finding the nearest point on all vertices 
and a positive phase is defined if the mesh point is located inside the object, otherwise a negative phase is marked. 
Then, 
the particle evolution is conducted for a number of steps following the strategy proposed by Fu et al. \cite{fu2019isotropic} (see Section 5.2 in their work).
Note that in this work a constant particle smoothing length and constant background pressure and density are used. 
Also note that the singularities are not taken into consideration, i.e.
the surface particles are only constrained on the geometry surface. 
Figure \ref{figs:biventricular} (b) shows the particle distribution for a biventricular heart after 5000 steps of relaxation with a background pressure of $p = 2.0$ and  a density of $\rho = 1.0$.  
As expected, an isotropic particle configuration is obtained and the geometry surface is reasonably well prescribed. 

Following the particle initialization, 
the fiber and sheet reconstructions are conducted. 
Assuming that the sheets are aligned with the transmural direction, 
the sheet direction can be approximated directly from the level-set function
\begin{equation}
\mathbf{s}_0 = \sign(\mathbf{N}, \mathbf{e}_y) \mathbf{N} ,
\end{equation}
where $\mathbf{N}$ is the normal direction obtained from 
\begin{equation}
\mathbf{N} = \frac{\nabla \phi}{|\nabla \phi|},
\end{equation}
and $\mathbf{e}_y$ is the normal vector parallel to the ventricular centerline, pointing from apex to base. 
For each particle, 
the sheet direction is interpolated from the level-set field by using the trilinear interpolation. 
Following the work of Quarterioni et al. \cite{quarteroni2017integrated}, 
the initial \textit{flat} fiber direction of each particle can be defined by 
\begin{equation}
\widetilde{\mathbf{f}} = \mathbf{s}_0  \times \mathbf{e}_y. 
\end{equation}
Then, the fiber direction $\mathbf{f}_0$ can be defined by rotating $\widetilde{\mathbf{f}}$ with respect to the $\mathbf{s}_0$ axis according to the following rotation formula
\begin{equation}
\mathbf{f}_0 =  \widetilde{\mathbf{f}} \cos\left( \theta \right) + \mathbf{s}_0 \times \widetilde{\mathbf{f}} \sin\left( \theta \right) + \mathbf{s}_0 \left(\mathbf{s}_0 \cdot \widetilde{\mathbf{f}} \right) \left[1 - \cos\left(\theta\right)\right], 
\end{equation}
where the rotation angle $\theta$ is computed from 
\begin{equation}
\theta = \left( \theta_{epi} - \theta_{endo} \right) \psi + \theta_{endo}. 
\end{equation}
Here $\theta_{epi} = -70^o$ and $\theta_{endo} = 80^o$ are the rotation angles at the epicardium and endocardium, respectively. 
The pseudo-distance $\psi$ is given by solving a simple Laplace equation of $\psi$ by imposing the boundary condition $\psi|_{epi} = 1$ and $\psi|_{endo} = 0$ \cite{quarteroni2017integrated}.
Figure \ref{figs:biventricular} shows the fiber direction of plane located at $y = -10 $ and $y = -30$ in epicardium (c) and endocardium (d), respectively. 
%
\subsubsection{Electrophysiology}
\label{subsubsec:biventricular-electrophysiology}
In this section, 
we consider the transmembrane potential propagation in the manner of a free pulse and a scroll wave 
for the generic biventricular heart with iso- and aniso-tropic material properties. 
The Aliev-Panfilow model is applied for monodomain equation with the constant parameters given in Table \ref{tab:ap-2} 
and the diffusion coefficients are set as  $d^{iso} = 1.0 \text{mm}^2 \cdot \text{ms}^{-1}$ and $d^{ani} = 0.1 \text{mm}^2 \cdot \text{ms}^{-1}$. 
\begin{table}[htb!]
	\centering
	\caption{Parameters for the Aliev-Panfilow model. }
	\begin{tabular}{cccccc}
		\hline
		k	& a & b  & $\epsilon_0$ & $\mu_1$ & $\mu_2$  \\
		\hline
		8.0	& 0.01   & 0.15      & 0.002 & 0.2 & 0.3  \\
		\hline	
	\end{tabular}
	\label{tab:ap-2}
\end{table}

\begin{figure}[htb!]
	\centering
	\includegraphics[trim = 2mm 2mm 2mm 2mm, clip, width=.995\textwidth]{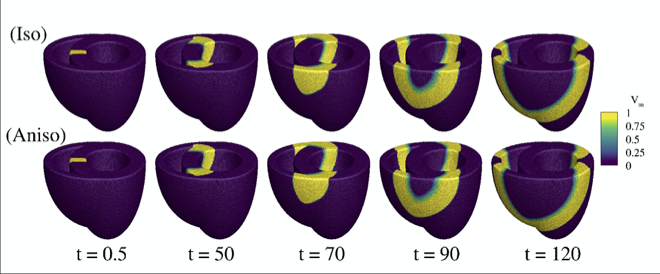}
	\caption{Generic biventricluar heart: transmembrane potential propagates in the heart in the free-pulse pattern.
			The snapshots depict contours of the transmembrane potential $V_m$.}
	\label{figs:biventricular-excitation}
\end{figure}

\begin{figure}[htb!]
	\centering
	\includegraphics[trim = 2mm 2mm 2mm 2mm, clip, width=.85\textwidth]{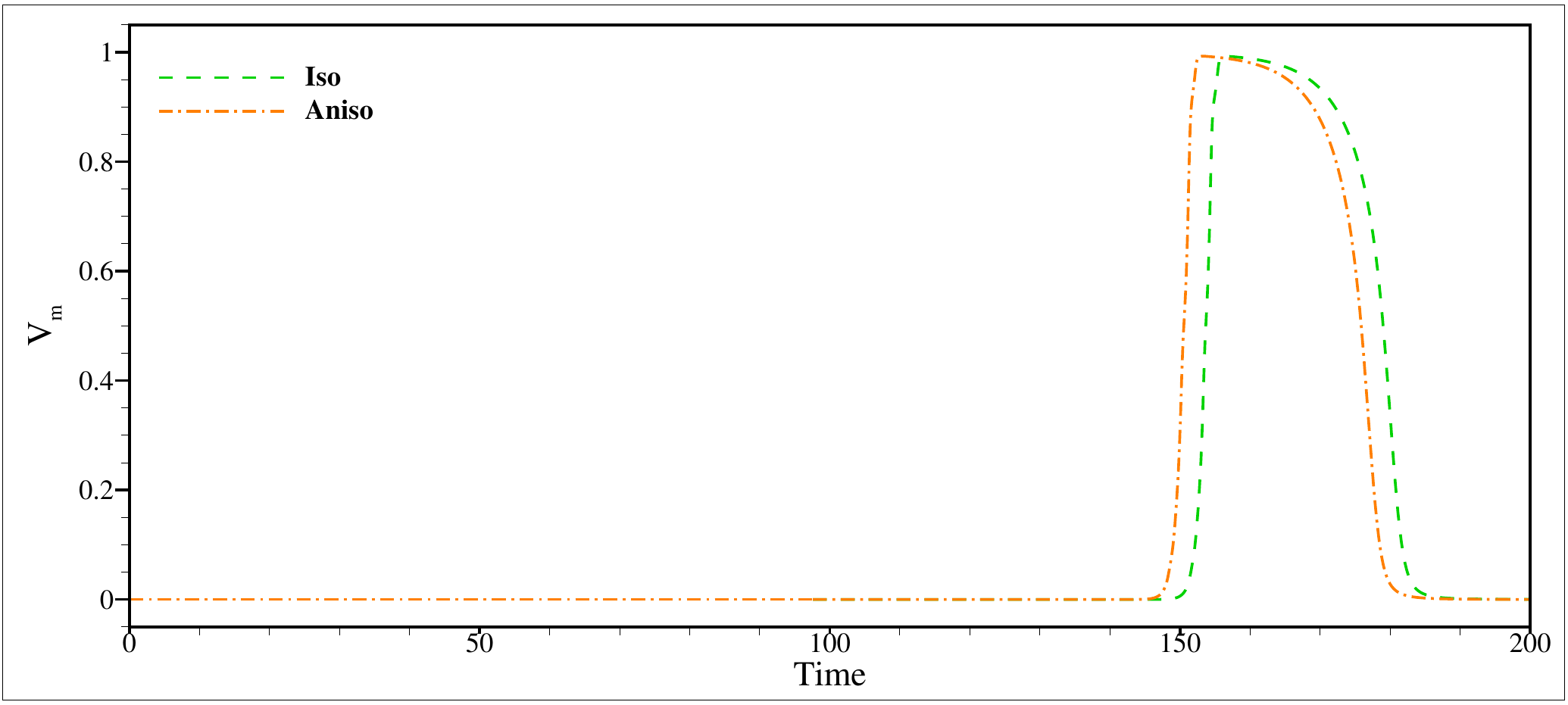}
	\caption{Generic biventricluar heart: time evolution of the transmembrane potential recorded on the apex.}
	\label{figs:biventricular-excitation-data}
\end{figure}

In the first test, 
the transmembrane potential travels in the heart in the free-pulse pattern.
One stimulus, 
termed as $S1$, 
is initiated by externally stimulating the particles 
located at the upper part of the septum (wall separating the ventricles) 
as indicated by the partially depolarized region at $t = 0.5$ in Figure \ref{figs:biventricular-excitation} with an stimulation $V_m = 0.92$. 
The stimulus generates the depolarization through the heart as shown in Figure \ref{figs:biventricular-excitation}. 
It can be observed that the transmembrane potential propagates in the similar pattern for both iso- and aniso-tropic material model.
For more comprehensive comparison, 
the transmembrane potentials recorded at apex are plotted in Figure \ref{figs:biventricular-excitation-data}. 
As expected, the transmembrane potential propagates in anisotropic model faster than that in isotropic model. 
For both iso- and aniso-tropic materials, 
the transmembrane potential profiles show a good agreement with the results reported in Ref. \cite{aliev1996simple}. 

As mentioned in Section \ref{sec:2dspiralwave},
the present method shows good accuracy in reproducing two-dimensional spiral waves.
In this section, 
we will demonstrate the present method's ability to reproduce the formation of scroll waves in a more complex biventricular heart.  
We consider two tests: one with single scroll wave and another with two scroll waves interacting with each other when propagating. 
To generate the scroll wave, the S1-S2 protocol, 
where a second broken stimulus (S2) is triggered during the repolarization phase of the S1 wave, 
is applied. 

For a single scroll wave, 
the S2 stimulus is initiated at a small region of
\begin{equation}
\Omega = \{\left(x, y \right) | 0 \leq x \leq 6  \land -6 \leq y \leq 0  \}
\end{equation}
located at the anterior ventricular wall from time $t = 105$ to $t = 105.2$ with an external stimulation $V_m = 0.95$. 
Figure \ref{figs:biventricular-excitation-twowave} shows the formation and evolution of the vortex wave re-entry 
in both iso- and aniso-tropic material models.  
It can be observed that the combination of the complex biventricular geometry, 
the non-symmetric perturbation and the inhomogeneous fiber and sheet orientation clearly triggers 
a chaotic non-stationary wave pattern with the center of the scroll moving in the septal basal region. 
Figure \ref{figs:biventricular-excitation-data-singlewave} gives the recorded profiles of the transmembrane potential at the apex. 
After the depolarization of the S1 wave, 
self-oscillatory transmembrane potential is noted due to the propagation of scroll waves. 
Note that the self-oscillatory transmembrane potential shows higher frequency in anisotropic model compared to the isotropic model.
\begin{figure}[htb!]
	\centering
	\includegraphics[trim = 2mm 2mm 2mm 2mm, clip, width=.995\textwidth]{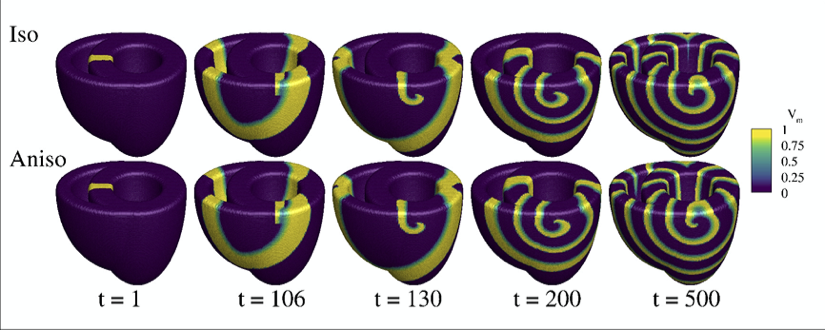}
	\caption{Generic biventricluar heart: transmembrane potential propagation on both iso- and aniso-tropic models.
		The snapshots depict contours of the transmembrane potential $V_m$.}
	\label{figs:biventricular-excitation-singlewave}
\end{figure}
\begin{figure}[htb!]
	\centering
	\includegraphics[trim = 2mm 2mm 2mm 2mm, clip, width=.85\textwidth]{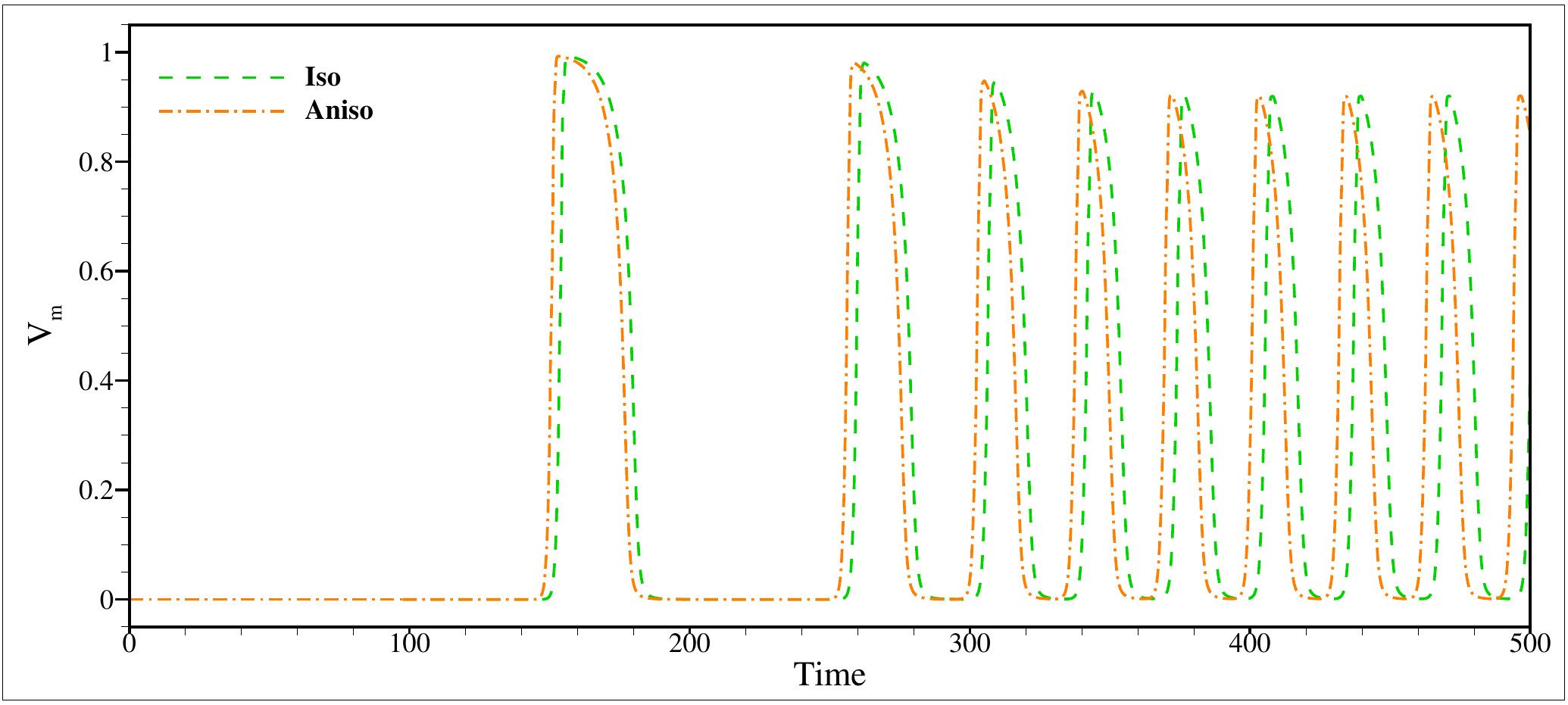}
	\caption{Generic biventricluar heart: time evolution of the transmembrane potential recorded at the apex.}
	\label{figs:biventricular-excitation-data-singlewave}
\end{figure}

For the two scroll waves,  
the initiated region of S2 is extended to
\begin{equation}
\Omega = \{\left(x, y \right) | 0 \leq x \leq 12 \land -24 \leq y \leq 0  \}, 
\end{equation}
and the initiated time is changed to the interval between $t = 125$ and $t = 125.2$.
Figure \ref{figs:biventricular-excitation-twowave} shows the formation and evolution of two vortex waves re-entry in both iso- and aniso-tropic material models.  
Compared with the previous single wave model,
a more complex chaotic non-stationary wave pattern is generated. 
Figure \ref{figs:biventricular-excitation-data-singlewave} gives the recorded profiles of the transmembrane potential at the apex. 
As expected, 
the self-oscillatory transmembrane potential shows higher frequency in anisotropic model than isotropic model. 
These two tests demonstrate the ability of the proposed method to reproduce the evolution of the re-entrant scroll waves on complex cardiac geometries.
\begin{figure}[htb!]
	\centering
	\includegraphics[trim = 2mm 2mm 2mm 2mm, clip, width=.995\textwidth]{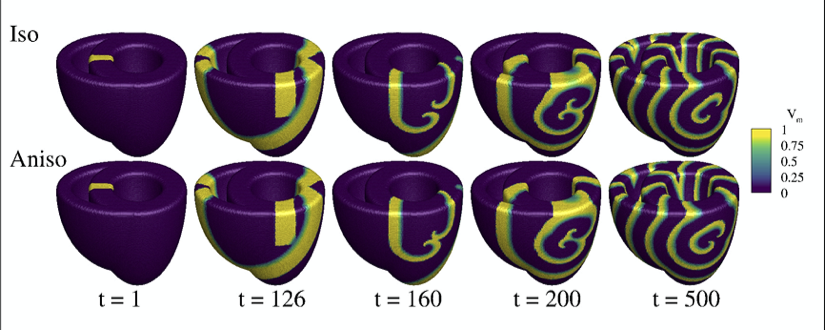}
	\caption{Generic biventricluar heart: transmembrane potential propagation on both iso- and aniso-tropic models.
		The snapshots depict contours of the transmembrane potential $V_m$.}
	\label{figs:biventricular-excitation-twowave}
\end{figure}
\begin{figure}[htb!]
	\centering
	\includegraphics[trim = 2mm 2mm 2mm 2mm, clip, width=.85\textwidth]{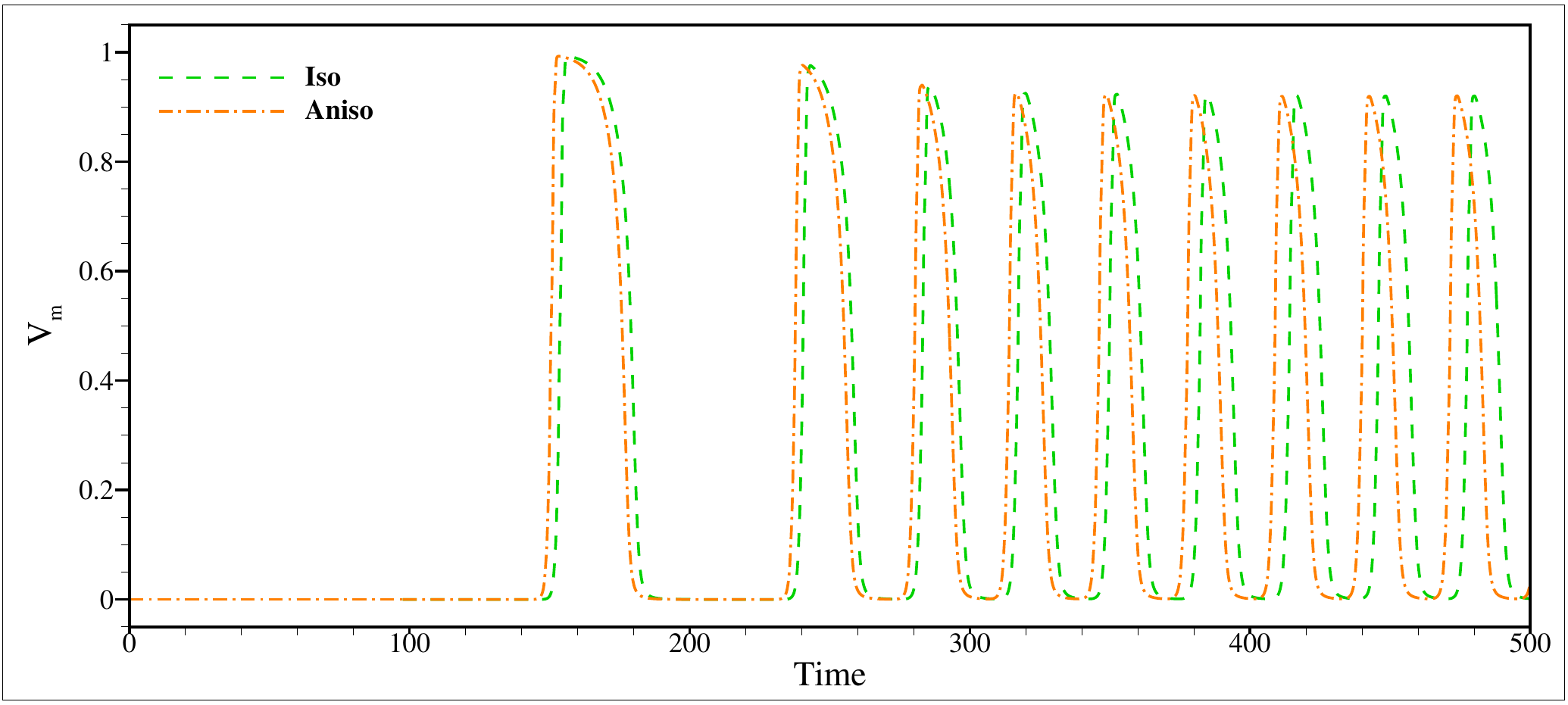}
	\caption{Generic biventricluar heart: time evolution of transmembrane potential recorded at the apex.}
	\label{figs:biventricular-excitation-data-twowave}
\end{figure}
%
\subsubsection{Excitation-contraction}
\label{subsubsec:biventricular-electromechanics}
In this section, 
we demonstrate that the basic feature of the cardiac function can be captured by the present SPH framework in a reasonable manner 
by modeling the excitation-contraction of the biventricular heart through electromechanical coupling. 
Three different excitations, including
not only the free pulse but also the scroll waves represented in Section \ref{subsubsec:biventricular-electrophysiology}, 
are considered. 
As a matter of fact, 
the scroll wave may correspond to pathological heart diseases, 
i.e. cardiac arrhythmias can be related to the presence of wavefront spirals which lead to an irregular contraction of the cardiac muscle. 
Therefore, 
reproducing the excitation-contraction under the scroll wave may extend human understanding of cardiac arrhythmias. 
For simplicity, 
the displacement degrees of freedom on the top base are constrained and the whole heart surface is assumed to be flux-free. 
To record the heart displacement, 
three nodes, namely A located at $(0, -30, 26)^T \text{mm}$, B at $(0, -70, 0)^T \text{mm}$ and C at $(-30, -50, 0)^T \text{mm}$, are used. 
Moreover, 
the constant parameters of  Holzapfel-Ogden model are given in the Table \ref{tab:ho-model1} and the active contraction stress is $T_a = 0.15 ~\text{kPa}$. 

\begin{figure}[htb!]
	\centering
	\includegraphics[trim = 1mm 1.5cm 1mm 1.5cm, clip, width=.975\textwidth]{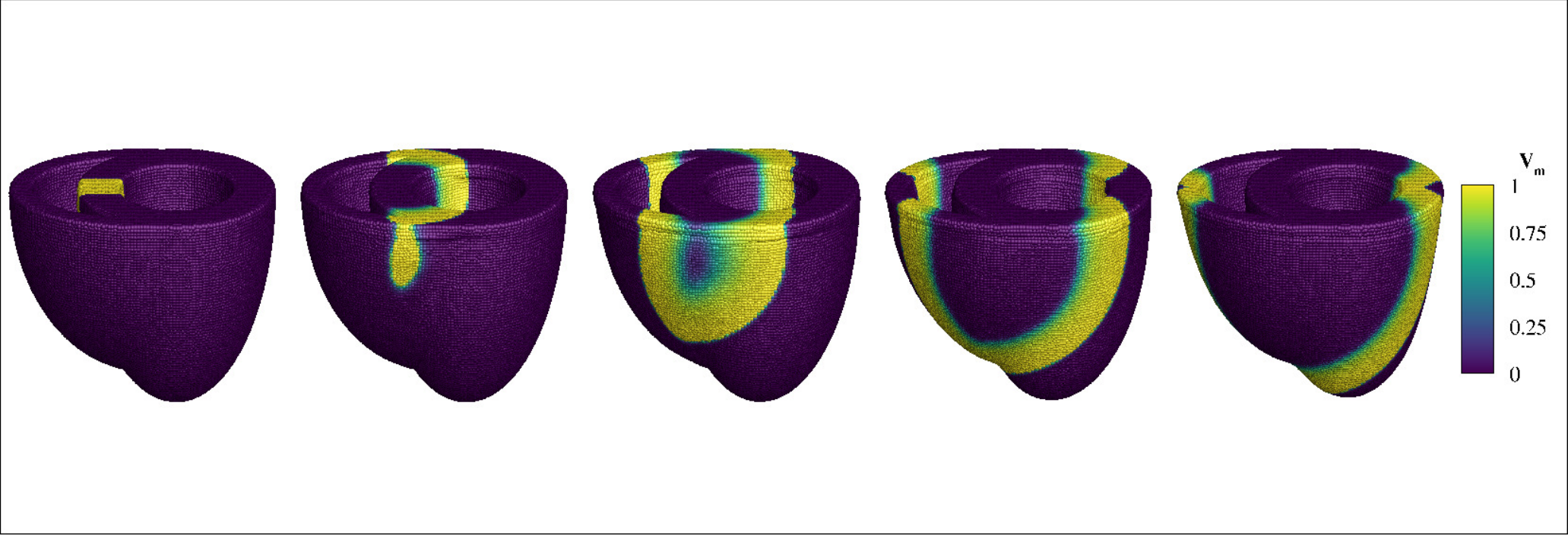}
	\includegraphics[trim = 1mm 1cm 1mm 1.5cm, clip, width=.975\textwidth]{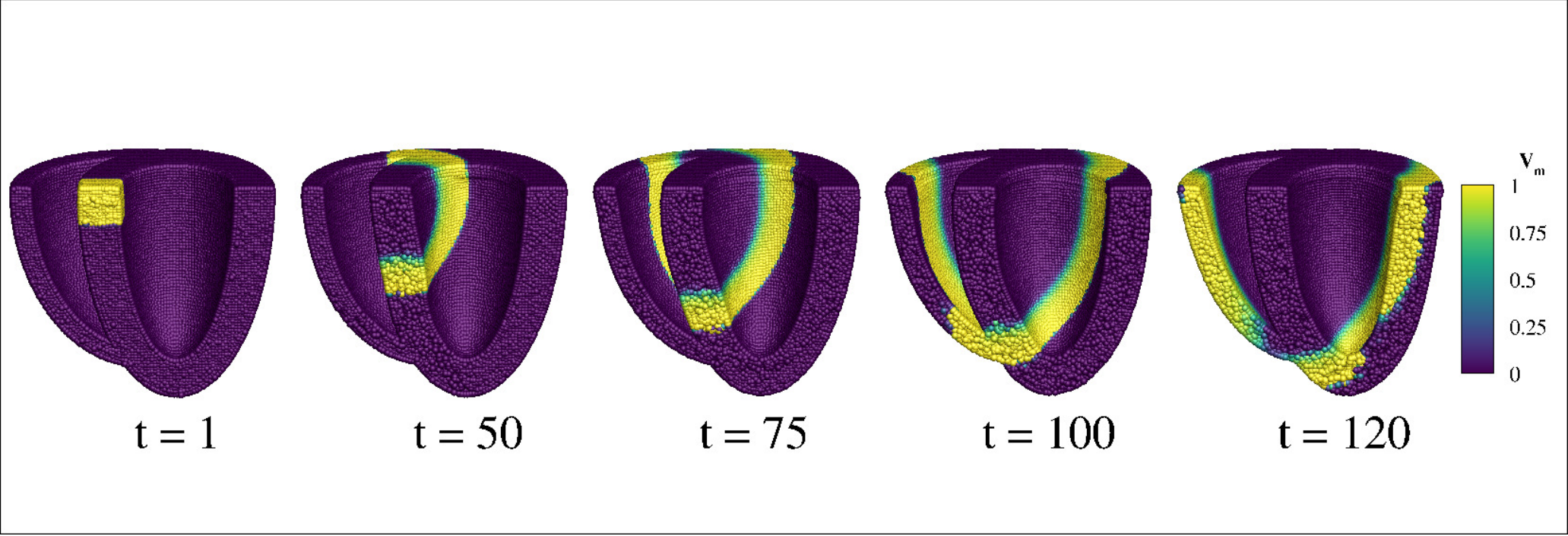}
	\caption{Generic biventricular heart: coupled excitation-contraction induced by the transmembrane potential as a free pulse. 
		Snapshots of the deformed body depict the transmembrane potential contours at different stages of the depolarization and the corresponding cross sections.}
	\label{figs:biventricular-excitation-contraction}
\end{figure}

\begin{figure}[htb!]
	\centering
	\includegraphics[trim = 2mm 2mm 1mm 2mm, clip, width=.85\textwidth]{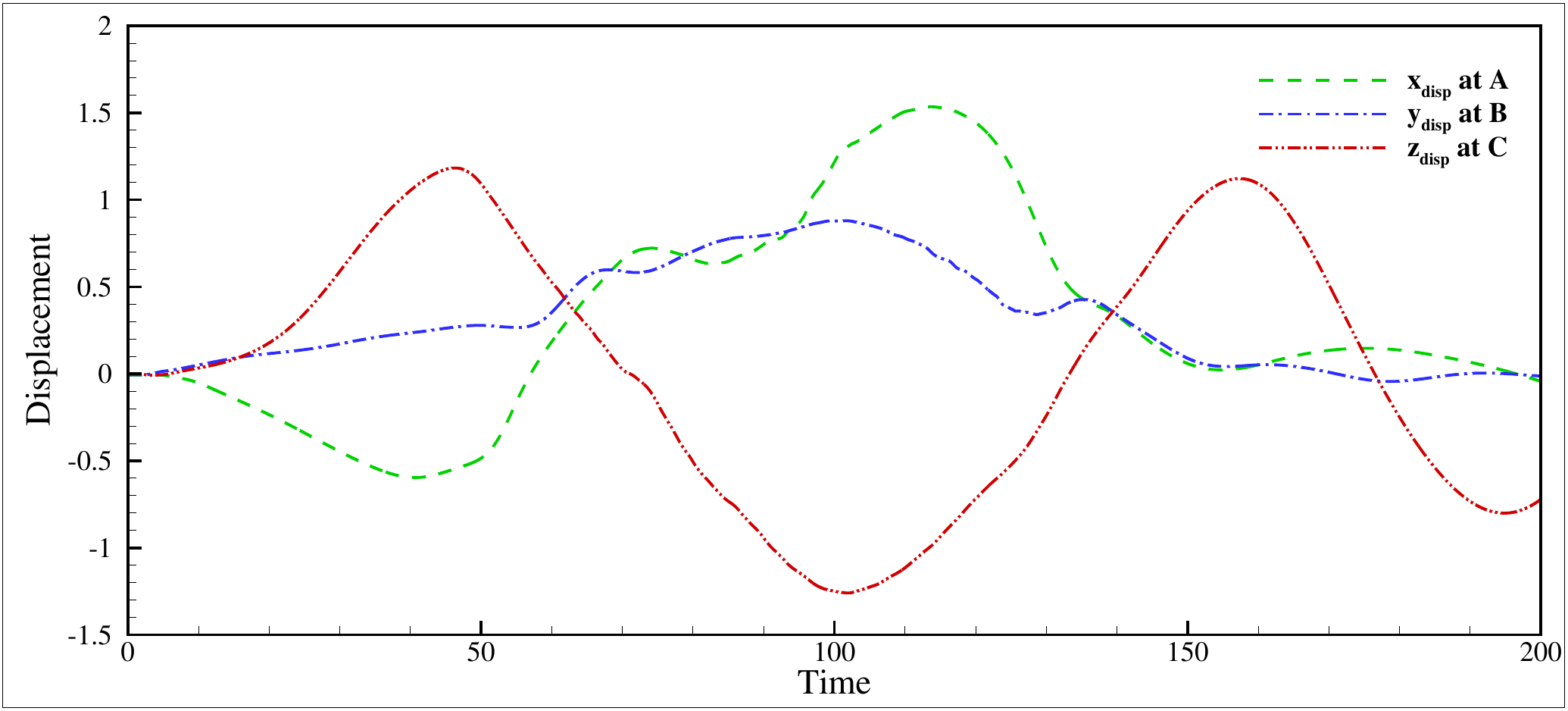}
	\caption{Generic biventricular heart: coupled excitation-contraction induced by the transmembrane potential as a free pulse.
					The time histories of displacement at nodes A, B and C.}
	\label{figs:biventricular-excitation-contraction-data}
\end{figure}

In the first test, 
we consider excitation-contraction under the transmembrane potential propagation as a free-pulse. 
Figure \ref{figs:biventricular-excitation-contraction} shows the resulting excitation-contraction of the heart 
with the transmembrane potential contours and the corresponding cross sections. 
It can be observed that excitation-contraction gives rise to the upward motion of the apex 
as the depolarization front traveling through the heart. 
Also, 
the apex's upward motion is accompanied by 
the physiologically observed wall thickening and the overall torsional motion of the heart 
as shown in the cross sections of Figure \ref{figs:biventricular-excitation-contraction}. 
This physiologically active response through the non-uniform contraction of myofibers is due to the inhomogeneous myocyte orientation distribution incorporated with the anisotropic material model. 
Figure \ref{figs:biventricular-excitation-contraction-data} shows the time evolution of the $x$, $y$ and $z$ components of the displacement at points A, B and C, respectively. 
At the end of the depolarization process, 
the reference configuration is recovered. 

\begin{figure}[htb!]
	\centering
	\includegraphics[trim = 1mm 2cm 1mm 1.5cm, clip, width=.975\textwidth]{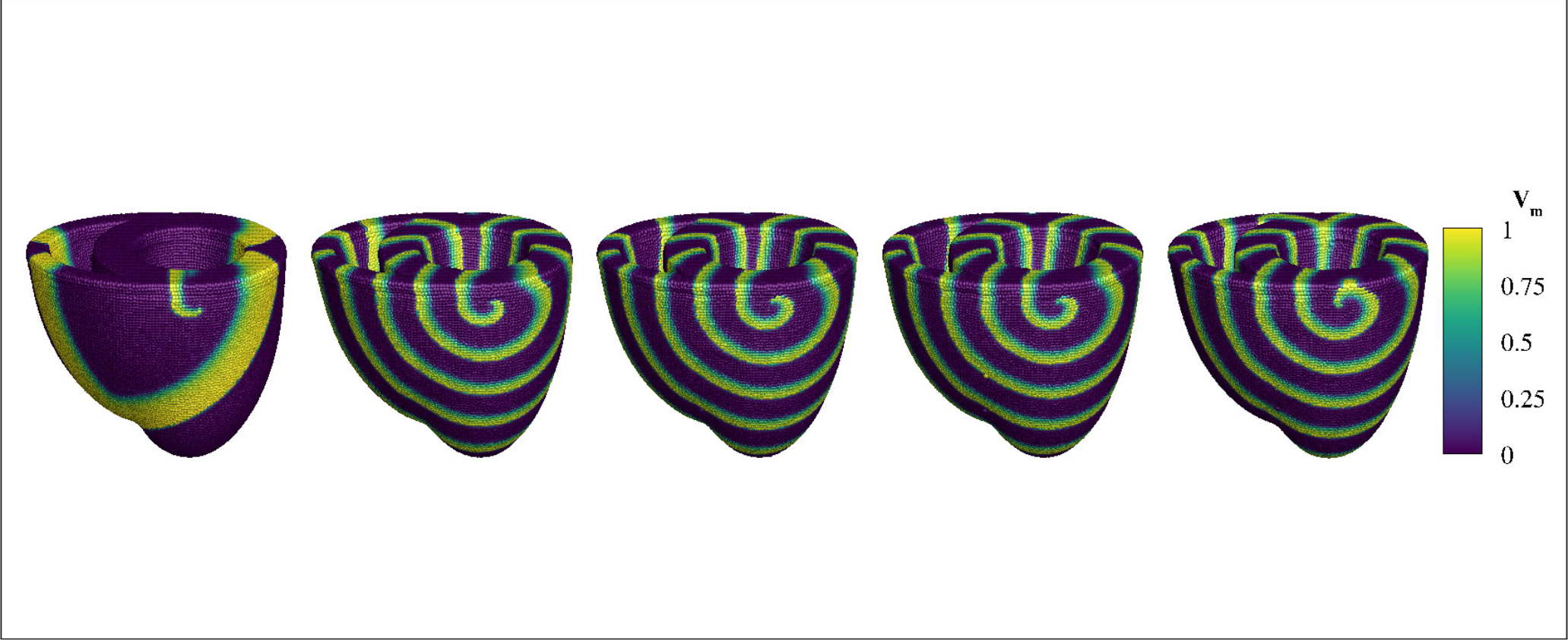}
	\includegraphics[trim = 1mm 1cm 1mm 2.5cm, clip, width=.975\textwidth]{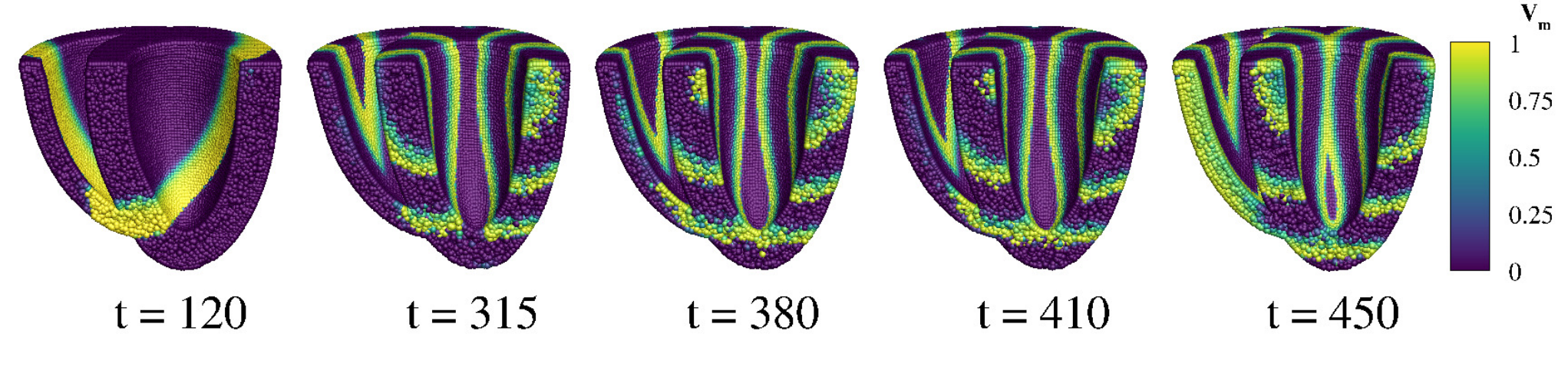}
	\caption{Generic biventricular heart: coupled excitation-contraction induced by the transmembrane potential as a single scroll wave. 
					Snapshots of the deformed body depict the transmembrane potential contours at different stages of the depolarization and the corresponding cross sections.}
	\label{figs:biventricular-excitation-contraction-singlewave}
\end{figure}

\begin{figure}[htb!]
	\centering
	\includegraphics[trim = 2mm 2mm 1mm 2mm, clip, width=.85\textwidth]{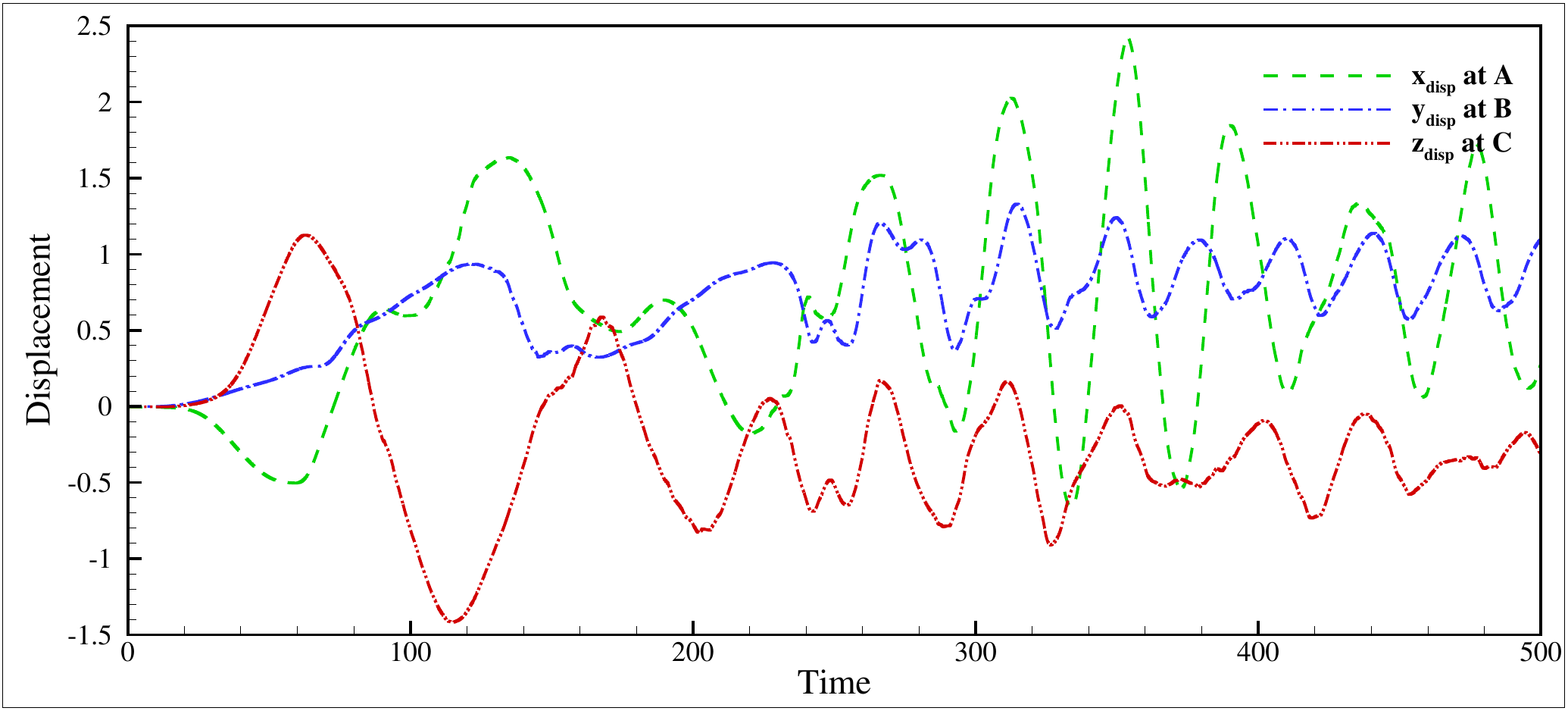}
	\caption{Generic biventricluar heart: coupled excitation-contraction induced by the transmembrane potential as a single scroll wave.
		The time histories of displacement at nodes A, B and C.}
	\label{figs:biventricular-excitation-contraction-singlewave-data}
\end{figure}

In the second test, 
we consider the excitation-contraction corresponding to the single scroll wave. 
The resulting excitation-contraction of the heart is shown in Figure \ref{figs:biventricular-excitation-contraction-singlewave}. 
As the propagation of the scroll wave, 
the myocytes show oscillatory excitation-contraction and the heart is under contracted state during the simulation. 
Figure \ref{figs:biventricular-excitation-contraction-singlewave-data} shows the time evolution of the $x$, $y$ and $z$ components of the displacement at points A, B and C, respectively. 
Different from the previous results for a free pulse, 
the motion of the observed points are highly oscillatory and non-recoverable. 

\begin{figure}[htb!]
	\centering
	\includegraphics[trim = 1mm 2cm 1mm 1.5cm, clip, width=.975\textwidth]{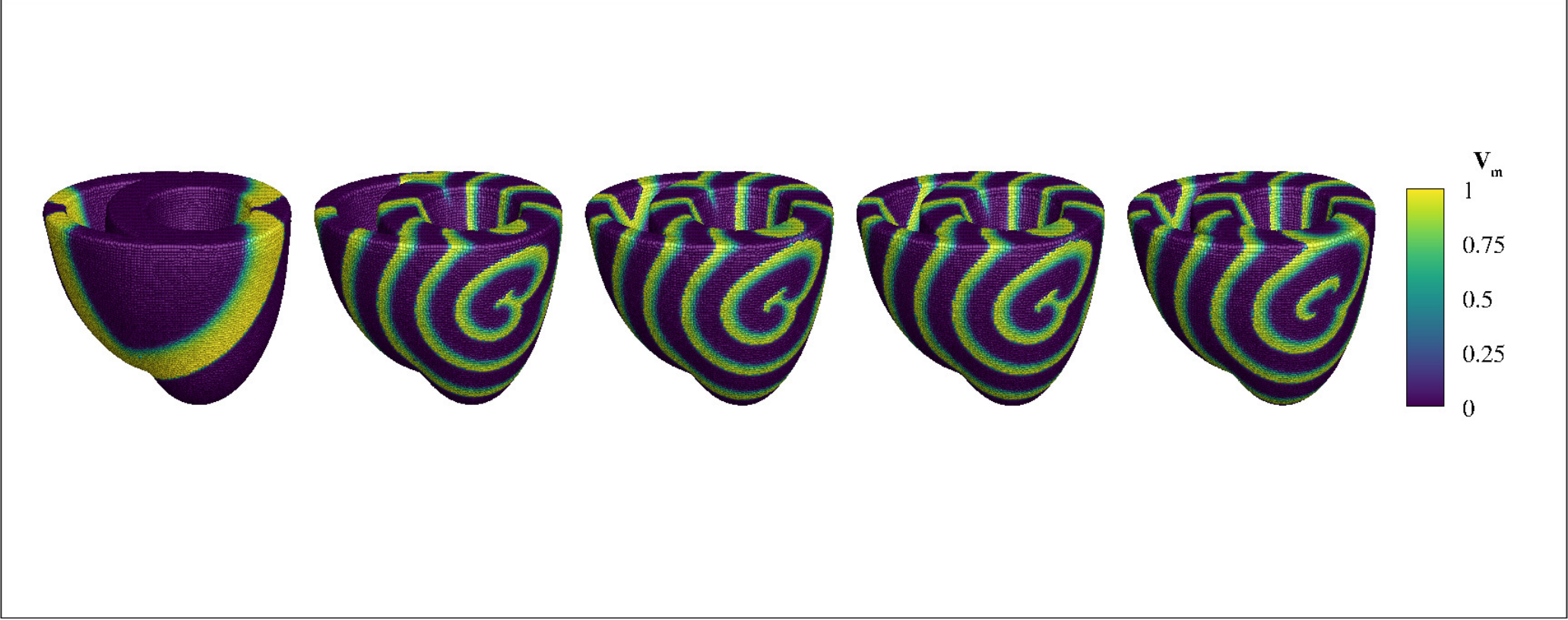}
	\includegraphics[trim = 1mm 1cm 1mm 2.5cm, clip, width=.975\textwidth]{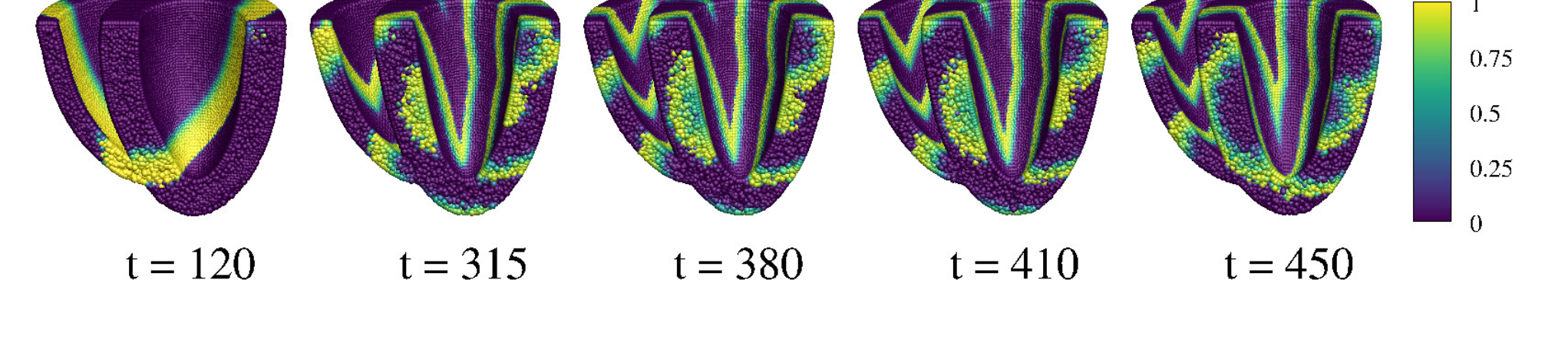}
	\caption{Generic biventricluar heart: coupled excitation-contraction induced by the transmembrane potentials double scroll waves. 
					Snapshots of the deformed body depict the transmembrane potential contours at different stages of the depolarization and the corresponding cross sections.}
	\label{figs:biventricular-excitation-contraction-twowave}
\end{figure}

\begin{figure}[htb!]
	\centering
	\includegraphics[trim = 2mm 2mm 1mm 2mm, clip, width=.85\textwidth]{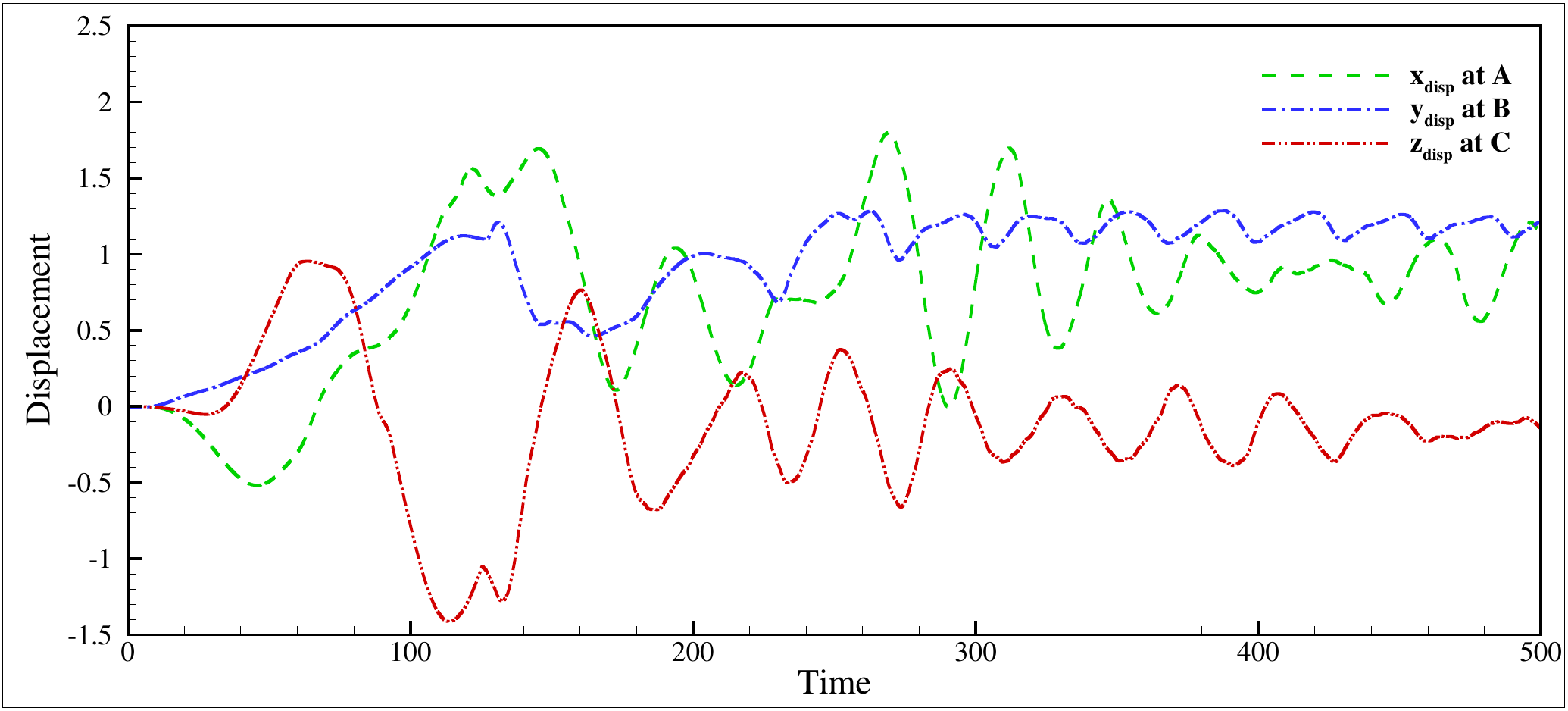}
	\caption{Generic biventricluar heart: coupled excitation-contraction induced by the transmembrane potential as two scroll waves. 
		The time histories of displacement at nodes A, B and C.}
	\label{figs:biventricular-excitation-contraction-twowave-data}
\end{figure}

In the third test,
the two scroll wave excitation-induced contraction is investigated and the results 
are shown in Figure \ref{figs:biventricular-excitation-contraction-twowave} with the transmembrane potential field and the corresponding cross sections.
Also here, 
the myocyte shows oscillatory excitation-contraction resulting the contracted state of the heart. 
As the propagation of the scroll wave, 
the myocytes show oscillatory excitation-contraction and the heart is under contracted state during the simulation. 
Figure \ref{figs:biventricular-excitation-contraction-twowave-data} shows the time evolution of the $x$, $y$ and $z$ components of the displacement  at points A, B and C, respectively. 
Different from the previous results for a single scroll wave, 
the amplitude of the oscillation is decreased while the frequency is slightly increased.
%
%
%
\section{Concluding remarks}\label{sec:conclusion}
As a realistic starting-point for developing a unified SPH approach for simulating total heart function, 
this paper address the numerical modeling of  many challenging aspects of heart function, 
including cardiac electrophysiology, passive mechanical response and the electromechanical feedback. 
For electrophysiology, 
we solve the monodomain equation by introducing a splitting reaction-by-reaction method combined with quasi-steady-state (QSS) solver to capture the stiff waves. 
For stable prediction of the large deformations and the strongly anisotropic behavior of the myocardium, 
we employ the total Lagrangian SPH formulation. 
Then, 
the coupling of electrophysiology and tissue mechanics for electromechanical feedback is conducted 
in unified SPH framework. 
A comprehensive and rigorous study of iso- and aniso-tropic diffusion process,  
transmembrane potential propagation in the free-pulse and spiral wave pattern, 
passive and active responses of myocardium, 
electrophysiology and electromechanics in a generic biventricular heart model
has been conducted. 
The results demonstrate the robustness, accuracy and feasibility of the proposed SPH framework 
for cardiac electrophysiology and electromechanics. 

The SPH methods developed in this work are the main components of an unified meshless approach 
for multi-physics modeling of total cardiac function. 
In the future work, an open-heart simulator based on our opensource SPHinXsys library will be developed for total human heart modeling. 
In particular, 
fully coupled fluid-electro-structure interactions, which consist of four chambers and four valves as electrically excitable, 
deformable and electroactive bodies interacting with the blood flows will be modeled. 
These studies will be beneficial for understanding fundamental mechanisms of the total cardiac function.
%
%
\section{Acknowledgement}
The authors would like to thank Dr. Luca Ratti for sharing the dataset used in Section \ref{sec:actionpotential-simple} 
to validate our results and express their gratitude to Deutsche Forschungsgemeinschaft for their sponsorship 
of this research under grant number DFG HU1572/10-1 and DFG HU1527/12-1. 
%
%
\section*{References}
\bibliography{mybibfile}
%
%
\section*{Appendices}
\subsection*{Appendix A : Algorithms of the SPH framework for the simulation of cardiac functions.}
\label{app_a}
\begin{algorithm}[htb!]
	Setup parameters and initialize the simulation\;
	Compute the correction matrix $\mathbb{B}$ for each particle\;
	\While{simulation is not finished}
	{
		Compute time-step size $\Delta t_p$\;
		Integrate the ODE $ \dot{V}_m  = \frac{1}{C_m } I_{ion}$ for half time step $\frac{1}{2} \Delta t_p$\;
		Integrate the ODE $\dot{w}  = g$ for  half time step $\frac{1}{2} \Delta t_p$\;
		Integrat the diffusive operator $\dot{V}_m  = \frac{1}{C_m } \nabla \cdot (\mathbb{D} \nabla V_m)$ for a time step $\Delta t_p$\;
		Integrate the ODE $\dot{w}  = g$ for  half time step $\frac{1}{2} \Delta t$\;
		Integrate the ODE $\dot{V}_m  = \frac{1}{C_m } I_{ion}$ for  half time step $\frac{1}{2} \Delta t_p$\;
		\uIf{active response is considered}
		{
			Compute time-step size $\Delta t_m$  and choose the time step $\Delta t = \text{min}\left(\Delta t_p, \Delta t_m \right)$\;
			Integrate the ODE $\dot{T_a} = f$ with the QSS method for a time step $\frac{1}{2} \Delta t$\;
			Compute the active first Piola-Kirchhoff stress\;
			Integrate the elastic equations\;
		}
	}
	Terminate the simulation\;
	\caption{Algorithms of the newly proposed SPH framework for the simulation of cardiac functions.}
	\label{al:algorithm1}
\end{algorithm}

%
%
\end{document}